\documentclass[a4paper,11pt]{article}
\usepackage{jcappub}
\usepackage{amsmath, amssymb}
\usepackage{graphicx}
\usepackage{xcolor} 
\usepackage{subcaption}
\usepackage{graphicx}
\usepackage{multirow}
\usepackage{subcaption}
\usepackage{pgffor}
\usepackage{booktabs}
\usepackage{enumitem}
\usepackage{array}
\usepackage{hyperref}
\newlength{\imgcolwidth}
\newcommand{\graphicwidth}{0.24\textwidth}

\title{Deep Learning for CMB Foreground Removal and Beam Deconvolution: A U-Net GAN Approach}

\author[a]{Obasho M,}
\affiliation[a]{School of Physical Sciences, Indian Institute of Technology Mandi, India}
\emailAdd{obashosolitude@gmail.com}
\author[a,b]{Shambhavi Jaiswal,}
\affiliation[b]{Department of Physics, Indian Institute of Technology Delhi, India}
\emailAdd{sjaiswal.phy@gmail.com}
\author[c]{Santanu Das,}
\affiliation[c]{Oridigm Inc., 43575 Mission Blvd  716, Fremont, CA 94539, USA}
\emailAdd{santanu@cosmocommunity.in}
\author[a]{Krishna Mohan Parattu}
\emailAdd{krishna@iitmandi.ac.in}

\abstract{
Extracting cosmological information from microwave sky observations requires accurate estimation of the underlying Cosmic Microwave Background (CMB) by removing foreground contamination, instrumental noise, and the effects of beam convolution. In this work, we develop a machine learning–based approach for CMB reconstruction using a generative adversarial network (GAN) architecture, where the generator is modeled as a U-Net–based convolutional neural network. To train the network, we generate realistic microwave sky maps by simulating Planck-like observations: scanning HEALPix-simulated skies with real Planck beam profile, actual scan patterns, and anisotropic noise consistent with Planck data. Our method achieves high-fidelity reconstruction, with the difference between the input and recovered maps being less than $1\%$ (approximately $2\mu\mathrm{K}$ for temperature and less than $0.5\mu\mathrm{K}$ for polarization) outside the Galactic region. Even within the Galactic plane, the reconstruction error stays below $2$–$3\%$ for temperature maps across most regions, and is even smaller for polarization, apart from a few isolated pixels.. Most importantly, we demonstrate, for the first time, that a GAN-based method can effectively correct for foreground contamination, the systematic effects of non-circular beams and the asymmetric Planck scan pattern for both T and E-mode skymaps. Our results demonstrate the effectiveness of our method for robust and accurate recovery of the CMB signal, even in the presence of strong astrophysical foregrounds and instrumental systematics.
}


\begin{document}
\maketitle
\flushbottom

\section{Introduction}

Among the various tools that we have at our disposal for understanding the early evolution and current composition of our universe, probably none has been as successful as observations of the CMB \cite{durrer2020cosmic,Planck:2018nkj}. This faint afterglow of the hot early phase of our universe allows us to probe the universe roughly 380,000 years after the Big Bang, offering invaluable insights into its composition, evolution, and the seeds of cosmic structure. Precise measurements of CMB temperature and polarization anisotropies enable us to constrain cosmological parameters, search for signatures of primordial gravitational waves, and test key cosmological paradigms, including inflationary models and dark matter properties. However, the observed CMB signal is heavily contaminated by astrophysical foreground emissions, including synchrotron, free-free, and thermal dust radiation, as well as instrumental effects such as the telescope’s non-ideal beam response, complex scanning strategies, instrumental noise, etc \cite{Planck:2018nkj}. These contaminants, spanning a wide range of frequencies and spatial scales, introduce significant challenges in accurately reconstructing the intrinsic CMB signal. 

The spectral characteristics of galactic foreground emissions differ markedly from those of the intrinsic CMB signal. This distinction serves as the foundation for separating foreground contamination from the CMB. Traditional component separation methods, such as Internal Linear Combination (ILC)~\cite{1992ApJ...396L...7B,Eriksen:2004jg,ILC}, its extension Needlet ILC~\cite{NILC}, and various Bayesian approaches~\cite{colombo2023beyondplanck,grumitt2020hierarchical}, have been widely employed for foreground cleaning. ILC methods minimize variance in multi-frequency maps while preserving the CMB signal. While effective to some extent, these techniques often struggle with the complex, non-Gaussian nature of foregrounds, which violates assumptions made by many statistical techniques, and require detailed prior knowledge of their statistical properties. Furthermore, the combined effects of the telescope's beam and its complex scanning strategy introduce significant spatial distortions and noise correlations, further complicating CMB map reconstruction. Correcting for these effects using conventional deconvolution techniques can be computationally expensive and prone to systematics~\cite{keihanen2017application}.

In recent years, machine learning (ML) has emerged as a powerful tool for astrophysical data analysis \cite{Kembhavi:2022,KPSingh:2022}, particularly in problems involving high-dimensional and non-linear features. Deep learning architectures, such as convolutional neural networks (CNNs) and U-Nets, have demonstrated remarkable success in image denoising and pattern recognition tasks. A graph-based Bayesian convolutional neural network based on the U-Net architecture can predict cleaned CMB with pixel-wise uncertainty estimates. The U-Net, with its unique encoder-decoder structure and skip connections, excels at capturing both local and global features, making it ideally suited for the complex task of disentangling foregrounds from the CMB~\cite{Krachmalnicoff_2021,Petroff_2020,Aylor2020,Adak:2025iyj}. By leveraging this capability, U-Nets can effectively learn the complex mapping between contaminated and clean CMB maps, simultaneously mitigating foreground contamination and correcting for beam distortions.

This work presents a Python-based pipeline where we explore generative adversarial network (GAN)-based approach built upon a U-Net architecture for the simultaneous removal of foreground contamination and deconvolution of beam effects in CMB temperature and polarization maps. We begin by generating realistic CMB simulations incorporating the non-circular beam and complex scanning patterns of the Planck satellite. These simulations are further contaminated with realistic foreground emissions and instrumental noise to create synthetic observations that closely resemble real CMB data. The U-Net model is then trained to reconstruct the clean CMB signal from these contaminated maps. Preliminary results indicate that our approach successfully suppresses foregrounds and mitigates beam distortions while preserving the crucial statistical properties of the CMB, and retrieving the original CMB map signal with high accuracy.

To simulate realistic foregrounds, we utilize the Python Sky Model (PySM), a widely used tool for generating full-sky simulations of Galactic emissions in intensity and polarization~\cite{pysm}. PySM allows us to model various components, including thermal dust, synchrotron, and free-free emission, using publicly available data and phenomenological models. By incorporating PySM, our simulations capture the complexity of the foregrounds, providing a more realistic test for our U-Net-based cleaning method.

While several CMB foreground cleaning methods exist, each comes with its own limitations and systematic biases. A comparative analysis of results from multiple independent techniques, including our U-Net approach, is crucial for validating inferences and quantifying systematic uncertainties, ultimately strengthening the reliability of cosmological inferences. Our analysis demonstrates that the GAN-based model not only suppresses foreground contamination and instrumental noise, but also performs effective beam deconvolution. In particular, it corrects for the statistical isotropy (SI) violation caused by noncircular beams and the asymmetric scan pattern—systematic effects that conventional foreground-cleaning methods are fundamentally unable to remove—thereby yielding a more accurate recovery of the true CMB sky.
At present we have explored our analysis only on the CMB temperature and E-mode polarization skymaps. In future, this method can be extended to B-mode polarization studies and applied to other cosmological datasets, such as HI intensity mapping etc. The computational efficiency of deep learning approaches also makes them attractive for large-scale cosmological surveys, where traditional methods may be prohibitively expensive.

This paper has been organized as follows. In Section~\ref{Sec:SimulationPipeline}, we describe the simulation pipeline used to generate the realizations of the observed sky maps. These realizations are later used to train our network. Section~\ref{Section3:NeuralNetwork} provides details of the neural network that we use for the foreground removal. The results from our analysis are presented in Section~\ref{sec:results-discussion}. We highlight the potential of this approach to advance CMB data analysis and also discuss challenges and limitations in Section~\ref{sec:discussion}. The final section, Section~\ref{sec:conclusion}, presents our conclusions.


\section{Simulation Pipeline for Generating Observed CMB Sky Maps}
\label{Sec:SimulationPipeline}

Training neural networks demands large and diverse datasets. However, only a single realization of the CMB sky is available to us, making direct data-driven training infeasible. 
To address this, we develop a robust simulation pipeline. The objective is to generate synthetic CMB sky maps that closely resemble real observational data, including astrophysical foregrounds, instrumental noise, and beam effects. This provides us with a realistic and controlled environment for training the neural network and rigorously testing the performance of its cleaning and deconvolution methods.

\subsection{Sky Model and Foregrounds}
\label{sec:2.1}
Our simulation pipeline for CMB map generation proceeds in several interconnected stages designed to replicate the complexities of real observational data. We begin by using CAMB~\cite{Lewis:1999bs,Howlett:2012mh}\footnote{\url{https://camb.info/}} to compute the theoretical angular power spectrum of the CMB, from which we generate multiple realizations of the CMB sky using HEALPix~\cite{Gorski:2004by}\footnote{\url{https://healpix.sourceforge.io/}}.
In addition to the temperature anisotropy spectrum $C_{\ell}^{TT}$, the polarization spectra $C_{\ell}^{EE}$ is also used to generate the corresponding Stokes $Q$ and $U$ maps. In order to simplify the validation of the polarization response in our simulations, we set the $C_{\ell}^{BB}$ and $C_{\ell}^{EB}$ spectrum to zero.

From Stokes $Q$ and $U$ maps, we construct the scalar $E$-mode polarization maps by performing a spin-2 spherical harmonic decomposition of the polarization field. This transformation is implemented using the \texttt{map2alm} routine from the HEALPix package, which takes as input the Stokes maps $\{I, Q, U\}$ and computes the corresponding harmonic coefficients $(a_{\ell m}^{T}, a_{\ell m}^{E}, a_{\ell m}^{B})$.

The $E$-mode maps are then obtained by transforming $a_{\ell m}^{E}$ back to real space using \texttt{alm2map}. In our simulations, the $B$-mode power spectrum is set to zero ($C_\ell^{BB} = 0$), so only the $E$-mode component contributes to the polarization signal. Same procedure applied to foreground maps as well.

To mimic foreground contamination present in actual CMB observations, we simulate astrophysical foreground emissions using the Python Sky Model (PySM)~\cite{Panexp_2025,Zonca_2021,pysm}~\footnote{\url{https://github.com/galsci/pysm}}. PySM generates realistic, frequency-dependent emission maps across six frequency bands \--- 70, 100, 143, 217, 353, and 545 GHz \--- capturing the key foreground components, namely thermal dust emission, synchrotron radiation, and free-free (bremsstrahlung) emission. These foregrounds vary both spatially and spectrally, introducing non-trivial structure that complicates CMB signal recovery.
For each frequency, multiple sets of foreground maps are generated in the Stokes I, Q, and U components to account for both intensity and polarization of the emission.
We combine the synthetic CMB signal with these foregrounds, and create multi-frequency observations that closely resemble the real CMB sky. 

\begin{figure}
    \centering
    \includegraphics[width=0.9\linewidth, trim={150pt 180pt 380pt 140pt}, clip]{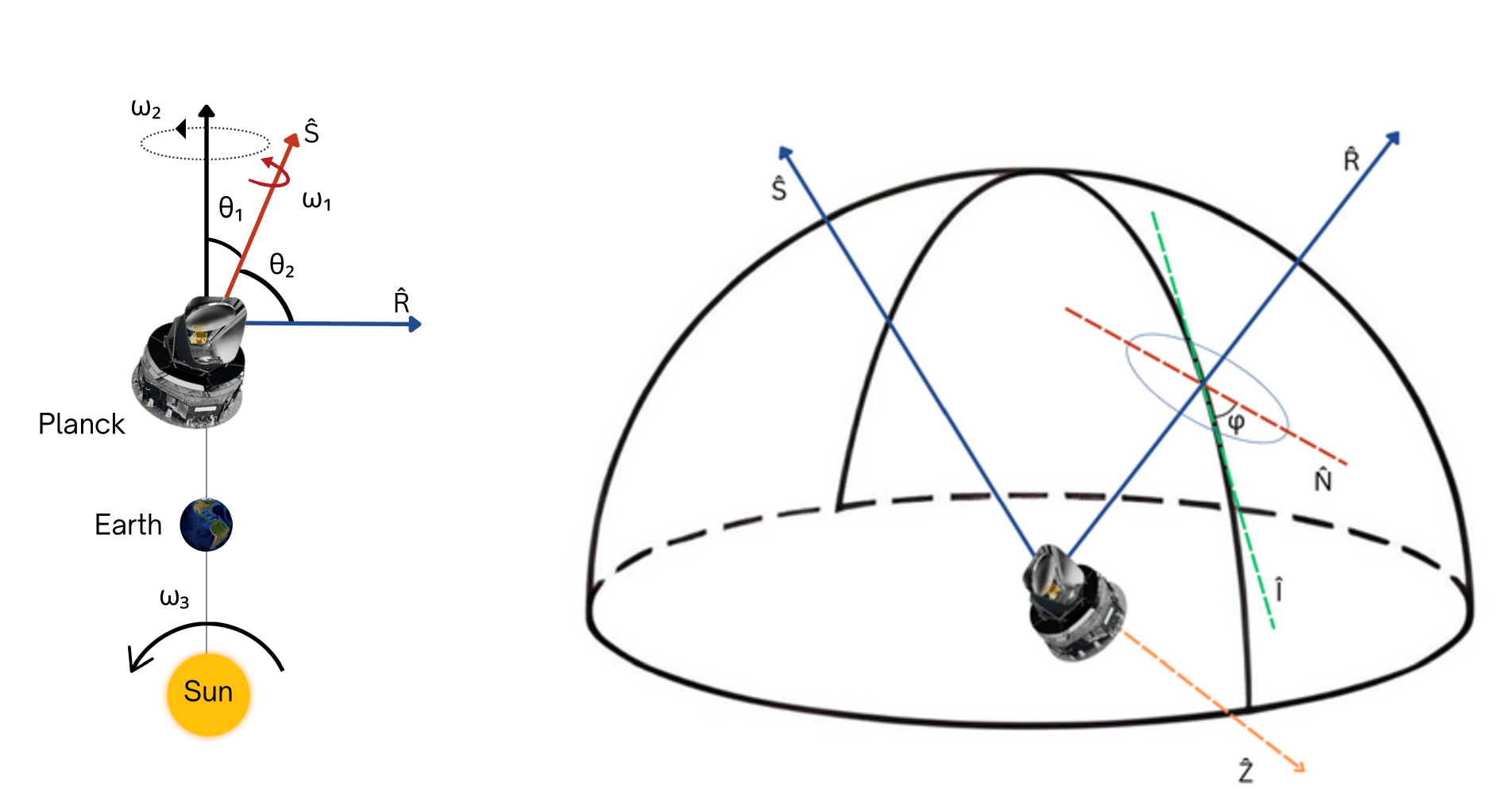}\caption{Illustration of Planck satellite scanning geometry. (left) Blue arrow indicates line of sight of the beam, red arrow the satellite's symmetry axis,
    and black arrow the Sun–Planck direction. Note that the axes do not all lie in the same plane. (right)  We define $\hat{Z} = \hat{R} \times \hat{S}$, perpendicular to both the beam and spin axes, and
 $\hat{I} = \hat{R} \times \hat{Z}$, orthogonal to $\hat{R}$ in the beam scan plane. The axis $\hat{N}$ lies in the $\hat{I}\hat{Z}$ plane perpendicular to the beam direction $\hat{R}$, with orientation angle $\varphi$ measured from $\hat{I}$.} 
    \label{fig:planck_scan}
\end{figure}

\subsection{Planck Scan Pattern\label{sec:planckScan}}
The Planck satellite scans the sky using a single beam that directly measures temperature, unlike the WMAP satellite, which performed differential measurements. The beam is positioned approximately $85^\circ$ off the satellite’s symmetry axis, with a precession angle of around $7.5^\circ$. The satellite spins at a rate of $180^\circ$/min, and its precession traces a complete revolution over six months. The scan pattern for the Planck scan~\cite{Planck_prelaunch_2010,{Das:2013nfa,Das:2012ft}} is shown in Figure~\ref{fig:planck_scan}.

The key angular parameters and rotational velocities involved in the scan geometry (see Figure~\ref{fig:planck_scan}) are
\begin{itemize}[label={}, noitemsep, topsep=0pt]
\item $\theta_1 = 7.5^\circ$ \hfill (precession angle)
\item $\theta_2 = 85^\circ$ \hfill (beam tilt from the spin axis)
\item $\omega_1 = 2\pi$ rad/min \hfill (spin rate)
\item $\omega_2 = 2\omega_3 = 2.3908 \times 10^{-5}$ rad/min \hfill (precession rate of spin axis)
\item $\omega_3 = 1.1954 \times 10^{-5}$ rad/min \hfill (annual orbit rate)
\end{itemize}
Due to this scanning configuration, pixels near the ecliptic poles are visited more frequently than those near the equator, resulting in non-uniform sky coverage~\cite{Planck_Early_2011,{Das:2013nfa,Das:2012ft}}.

Let $\hat{R}(t)$ denote the line-of-sight (LOS) direction of the beam in the ecliptic coordinate system. It is expressed as a time-dependent composite rotation:
\begin{equation}
\hat{R}(t) = R_z(\omega_3 t) \, R_x(\omega_2 t) \, R_y(\theta_1)
\begin{bmatrix}
\cos \theta_2 \\
\sin \theta_2 \cos(\omega_1 t) \\
\sin \theta_2 \sin(\omega_1 t)
\end{bmatrix},
\end{equation}

\noindent Here, $R_y(\theta_1)$, $R_x(\omega_2 t)$, and $R_z(\omega_3 t)$ are rotation matrices representing precession, spin, and orbital motion, respectively. Note that the subscripts $x$, $y$, and $z$ refer to the axes of the local coordinate system. In the context of the $3 \times 3$ rotation matrix, these subscripts indicate the axis about which the rotation is performed, i.e. they are given by


\begin{gather}
R_z(\omega_3 t) =
\begin{bmatrix}
\cos \omega_3 t & \sin \omega_3 t & 0 \\
-\sin \omega_3 t & \cos \omega_3 t & 0 \\
0 & 0 & 1
\end{bmatrix}, \qquad\qquad
R_x(\omega_2 t) =
\begin{bmatrix}
1 & 0 & 0 \\
0 & \cos \omega_2 t & \sin \omega_2 t \\
0 & -\sin \omega_2 t & \cos \omega_2 t
\end{bmatrix},  \nonumber\\[6pt]
R_y(\theta_2) =
\begin{bmatrix}
\cos \theta_1 & 0 & \sin \theta_1 \\
0 & 1 & 0 \\
-\sin \theta_1 & 0 & \cos \theta_1
\end{bmatrix}.
\end{gather}

The satellite’s symmetry axis, $\hat{S}(t)$, is defined by rotating a reference vector through the same transformation without the final spherical component:
\begin{equation}
\hat{S}(t) = R_z(\omega_3 t) \, R_x(\omega_2 t) \, R_y(\theta_1)
\begin{bmatrix}
1 \\
0 \\
0
\end{bmatrix} \,. 
\end{equation}

To estimate the scanning duration per pixel, we consider the full sky solid angle $4\pi$ steradians divided among $N_{\text{pix}}$ pixels. Each pixel subtends:
\begin{equation}
A_{\text{pix}} = \frac{4\pi}{N_{\text{pix}}} \,.
\end{equation}

\noindent Assuming uniform beam motion, the time spent on a pixel is approximately:
\begin{equation}
\text{Scan time} = \frac{1}{\omega_1} \sqrt{\frac{4\pi}{N_{\text{pix}}}}\,.
\end{equation}

We compute $\hat{R}(t)$ at intervals of this scan time, and then convert each $\hat{R}(t)$ vector into a corresponding HEALPix pixel index. By accumulating the number of scans per pixel over a full year, we generate a simulated hit-count map representing the Planck scanning strategy.

\begin{figure}
    \centering
    \includegraphics[width=0.9\linewidth]{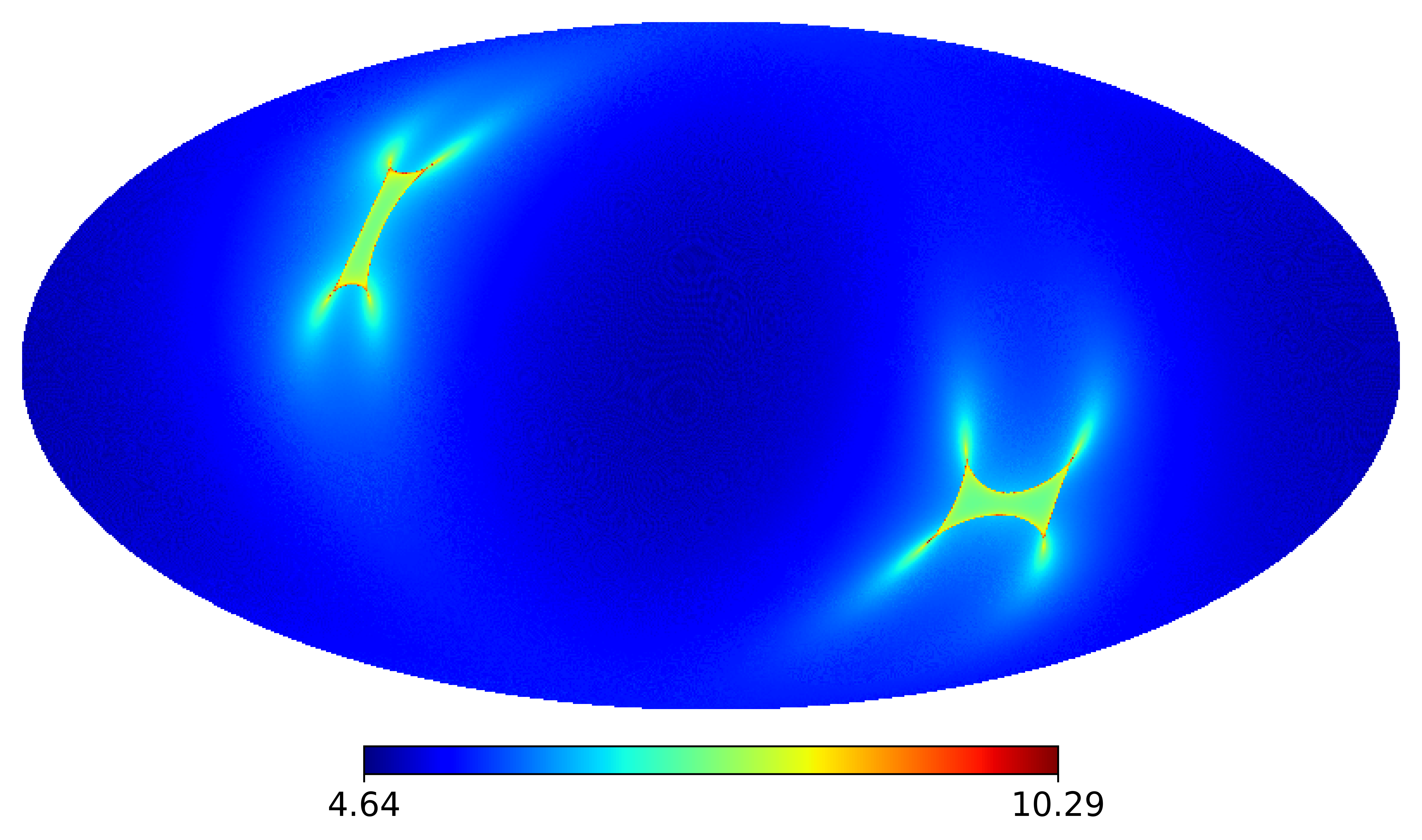}
    \caption{Logarithmic map of the number of observations per pixel over a 12-month Planck scan in the Galactic coordinate system. The scan strategy results in significantly higher hit counts near the ecliptic poles compared to the equatorial regions, leading to strongly non-uniform sky coverage.}

    \label{fig:hitcount}
\end{figure}

As discussed, this scanning geometry leads to more frequent observations near the ecliptic poles. Figure~\ref{fig:hitcount} shows the resulting hit-count map on a logarithmic scale from a 12-month Planck scan, displayed in galactic coordinates. This hit-count map is later used to compute the spatially varying variance of the instrumental noise.


\subsection{Beam Convolution and Polarized Detector Response \label{sec:beamConv}}
\subsubsection{Temperature Convolution \label{sec:TempbeamConv}}
The observed skymap is a convolution of the intrinsic sky with the beam function. 

\begin{equation}
    T_{\mathrm{obs}}(\hat{n}) = \int B(\hat{n},\hat{n}') T(\hat{n}') d\Omega_{\hat{n}'} + T_n(\hat{n})
\end{equation}

\noindent Here $T(\hat{n}')$ is the intrinsic sky temperature.  \( B(\hat{n}, \hat{n}') \) denotes the beam sensitivity between the central direction \( \hat{n} \) and another direction \( \hat{n}' \). $T_n(\hat{n})$ is the instrumental noise. 

To simulate the observational response from the Planck satellite, we convolved the HEALPix maps with the real Planck beams and scan pattern. Our analysis accounts for both the asymmetric shape of the instrument beam and its orientation at each sky location. 

At each pointing direction \( \hat{n} \), the observed temperature \( T_{\mathrm{obs}}(\hat{n}) \) is computed as a weighted average over neighboring pixel temperatures, using a direction-dependent beam response function as the weighting kernel:
\begin{equation}
T^s_{\mathrm{obs}}(\hat{n}) = \frac{\sum_{i\in \mathcal{S}} B^s(\hat{n}, \hat{n}'_i)\, T(\hat{n}'_i)}{\sum_{i\in \mathcal{S}} B^s(\hat{n}, \hat{n}'_i)} \,,
\end{equation}

\noindent where $\hat{n}$ is the pointing direction, $\hat{n}'_i$ is the central direction of the $i$th pixel and $\mathcal{S}$ is the set of all the pixels for which $B(\hat{n}, \hat{n}'_i)$ is non-zero. In the previous section we discussed that during the Planck scan each of the pixels get scanned multiple number of times. Here the superscript $T^s_{\mathrm{obs}}(\hat{n})$ indicates the observed temperature of a pixel along $\hat{n}$ direction during $s$th scan (note that $\hat{n}$ can be any direction within the pixel and does not necessarily correspond to the direction of the pixel center). As each pixel is scanned multiple times, the final observed sky temperature for a pixel centered at direction $\hat{n}_j$ can be obtained by averaging the temperature values measured along all the directions that fall within that pixel across different scans, i.e. $T_{\mathrm{obs}}(\hat{n}_j) =\frac{1}{h_j} \sum_sT^s_{\mathrm{obs}}(\hat{n})$. Here $h_j$ is the hitcount of the pixel centered along $\hat{n}_j$, and $\hat{n}$ represents all the directions that fall on that pixel. We can rearrange the above equation to get 

\begin{equation}
T_{\mathrm{obs}}(\hat{n}_j) = \sum_{i\in \mathcal{S}}\left[\frac{1}{h_j} \sum_s \frac{ B^s(\hat{n}, \hat{n}'_i)\, }{\sum_{i\in \mathcal{S}} B^s(\hat{n}, \hat{n}'_i)} \right]T(\hat{n}'_i) = \sum_{i\in \mathcal{S}}\widetilde{B}(\hat{n}_j, \hat{n}'_i) T(\hat{n}'_i)\,,
\end{equation}

\noindent Here $\widetilde{B}(\hat{n}_j, \hat{n}'_i)$ is an effective beam function. Since different pixels are scanned a varying number of times and from different orientations, the effective beam function differs from pixel to pixel. Generating thousands of independently scanned maps for training the network would be computationally expensive and time-consuming. To address this, we precompute and store the effective beam at each pixel. This effective beam is then multiplied with maps to obtain the scanned skymaps. 


For generating the effective beam, we need to know the orientation of the beam at every time step. To define the beam orientation in the local sky patch, we construct an orthonormal coordinate system at each time step using the scan and beam directions (refer to Figure~\ref{fig:planck_scan}):

\begin{itemize}[label={}, noitemsep, topsep=0pt]
\item $\hat{Z} = \hat{R} \times \hat{S}$ \hfill (perpendicular to both beam and spin axis)
\item $\hat{I} = \hat{R} \times \hat{Z}$ \hfill (orthogonal to $\hat{R}$ in the beam scan plane)
\end{itemize}

\noindent The $\hat{I} - \hat{Z}$ plane is a plane perpendicular to the beam pointing direction and is the plane on which the  beam is located.  The scan axis of the non-circular beam, $\hat{N}$, is given by:
\begin{equation}
\hat{N} = \cos \varphi \, \hat{I} + \sin \varphi \, \hat{Z} \,.
\end{equation}

\noindent where $\varphi$ is the angle of the scan axis of the beam relative to the $\hat{I}$ axis.\\

To evaluate \( B(\hat{n}_j, \hat{n}_i') \), we project the angular displacement between the central pixel and its neighbors into a local 2D Cartesian coordinate system defined by the scan orientation at each time step. For a given observation time, the line-of-sight (LOS) direction of the beam is represented by the unit vector \( \vec{R}(t) \), and the satellite’s spin axis is denoted by \( \vec{S}(t) \). These vectors define a local orthonormal frame, which is used to orient the beam response function according to the current scan geometry. In our implementation, we fix the beam orientation angle \( \varphi = 0 \), so that the effective scan-aligned beam direction \( \vec{N} \) is aligned with \( \vec{I} \), simplifying the coordinate transformation. 

For each central pixel, we identify neighboring pixels within a radius of 99 arcminutes using the \texttt{query\_disc} function from the HEALPix library (After 99 arc min most of the pixels have 0 values). This typically results in approximately 2800 neighboring pixels . For each neighbor \( i \), we compute the angular separation \( \theta_i \) between the pointing direction ($\hat{n}$) and the neighbor ($\hat{n}_i$), as well as the angle \( \alpha_i \) between the scan direction \( \vec{N} \) and the great-circle arc connecting the two pixels. These spherical coordinates are then converted to Cartesian offsets as
\begin{equation}
x_i = \theta_i \cos\alpha_i, \quad y_i = \theta_i \sin\alpha_i \,.
 \label{eqn:xy}
\end{equation}

\begin{figure}
    \centering
    \includegraphics[width=0.9\linewidth]{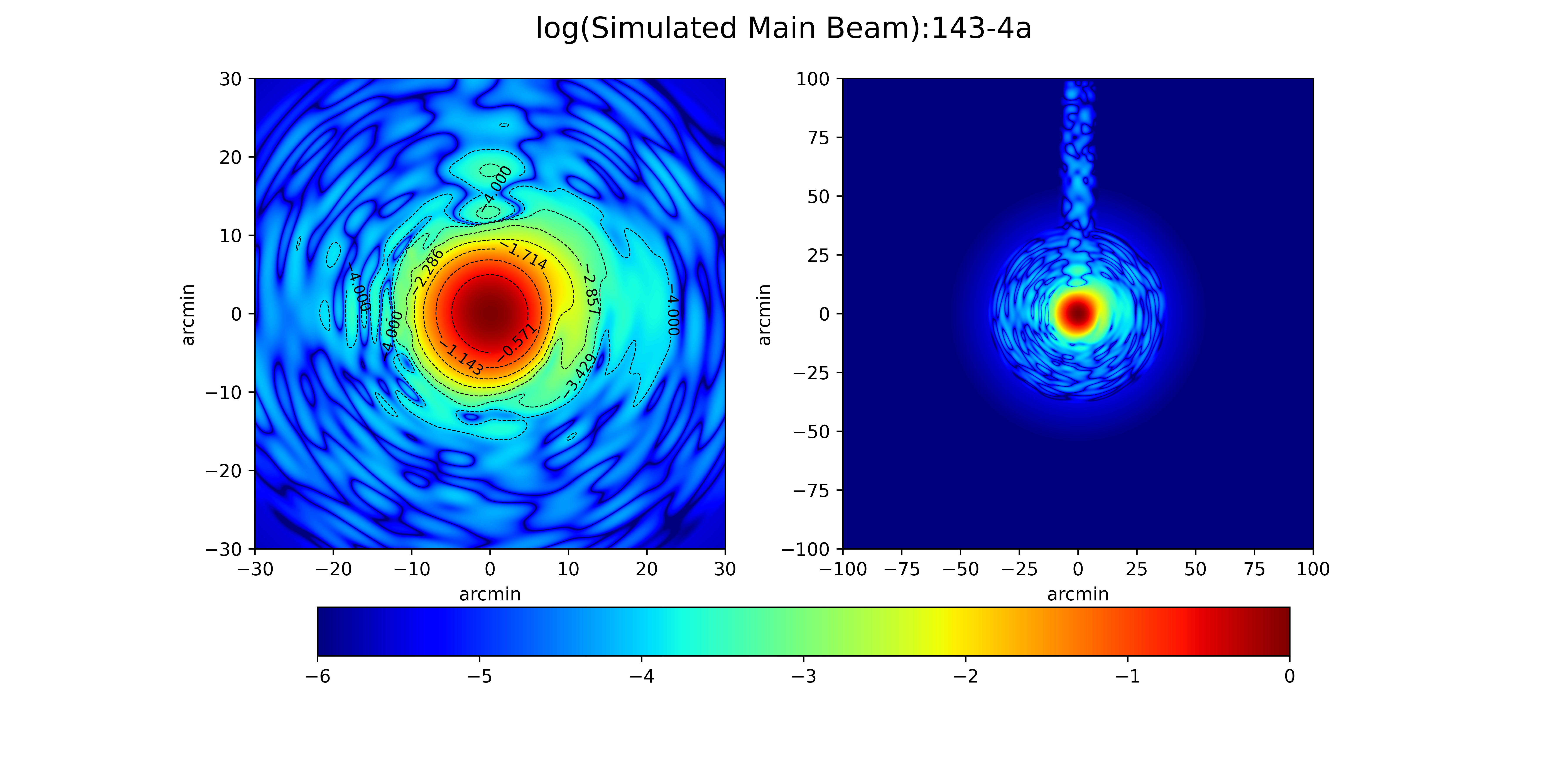}
    \caption{Logarithmic intensity plots of the simulated main beam for the 143 GHz Planck channel (beam 143-4a). The left panel shows the central beam region with fine structure, while the right panel displays a wider field including far sidelobes.}

    \label{fig:beam}
\end{figure}

The coordinates \( (x_i, y_i) \) are mapped to discrete indices on a precomputed 2D beam grid \( B(x, y) \), which is centered on the beam and typically spans a \( 100' \times 100' \) region with 2 arcsecond resolution. This beam map, shown in Figure \ref{fig:beam}, derived from Planck's measured scan beams (e.g., \texttt{HFI\_ScanBeam\_143-1a\_R2.00.fits}), encodes the instrument’s angular sensitivity as a function of beam-centered displacement. The corresponding value \( B(x_i, y_i) \) is then used for calculating  the effective beam matrix, $\widetilde{B}(\hat{n}_j,\hat{n}'_i)$. 
This effective beam is then used for convolving all the skymaps. 

While the above formulation describes the temperature response of the instrument, many Planck detectors are also sensitive to linear polarization. The detector response must therefore be extended to include the Stokes parameters $Q$ and $U$.

\subsubsection{Polarization Response}
In addition to temperature anisotropies, several channels of the Planck High Frequency Instrument also measure the linear polarization of the microwave sky. The polarization state of the radiation field is described by the Stokes parameters $Q(\hat{n})$ and $U(\hat{n})$, which represent the amplitude and orientation of the linear polarization. Unlike temperature, which is a scalar field on the sphere, $Q$ and $U$ depend on the choice of coordinate axes in the local tangent plane.

A polarization–sensitive detector measures the projection of the incident radiation along its polarization-sensitive axis. The signal recorded by a detector pointing toward the direction $\hat{n}$ can therefore be written as

\begin{equation}
d(\hat{n}) =\left[ I(\hat{n}) , \, \rho  Q(\hat{n}) \cos(2\psi_t) ,  \, \rho U(\hat{n}) \sin(2\psi_t) \right],
\end{equation}

\noindent where $I(\hat{n})$ is the total intensity, $Q(\hat{n})$ and $U(\hat{n})$ are the Stokes parameters, $\rho$ is the polarization efficiency of the detector, and $\psi_t$ is the effective polarization orientation of the detector projected onto the sky at time $t$. In the Planck instrument model, the polarization efficiency is related to the polarization leakage parameter $\epsilon$ through $\rho = 1 - \epsilon$. The intrinsic detector polarization angles $\psi_{\mathrm{pol}}$ and leakage parameters $\epsilon$ used in our simulations are obtained from the Planck HFI Reduced Instrument Model (RIMO).

For the LFI 70\,GHz channel, the Planck beam model directly provides Stokes beam components, which describe the full polarized beam response.

The polarization orientation $\psi_t$ varies during the scan as the spacecraft rotates. In our implementation, following the FEBeCoP formalism, $\psi_t$ is obtained by projecting the detector polarization direction onto the local sky basis at each pointing direction~\cite{Mitra_2011}. The detector polarization direction in the local beam frame is given by:
\begin{equation}
\vec{P} = \cos(\psi_{\mathrm{pol}})\,\vec{I} + \sin(\psi_{\mathrm{pol}})\,\vec{Z},
\end{equation}

\noindent where $\vec{I}$ and $\vec{Z}$ are the local orthonormal basis vectors defined in Section~\ref{sec:TempbeamConv}. The local spherical basis vectors $(\hat{e}_\theta, \hat{e}_\phi)$ define the tangent plane at the pointing direction $\hat{R}$. These are constructed from the unit pointing vector $\hat{R}$ as
\begin{equation}
\hat{e}_\phi = \frac{\hat{z} \times \hat{R}}{|\hat{z} \times \hat{R}|}, \quad
\hat{e}_\theta = \hat{e}_\phi \times \hat{R},
\end{equation}
where $\hat{z}$ is the unit vector along the global $z$-axis. The effective polarization angle on the sky is then computed as
\begin{equation}
\psi_t = \tan^{-1} \left( \frac{\vec{P} \cdot \hat{e}_\phi}{\vec{P} \cdot \hat{e}_\theta} \right),
\end{equation}
This formulation directly provides the physically relevant detector orientation on the sky.

The trigonometric projections $\cos(2\psi_t)$ and $\sin(2\psi_t)$ determine how the Stokes parameters $Q$ and $U$ project onto the detector polarization axis.
To incorporate polarization into the beam convolution, the detector response is applied directly during the effective beam construction. For each neighboring pixel $i$, the beam response weight $B(\hat{n}, \hat{n}'_i)$ is multiplied by the corresponding polarization projection factors, producing three sets of weights corresponding to the temperature and polarization components:

\begin{equation}
\begin{aligned}
B_I(\hat{n}, \hat{n}'_i) &= B(\hat{n}, \hat{n}'_i), \\
B_Q(\hat{n}, \hat{n}'_i) &= \rho \, B(\hat{n}, \hat{n}'_i) \cos(2\psi_t), \\
B_U(\hat{n}, \hat{n}'_i) &= \rho \, B(\hat{n}, \hat{n}'_i) \sin(2\psi_t).
\end{aligned}
\label{eqn:b}
\end{equation}

The observed Stokes parameters are obtained by applying the corresponding effective beam kernels to each component separately:

\begin{equation}
\begin{aligned}
I_{\mathrm{obs}}(\hat{n}) &= \sum_i B_I(\hat{n}, \hat{n}'_i)\, I_i, \\
Q_{\mathrm{obs}}(\hat{n}) &= \sum_i B_Q(\hat{n}, \hat{n}'_i)\, Q_i, \\
U_{\mathrm{obs}}(\hat{n}) &= \sum_i B_U(\hat{n}, \hat{n}'_i)\, U_i.
\end{aligned}
\end{equation}

In our implementation, these three sets of weights are stored separately during the effective beam construction stage and later applied independently to the input $I$, $Q$, and $U$ maps to generate the corresponding observed components. 

The resulting convolution therefore accounts simultaneously for beam asymmetry, scan geometry, and detector polarization response, producing simulated observations that closely reproduce the characteristics of Planck polarization measurements.

The polarization convolution follows the same beam--sampling procedure described in Section~2.3.1. For each central pixel, neighboring pixels within a fixed angular radius are identified using the \texttt{HEALPix} \texttt{query\_disc} routine. For the polarization kernels we restrict the neighbor search to a radius of $50$~arcminutes, which captures the dominant beam response while reducing the computational cost and storage requirements associated with the polarization beam matrices.

The angular displacement between the central pixel and each neighbor is projected onto the local scan-aligned coordinate system to obtain the beam-centered offsets $(x_i, y_i)$. These coordinates are then mapped onto the precomputed beam grid $B(x,y)$, which encodes the instrument response. The resulting beam weight $B(\hat{n}, \hat{n}_i')$ is subsequently multiplied by the polarization projection factors given in Eqs.~(\ref{eqn:b}) to obtain the effective polarization kernels $B_Q$ and $B_U$.
\begin{table}[h]
\centering
\caption{Planck detectors and beam models used in the simulations. The detector polarization angle $\psi_{\mathrm{pol}}$, polarization leakage $\epsilon$, and polarization efficiency $\rho = 1-\epsilon$ are taken from the Planck RIMO. Temperature beams correspond to representative detectors at each frequency, while polarization simulations use the detectors listed below.}
\begin{tabular}{|c|c|c|c|c|c|}
\hline
Frequency (GHz) & Detector (Temp) & Detector (Pol) & $\psi_{\mathrm{pol}}$ (deg) & $\epsilon$ & $\rho = 1-\epsilon$ \\
\hline
70  & 18M & 18M   & --      & --     & --     \\
217 & 217-1  & 217-6b & -44.100 & 0.0236 & 0.9764 \\
353 & 353-1  & 353-3b & -46.100 & 0.0413 & 0.9587 \\
545 & 545-1  & -- & -- & -- & -- \\
\hline
\end{tabular}
\end{table}
\subsection{Generating the final skymaps \label{Sec:NoiseBeam}}

In Sections~\ref{sec:planckScan} and~\ref{sec:beamConv}, we describe how we convolve the real beam to generate realistic simulated skymaps. 
For training and testing our neural network we have generated 2 sets of simulated skymaps: 

\begin{itemize}
    \item Circular beam convolution: Each frequency map is smoothed using a circular Gaussian kernel. This serves as a useful baseline for comparison.
    \item Real beam convolution: Precomputed beam response maps from Planck data are used,  as described in Section~\ref{sec:beamConv}.
\end{itemize}

For the circular simulation, we apply a Gaussian smoothing to each frequency channel using its corresponding full-width at half-maximum (FWHM) value.
This is implemented using the \texttt{healpy.sphtfunc.smoothing} function from the HEALPix package, which smooths a map with a symmetric Gaussian beam. The FWHM and the detector noise values used for each frequency are summarized in Table~\ref{tab:beam_sensitivity}. 

\begin{table}[h]
\centering
\caption{Frequency channels with corresponding Gaussian beam FWHM (used in circular convolution) and noise standard deviation ($\sigma$)}
\label{tab:beam_sensitivity}
\begin{tabular}{|c|c|c|}
\hline
\textbf{Frequency [GHz]} & \textbf{FWHM [arcmin]} & \textbf{Noise $\sigma$ [$\mu K$] } \\
\hline
70  & 13.31 & 1.31 \\
100 & 9.66  & 1.15 \\
143 & 7.27  & 1.78 \\
217 & 5.01  & 1.91 \\
353 & 4.86  & 4.66 \\
545 & 4.84  & 7.99 \\
\hline
\end{tabular}
\end{table}

For the symmetric beam convolution maps we add uncorrelated Gaussian noise (white noise) to the  each pixel. As for the symmetric Gaussian beam convolution we have not considered the effect of the scanning, the noise variance at each pixel are kept the same. 

For the real beam convolution we use frequency-dependent Planck scan beam models corresponding to the detectors listed in Table 1. This allows the simulations to capture the variation of beam shapes across different observing frequencies.
As discussed  earlier Planck satellite follows non-uniform scan strategy, where regions near the ecliptic poles receive significantly more observations than those near the equator (see Figure~\ref{fig:hitcount}). As a result, pixels with high hit counts exhibit lower noise levels.
To account for spatially varying noise, we scale the pixelwise noise variance inversely with the hit count—i.e., the number of times a pixel was observed by the satellite. 

Combined, these noise and beam effects produce simulated maps that closely match the resolution and statistical properties of real Planck data. These simulations are essential for training the neural network and testing CMB 
foreground cleaning methods under conditions that realistically capture observational challenges.

\subsection{Input Data Preparation and Patching}

The full-resolution synthetic sky maps generated by our simulation pipeline are high-dimensional and span the full celestial sphere. However, standard convolutional neural networks are designed for flat Euclidean domains and do not directly support spherical convolutions. Implementing spherical CNNs would require a major code overhaul or the use of specialized frameworks such as DeepSphere~\cite{defferrard2020deepspheregraphbasedsphericalcnn}. To avoid this complexity, we adopt a patching strategy that enables training on planar projections.

Each map spans six frequency channels and is partitioned into 12 HEALPix regions, each represented as a square image of size $n_{side} \times n_{side} $. This enables the neural network to operate on localized sky regions, promoting more efficient training and inference.

The pixel values are indexed and arranged using the HEALPix nested ordering scheme. After patch extraction, the data is reshaped and stored in the ring ordering convention for compatibility with standard HEALPix-based tools. Each patch retains its correspondence to the original sky location and frequency band, ensuring spatial coherence and enabling evaluation of model performance both locally and globally.

This patching strategy enables the use of conventional CNNs without spherical convolution, while still preserving the spatial and multi-frequency structure of the input data.

\section{Neural Network Framework for CMB Map Reconstruction}
\label{Section3:NeuralNetwork}


In this work, we perform separate reconstructions for temperature and polarization. For temperature, the network takes six frequency channels—70, 100, 143, 217, 353, and 545~GHz—as input and reconstructs the CMB intensity map. For polarization, we use three frequency channels—70, 217, and 353~GHz—which have significant polarization sensitivity, and train the network to recover the scalar $E$-mode polarization map derived from the Stokes $Q$ and $U$ components (see Section~\ref{sec:2.1}).

To recover the underlying CMB signal from noisy multi-frequency observations, we adopt a generative adversarial framework consisting of a U-Net–based generator and a convolutional discriminator. The generator produces high-fidelity estimates of the CMB maps, while the discriminator guides the training by assessing the realism of these reconstructions. The architecture and training strategy are described below.

\subsection{GAN Architecture}

Generative Adversarial Networks (GANs), first introduced by Goodfellow et al. \citep{goodfellow2020generative}, are a class of machine learning models that learn to generate data by training two networks in competition: a generator and a discriminator. The generator aims to produce outputs that resemble real data, while the discriminator learns to distinguish between real and synthetic examples.  Through adversarial training, both networks improve iteratively, resulting in increasingly realistic generated outputs.  GANs have found widespread applications in various fields, including image synthesis \citep{karras2019style}, natural language processing \citep{zhang2017adversarial}, and even astrophysics, where they have been used for tasks like weak lensing map generation \citep{mustafa2019cosmogan}. While GANs offer powerful generative capabilities, they can be challenging to train due to issues like mode collapse and instability.

Our approach employs this adversarial framework to guide a U-Net-based generator towards producing CMB maps that are statistically consistent with real observations. Unlike standard GAN architectures that directly generate images from random noise, our generator takes noisy, foreground-contaminated multi-frequency sky maps as input. This allows the generator to learn the complex mapping from observed data to the underlying CMB signal. The generator architecture is based on the U-Net structure, which consists of an encoder and decoder network, connected via skip connections. The encoder progressively downsamples the input, extracting increasingly abstract features. The decoder then reconstructs the high-resolution CMB map, leveraging the features learned by the encoder and the skip connections to preserve fine-grained details.

The convolutional discriminator evaluates the realism of the generator's outputs by comparing them to ground truth CMB maps.  The discriminator acts as a learned loss function, guiding the generator to produce outputs that not only minimize pixel-wise reconstruction error but also adhere to the statistical properties of real CMB data. This interaction is formalized through a minimax objective: the generator aims to create outputs that the discriminator cannot distinguish from real maps, while the discriminator strives to correctly classify real versus generated maps. Through this process, the generator gradually improves to better approximate the true data distribution.

Our architecture leverages the encoder–decoder structure of U-Net for spatially precise reconstruction, while the discriminator enforces global and perceptual realism, resulting in more accurate and physically plausible outputs.

\begin{figure}[h]
    \centering
    \includegraphics[width=0.9\linewidth]{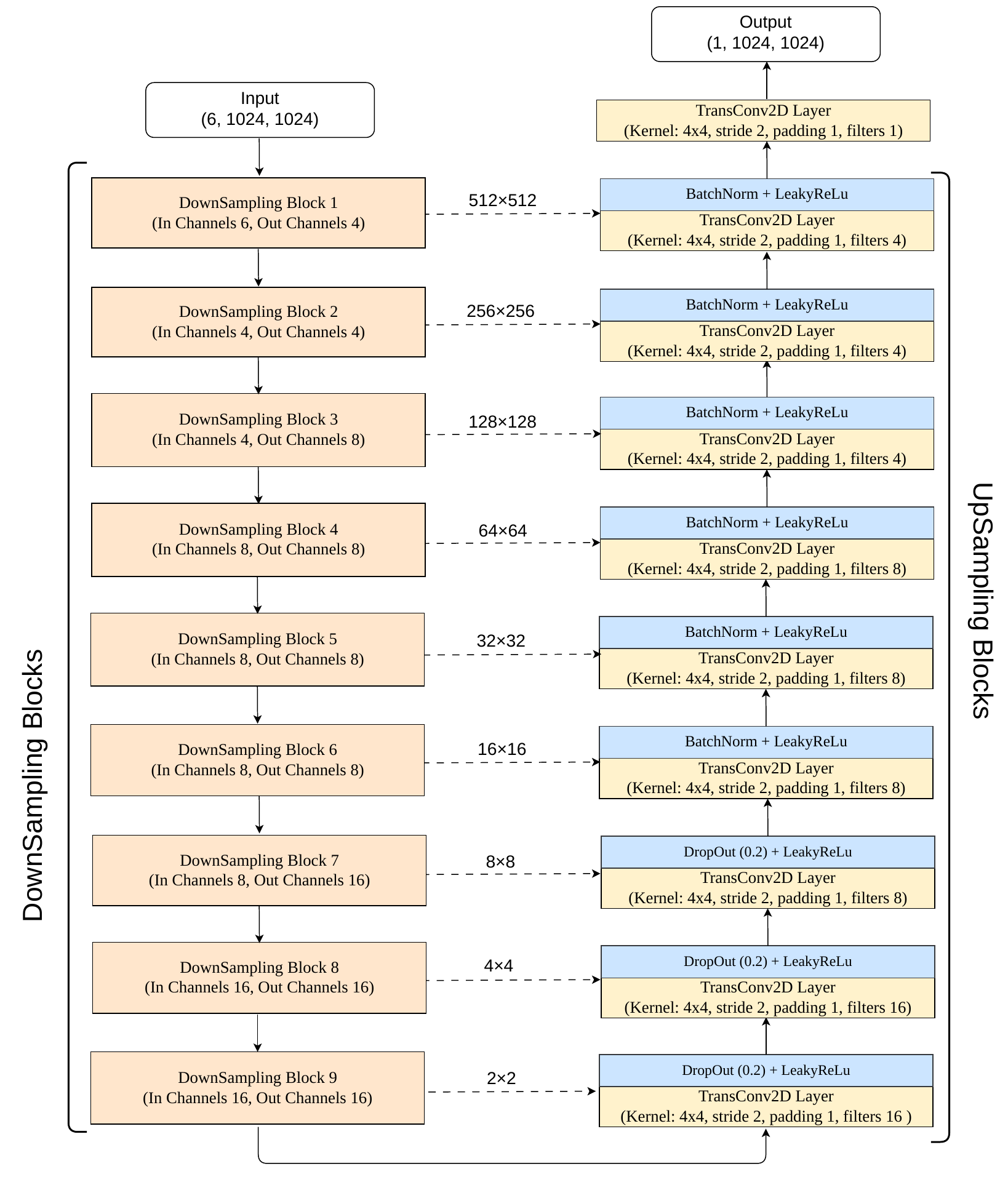}
    \caption{The figure illustrates the architecture of the generator network, which is based on a U-Net structure composed of an encoder and a decoder. Note that for polarization the input map consists of 3 frequency channels}
    \label{fig:unet}
\end{figure}

\subsection{U-Net Architecture for CMB Reconstruction}
The U-Net architecture, first introduced by Ronneberger \textit{et al.} in 2015, employs a symmetric encoder--decoder structure that has proven highly effective for biomedical image segmentation and, more recently, for astrophysical map reconstruction \cite{Ronneberger2015}. The encoder hierarchically extracts spatial features, and the decoder reconstructs the high-resolution outputs by integrating these features via skip connections. In the context of Cosmic Microwave Background (CMB) reconstruction, U-Net--based models excel at denoising and deconvolving observed maps, accurately recovering underlying temperature and polarization fields from noisy, foreground-contaminated inputs \cite{Costanza2024,Yan2023}. Figure ~\ref{fig:unet} shows the overall U-Net architecture that we have used for CMB reconstruction.  This particular U-Net is designed to take 6 frequency maps (3 frequency maps for polarization) with dimensions $n_{side} = 1024$ as input. 

\begin{figure}[h]
    \centering
    \begin{subfigure}[b]{0.48\linewidth}
        \centering
        \includegraphics[width=0.60\linewidth]{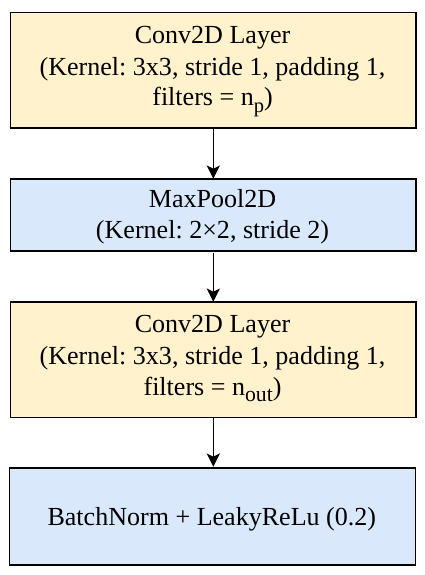}
        \caption{UNET Downsampling Block }
        \label{fig:unet_down}
    \end{subfigure}
    \begin{subfigure}[b]{0.48\linewidth}
        \centering
        \includegraphics[width=0.60\linewidth]{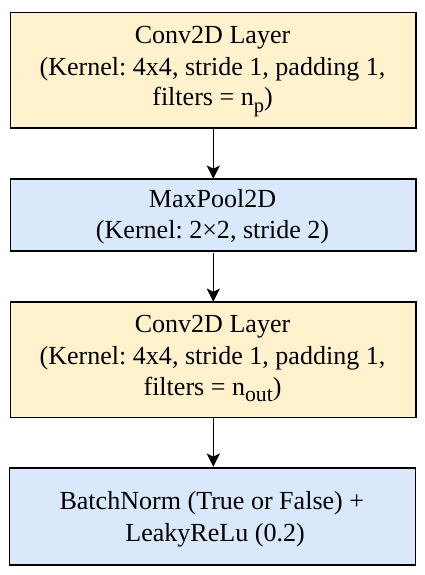}
        \caption{Discriminator Downsampling Block}
        \label{fig:disc_down}
    \end{subfigure}
    \caption{Detailed architecture of Downsampling blocks: (a) UNET and (b) Discriminator }
    \label{fig:down_blocks}
\end{figure}

\subsubsection{Encoder}
The encoder is structured as a sequence of nine down-sampling blocks, each designed to progressively reduce spatial resolution while increasing feature complexity. A schematic diagram of a single down-sampling block is shown in Figure~\ref{fig:unet_down}. 

Each down-sampling block starts with a 2D convolutional layer with a kernel size of $3 \times 3$, a stride of 1, and padding of 1, ensuring that the spatial dimensions remain unchanged. These layers operate with $n_{\text{in}}$ input channels and $n_{\text{p}}$ output channels.  The parameter $n_p$ acts as a control for the model size and is kept constant within each block.  Specifically, the input channels ($n_{in}$) start at 4 in the initial block, increase to 8 in intermediate blocks, and reach 16 in the final blocks. A max-pooling layer with a kernel size of $2 \times 2$ and a stride of 2 is applied immediately after the first convolutional layer, halving the spatial dimensions from $n_{side} \times n_{side}$ to $(n_{side}/2) \times (n_{side}/2)$.

Following max pooling, a second convolutional layer, also with a kernel size of $3 \times 3$, processes the $n_{\text{p}}$ input channels to generate $n_{\text{out}}$ output channels, further refining the feature representations. Batch normalization is applied after the second convolutional layer to standardize feature activations, stabilizing and accelerating training \cite{ioffe2015batch}. Finally, a leaky rectified linear unit (LeakyReLU) activation with a slope of 0.2 introduces nonlinearity, allowing the model to capture complex patterns in the input data \cite{maas2013rectifier}. The outputs of each down-sampling block are stored for skip connections in the decoding stage, ensuring that detailed spatial information from earlier stages is preserved and utilized in reconstruction \cite{Ronneberger2015}.


\subsubsection{Decoder}

The decoder reconstructs the high-resolution output map from the abstract features learned by the encoder. Each layer of the decoder, referred to as an \textit{upsampling block}, performs spatial upsampling while integrating feature information from the encoder through skip connections. 
The process begins with a transpose convolution layer, which increases the spatial dimensions of feature maps. This layer operates with a kernel size of $4 \times 4$, a stride of 2, and padding of 1, effectively doubling the spatial dimensions.  The transpose convolution transforms feature maps from $n_{\text{in}}$ input channels to $n_{\text{out}}$ output channels, where $n_{\text{in}}$ and $n_{\text{out}}$ decrease progressively as the feature map size increases. The architecture of the upsampling blocks are shown in Figure~\ref{fig:unet}.


In the first three upsampling blocks, batch normalization is omitted to preserve delicate spatial features and avoid potential artifacts during early-stage reconstruction. Instead, dropout with a rate of 0.2 is applied to prevent overfitting by randomly deactivating a subset of neurons during training \cite{gal2016dropout}. The remaining subsequent upsampling blocks incorporate batch normalization \cite{bjorck2018understanding} to ensure training stability. Each upsampling block concludes with a ReLU activation function to introduce non-linearity and support the reconstruction of complex spatial patterns \cite{agarap2018deep}.

Skip connections play a pivotal role in the decoder by linking feature maps from corresponding encoder layers to decoder layers. These connections concatenate the encoder features with the upsampled decoder feature maps, ensuring the preservation of fine-grained details. This integration allows the decoder to simultaneously utilize high-level contextual information from deeper layers and low-level spatial details from earlier layers \cite{Ronneberger2015}.

\subsubsection{Final Layer}
The final layer of the U-Net employs a transposed convolution to restore the spatial dimensions of the output map to match those of the input. This layer takes the upsampled feature maps from the final decoder block and produces the final cleaned multi-channel CMB map. It uses a kernel size of $4 \times 4$, stride of 2, and padding of 1, ensuring that the output resolution aligns precisely with the original input.

\subsection{Discriminator Architecture }

\begin{figure}[h!]
    \centering
    \includegraphics[width=0.99\linewidth]{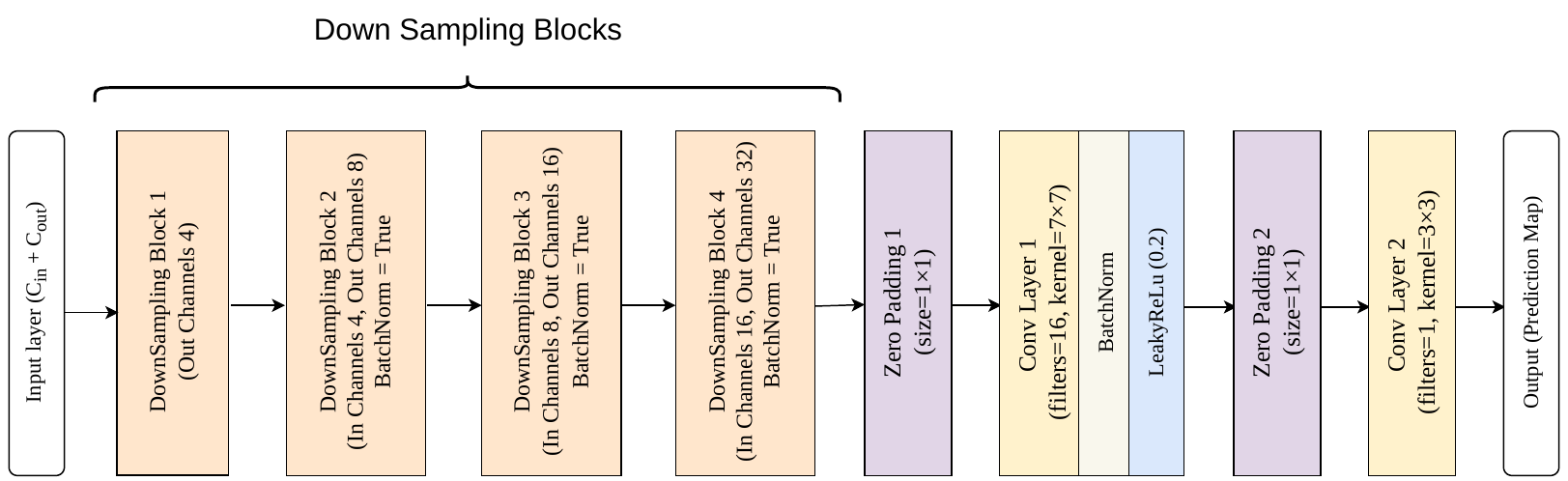}
    \caption{Schematic Diagram of the Discriminator Architecture.  This figure shows the sequence of downsampling blocks, convolutional layers, and zero padding. The input consists of the concatenation of the true and generated CMB maps.}
    \label{fig:discriminator_architecture}
\end{figure}

The discriminator network is designed to differentiate between real and reconstructed CMB maps, acting as a crucial component of the adversarial training framework. Its architecture consists of a series of downsampling blocks, similar to those in the U-Net encoder, followed by convolutional layers that compress input feature maps into a single scalar output representing the discriminator's assessment.  The input to the discriminator is formed by concatenating the true and generated CMB maps along the channel axis. Figure \ref{fig:discriminator_architecture} illustrates the overall structure, showing the sequence of downsampling operations, convolutional layers, and zero-padding stages.


The discriminator utilizes four downsampling blocks that are architecturally identical to the first four downsampling blocks described in the U-Net encoder (see Figure~\ref{fig:disc_down}). These blocks progressively reduce the spatial dimensions and increase the feature channel depth. The first dwnsampling block takes $C_{in}$ + $C_{out}$ input channels—the combined channels of the input and target maps—and subsequent blocks continue increasing the feature depth. Batch normalization is omitted in the first block to preserve low level input features but is applied in the remaining blocks to stabilize training. Each downsampling blocks concludes with a leaky rectified linear unit (LeakyReLU) activation function with a slope of 0.2, introducing non-linearity and aiding in gradient propagation.


Following the downsampling blocks, the output passed through a pair of convolutional layers, each separated by a one-pixel padding stage. The first convolutional layer has a kernel size of $7 \times 7$, a stride of 1, and no padding and maps the 32 input channels into 16 output channels. After this convolution, batch normalization and LeakyReLU are applied. The final convolutional layer uses a kernel size of $3 \times 3$, a stride of 1, and no padding and it maps the 16-channel feature map to a single output channel with shape of  $64\times 64$. This   output serves as the discriminator’s final judgment on whether diffrent portions of the input map is real or generated.

\begin{figure}[h!]
    \centering
    \includegraphics[width=1\columnwidth]{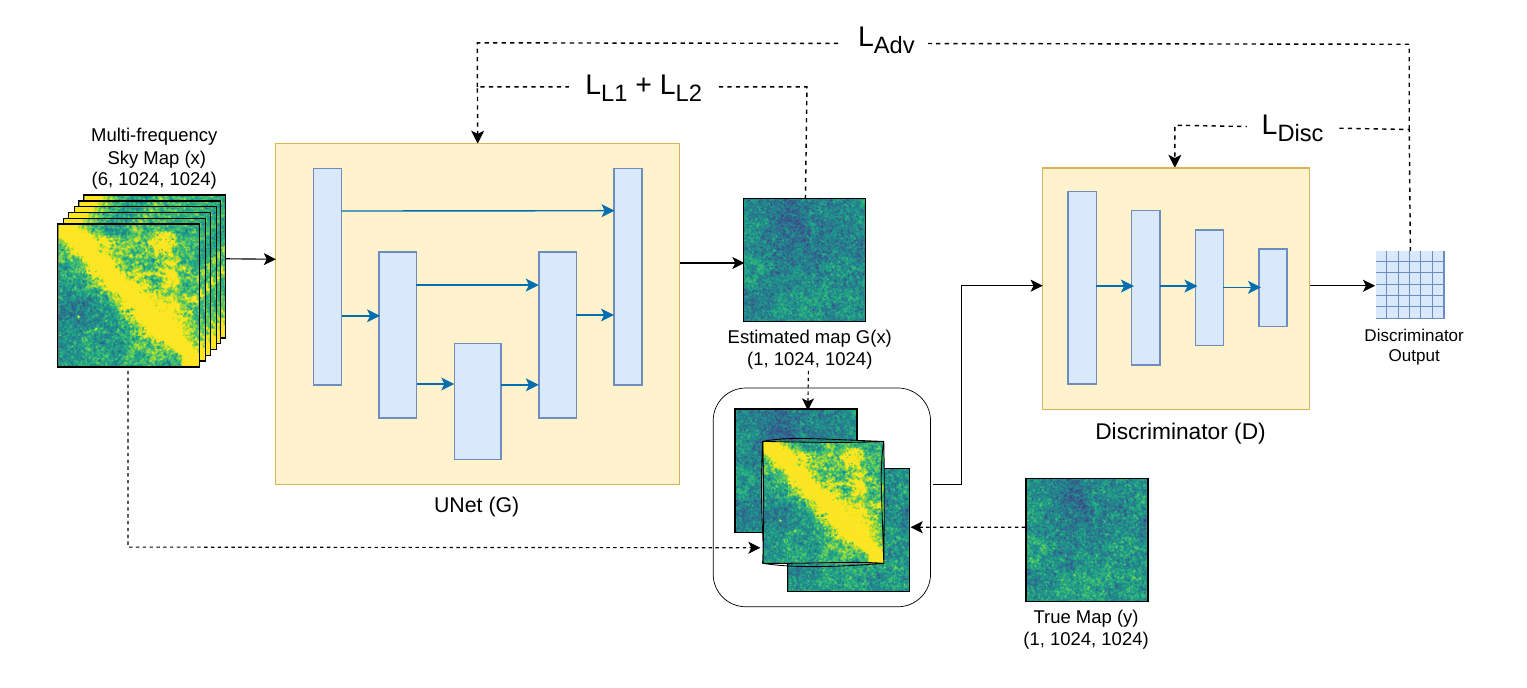}
    \caption{The figure provides an overview of the GAN architecture employed in our analysis. The real input comprises foreground-contaminated skymaps convolved with the instrument's beam and scan pattern, along with added instrumental noise. Maps from six different frequency bands are used as input. The generator is a U-Net-based neural network designed to reconstruct the intrinsic skymap. A discriminator network acts as a critic, distinguishing between real and generated data.}
    \label{fig:gan_architecture}
\end{figure}

\subsection{Model Workflow and the Loss functions}

The overall GAN workflow is illustrated in Figure ~\ref{fig:gan_architecture}.
The input to the generator consists of multi-frequency sky maps contaminated by astrophysical foregrounds and instrumental effects, including beam convolution and spatially varying noise. For temperature reconstruction, the input is given by
$
x = \{ T_{\mathrm{obs}}^{\nu} \}, \quad \nu \in \{70, 100, 143, 217, 353, 545\}\,\mathrm{GHz},
$
while for polarization reconstruction, the input consists of frequency-dependent polarization maps,
$
x = \{ E_{\mathrm{obs}}^{\nu} \}, \quad \nu \in \{70, 217, 353\}\,\mathrm{GHz}.
$
These maps are processed through the U-Net generator, which applies a series of downsampling and upsampling operations to produce a cleaned estimate of the underlying CMB sky. The encoder compresses the input into deep feature representations while storing intermediate outputs for skip connections. The decoder then reconstructs the high-resolution map, re-integrating spatial details via those skip connections. The discriminator evaluates whether the generator’s output resembles a real CMB map given the observed sky. 

The discriminator receives two types of input pairs, both formed by concatenating the multi-frequency input sky map---contaminated with astrophysical foregrounds and instrumental noise---with a CMB map: once with the true CMB map and once with the generated map, along the channel dimension. This setup enables the discriminator to learn the conditional distribution of the true CMB given the observed sky, rather than directly comparing the generated and true maps. It learns to distinguish between real samples $(T_{obs}, T_{cmb})$, where $ T_{cmb}$ is the true CMB, and fake samples $(T_{obs}, G(T_{obs}))$, where $G(T_{obs}(\hat{n}))$ is the output of the generator. This discriminator feedback is used to improve the generator through adversarial training. 

Training is driven by a combination of pixel-wise reconstruction losses and an adversarial loss. The reconstruction term comprises both mean squared error (\(L_2\)) and mean absolute error (\(L_1\)) components:

\begin{equation}
\mathcal{L}_{\mathrm{L2}} 
= \frac{1}{N_{\mathrm{pix}}} \sum_{i=1}^{N_{\mathrm{pix}}} \bigl(y_i - G(x)_i\bigr)^2,
\end{equation}

\begin{equation}
\mathcal{L}_{\mathrm{L1}} 
= \frac{1}{N_{\mathrm{pix}}} \sum_{i=1}^{N_{\mathrm{pix}}} \bigl|\,y_i - G(x)_i\bigr|.
\end{equation}

\noindent The \(L_2\) term penalizes large deviations, promoting overall pixel-wise accuracy, while the \(L_1\) term helps preserve sharp features and mitigates blurring. However, relying on pixel-wise losses alone typically limits recovery to large-scale (low \(l\)) structures (see Section~\ref{subsec:discriminator-comparison}).

To encourage the generator to produce outputs that are not only accurate but also perceptually realistic, an adversarial loss is added based on the discriminator’s feedback:

\begin{equation}
\mathcal{L}_{\mathrm{adv}} 
= -\,\frac{1}{N_{\mathrm{disc}}} \sum_{j=1}^{N_{\mathrm{disc}}} \log D\bigl(x, G(x)\bigr)_j.
\end{equation}

\noindent where $D(x, G(x))$ is the discriminator’s output when given the input sky map $x$ and the generated CMB map $G(x)$. The discriminator output is a two-dimensional activation map (e.g., $64 \times 64$) that provides spatially resolved realism scores across the map.

The total objective combines all terms:
\begin{equation}
\mathcal{L} 
= \lambda_2\,\mathcal{L}_{\mathrm{L2}}
+ \lambda_1\,\mathcal{L}_{\mathrm{L1}}
+ \lambda_{\mathrm{adv}}\,\mathcal{L}_{\mathrm{adv}},
\end{equation}
where \(\lambda_1\), \(\lambda_2\), and \(\lambda_{\mathrm{adv}}\) control the relative contributions of each term. 

These weights are chosen to bring the losses to a comparable numerical scale: specifically, \(\lambda_1 = 100\) for the \(L_1\) loss, \(\lambda_2 = 1\) for the \(L_2\) loss, and \(\lambda_{\mathrm{adv}} = 1000\) for the adversarial loss. For example, the unscaled \(L_2\) loss is typically in the thousands, whereas the raw adversarial loss is around 1.4. Scaling the adversarial term by 1000 ensures it contributes significantly during training.

\begin{equation}
\mathcal{L}_{\text{Disc}} = - \frac{1}{N_{disc}} \sum_{i=1}^{N_{disc}} \left[ \log D(x, y)_i + \log \left(1 - D(x, G(x))_i \right) \right]
\end{equation}
The discriminator is trained simultaneously to maximize its classification accuracy—distinguishing between real and generated samples—while the generator is trained to minimize the adversarial loss and fool the discriminator.
This adversarial training loop ensures that the generator not only reduces the reconstruction error but also learns to capture the distributional characteristics of real CMB maps.

In practice, we implement the adversarial losses using the \textit{BCEWithLogitsLoss} function in PyTorch. This formulation is numerically more stable than standard binary cross-entropy loss, as it combines a sigmoid activation and log loss in a single step. Since our discriminator does not apply a final sigmoid activation,\textit{BCEWithLogitsLoss} is appropriate for directly comparing its raw output logits to binary targets (1 for real, 0 for fake). 

Tables~\ref{tab:unet_spec} and~\ref{tab:disc_spec} summarize the architectures of the U-Net generator and the discriminator, respectively. These architectures are optimized for input maps with a resolution corresponding to \(n_{\text{side}} = 1024\), enabling high-fidelity recovery of the CMB signal from contaminated observations.
\begin{table}[h]
\centering
\caption{
U-Net generator architecture. 
$C_{\text{in}}$ and $C_{\text{out}}$ denote the input and output channels. 
Convolutions use $3 \times 3$ kernels with stride 1 ($s_1$) and padding 1 ($p_1$). 
Max pooling uses $2 \times 2$ kernels with stride 2 ($s_2$). 
Upsampling uses transposed convolutions with $4 \times 4$ kernels, stride 2 ($s_2$), and padding 1 ($p_1$). 
The input has 6 channels for temperature (3 for polarization) the output has 1 channel. 
Repeated channel sizes reflect skip connections.
(DS = Downsampling block; s1 = stride 1; s2 = stride 2; p1 = padding 1)
}
\label{tab:unet_spec}
 \resizebox{0.8\textwidth}{!}{
\begin{tabular}{|c|c|c|l|}
\hline
\textbf{Layer} & $C_{in}$ & $C_{out}$ & \textbf{Operation} \\
\hline

\multirow{2}{*}{DS Block 1} & \multirow{2}{*}{6} & \multirow{2}{*}{4} &
Conv2D(3$\times$3, s1, p1, out=$n_p$) + MaxPool2D(2$\times$2, s2) \\
 & & & Conv2D(3$\times$3, s1, p1, out=4) + BatchNorm + LeakyReLU(0.2) \\
\hline

\multirow{2}{*}{DS Block 2} & \multirow{2}{*}{4} & \multirow{2}{*}{4} &
Conv2D(3$\times$3, s1, p1, out=$n_p$) + MaxPool2D(2$\times$2, s2) \\
 & & & Conv2D(3$\times$3, s1, p1, out=4) + BatchNorm + LeakyReLU(0.2) \\
\hline

\multirow{2}{*}{DS Block 3} & \multirow{2}{*}{4} & \multirow{2}{*}{8} &
Conv2D(3$\times$3, s1, p1, out=$n_p$) + MaxPool2D(2$\times$2, s2) \\
 & & & Conv2D(3$\times$3, s1, p1, out=8) + BatchNorm + LeakyReLU(0.2) \\
\hline

\multirow{2}{*}{DS Block 4} & \multirow{2}{*}{8} & \multirow{2}{*}{8} &
Conv2D(3$\times$3, s1, p1, out=$n_p$) + MaxPool2D(2$\times$2, s2) \\
 & & & Conv2D(3$\times$3, s1, p1, out=8) + BatchNorm + LeakyReLU(0.2) \\
\hline

\multirow{2}{*}{DS Block 5} & \multirow{2}{*}{8} & \multirow{2}{*}{8} &
Conv2D(3$\times$3, s1, p1, out=$n_p$) + MaxPool2D(2$\times$2, s2) \\
 & & & Conv2D(3$\times$3, s1, p1, out=8) + BatchNorm + LeakyReLU(0.2) \\
\hline

\multirow{2}{*}{DS Block 6} & \multirow{2}{*}{8} & \multirow{2}{*}{8} &
Conv2D(3$\times$3, s1, p1, out=$n_p$) + MaxPool2D(2$\times$2, s2) \\
 & & & Conv2D(3$\times$3, s1, p1, out=8) + BatchNorm + LeakyReLU(0.2) \\
\hline

\multirow{2}{*}{DS Block 7} & \multirow{2}{*}{8} & \multirow{2}{*}{16} &
Conv2D(3$\times$3, s1, p1, out=$n_p$) + MaxPool2D(2$\times$2, s2) \\
 & & & Conv2D(3$\times$3, s1, p1, out=16) + BatchNorm + LeakyReLU(0.2) \\
\hline

\multirow{2}{*}{DS Block 8} & \multirow{2}{*}{16} & \multirow{2}{*}{16} &
Conv2D(3$\times$3, s1, p1, out=$n_p$) + MaxPool2D(2$\times$2, s2) \\
 & & & Conv2D(3$\times$3, s1, p1, out=16) + BatchNorm + LeakyReLU(0.2) \\
\hline

\multirow{2}{*}{DS Block 9} & \multirow{2}{*}{16} & \multirow{2}{*}{16} &
Conv2D(3$\times$3, s1, p1, out=$n_p$) + MaxPool2D(2$\times$2, s2) \\
 & & & Conv2D(3$\times$3, s1, p1, out=16) + BatchNorm + LeakyReLU(0.2) \\
\hline

US Block 1 & 32 & 16 & TransConv2D(4$\times$4, s2, p1) + Dropout(0.2) + LeakyReLU(0.2) \\
\hline
US Block 2 & 32 & 16 & TransConv2D(4$\times$4, s2, p1) + Dropout(0.2) + LeakyReLU(0.2) \\
\hline
US Block 3 & 24 & 8 & TransConv2D(4$\times$4, s2, p1) + Dropout(0.2) + LeakyReLU(0.2) \\
\hline
US Block 4 & 16 & 8 & TransConv2D(4$\times$4, s2, p1) + BatchNorm + LeakyReLU(0.2) \\
\hline
US Block 5 & 16 & 8 & TransConv2D(4$\times$4, s2, p1) + BatchNorm + LeakyReLU(0.2) \\
\hline
US Block 6 & 16 & 8 & TransConv2D(4$\times$4, s2, p1) + BatchNorm + LeakyReLU(0.2) \\
\hline
US Block 7 & 12 & 4 & TransConv2D(4$\times$4, s2, p1) + BatchNorm + LeakyReLU(0.2) \\
\hline
US Block 8 & 8 & 4 & TransConv2D(4$\times$4, s2, p1) + BatchNorm + LeakyReLU(0.2) \\
\hline
US Block 9 & 8 & 4 & TransConv2D(4$\times$4, s2, p1) + BatchNorm + LeakyReLU(0.2) \\
\hline
Final Layer & 4 & 1 & TransConv2D(4$\times$4, s2, p1) \\
\hline

\end{tabular}}
\end{table}
\begin{table}[h]
\centering
\caption{Layer Specifications for the Discriminator Architecture}
\label{tab:disc_spec}
 \resizebox{0.8\textwidth}{!}{
\begin{tabular}{|c|c|c|l|}
\hline
\textbf{Layer} & \textbf{$C_{in}$} & \textbf{$C_{out}$} & \textbf{Operation} \\
\hline

\multirow{2}{*}{DS Block1} & \multirow{2}{*}{$C_{in} + C_{out}$} & \multirow{2}{*}{4} &
Conv2D(4$\times$4, s1, p1, out=$n_p$) + MaxPool2D(2$\times$2, s2) \\
& & & Conv2D(4$\times$4, s1, p1, out=4) + LeakyReLU(0.2) \\
\hline

\multirow{2}{*}{DS Block2} & \multirow{2}{*}{4} & \multirow{2}{*}{8} &
Conv2D(4$\times$4, s1, p1, out=$n_p$) + MaxPool2D(2$\times$2, s2) \\
& & & Conv2D(4$\times$4, s1, p1, out=8) + BatchNorm + LeakyReLU(0.2) \\
\hline

\multirow{2}{*}{DS Block3} & \multirow{2}{*}{8} & \multirow{2}{*}{16} &
Conv2D(4$\times$4, s1, p1, out=$n_p$) + MaxPool2D(2$\times$2, s2) \\
& & & Conv2D (4$\times$4, s1, p1, out=16) + BatchNorm + LeakyReLU(0.2) \\
\hline

\multirow{2}{*}{DS Block4} & \multirow{2}{*}{16} & \multirow{2}{*}{32} &
Conv2D(4$\times$4, s1, p1, out=$n_p$) + MaxPool2D(2$\times$2, s2) \\
& & & Conv2D(4$\times$4, s1, p1, out=32) + BatchNorm + LeakyReLU(0.2) \\
\hline

ZeroPad1 & 32 & 32 & Zero Padding(1$\times$1) \\
\hline

ConvLayer1 & 32 & 16 & Conv2D(7$\times$7, s1, no padding) + BatchNorm + LeakyReLU(0.2) \\
\hline

ZeroPad2 & 16 & 16 & Zero Padding(1$\times$1) \\
\hline

Final Layer & 16 & 1 & Conv2D(3$\times$3, s1, no padding) \\
\hline

\end{tabular}}
\end{table}
\subsection{Training}


The dataset comprises 1000 simulated CMB maps, which were randomly split into a training set of 990 maps and a validation set of 10 maps. To model the galactic foregrounds accurately, we utilized the Python Sky Model (\texttt{PySM}) to simulate various dust and synchrotron emission models. Specifically, we employed 9 different dust foreground models (comprising 2 stochastic and 7 deterministic models) and 7 different synchrotron models (comprising 1 stochastic and 6 deterministic models). 

To capture the variability inherent in the stochastic foregrounds, we used different random initial seeds to generate 10 distinct maps for each stochastic model. For the deterministic models, a single map was generated per model. We explicitly allocated the \texttt{s1} and \texttt{s2} synchrotron models, along with the \texttt{d1}, \texttt{d2}, and \texttt{d4} dust models, to serve as the validation set for the foregrounds, reserving the remaining models for the training set. As detailed in Section~\ref{sec:2.1}, the \texttt{PySM} simulations initially yielded Stokes $I$, $Q$, and $U$ maps, which were subsequently converted into $T$ and $E$-mode maps.
These maps have a shape of (6, 1024, 1024), corresponding to 6 frequency channels at a resolution of \(n_{\text{side}} = 1024\) (3 frequency channels for polarization maps). To facilitate training, each map was divided into 12 patches of size (6, 1024, 1024) using HEALPix nested indexing.

Training was conducted on the Param Himalaya cluster, utilizing NVIDIA V100 GPUs with 32 GB of memory. The experiments involved using 2 GPUs on a node, allowing for the parallel training of two patches simultaneously. The GAN was implemented in \texttt{PyTorch}, and the Adam optimizer was employed for both the generator and discriminator. 
The Adam optimizer \cite{kingma2014adam} updates the parameters $\theta$ at iteration $t$ according to
\begin{align}
m_t &= \beta_1 m_{t-1} + (1 - \beta_1) g_t, \qquad
v_t = \beta_2 v_{t-1} + (1 - \beta_2) g_t^2, \nonumber\\
\hat{m}_t &= \frac{m_t}{1 - \beta_1^t}, \qquad
\hat{v}_t = \frac{v_t}{1 - \beta_2^t}, \qquad
\theta_t = \theta_{t-1} - \alpha \frac{\hat{m}_t}{\sqrt{\hat{v}_t} + \epsilon},
\end{align}
where $g_t = \nabla_\theta \mathcal{L}(\theta_{t-1})$ is the gradient of the loss function at step $t$. 
The exponential decay rates, $\beta_1$ and $\beta_2$ and numerical stability constant $\epsilon$ are set to the commonly used values
\[
\beta_1 = 0.9, \quad \beta_2 = 0.999, \quad \epsilon = 10^{-8}.
\]

\noindent During training, the learning rate ($\alpha$) for the generator was initially chosen to be of order $10^{-2}$ and $10^{-3}$ for the discriminator, 
slowly decreasing them to around orders of $10^{-5}$ and $10^{-7}$ respectively as the loss became lower and less stable.

Multiple training runs were performed for each patch to optimize the model parameters. The average training time for one patch was approximately 40 hours when simulating circular beam and extended to 4-5 days for real beam simulations. The maps convolved with real beam are trained against HEALPix skymaps convolved with an effective gaussian circular beam. During training, the loss curves exhibited instability, generally following a pattern of initial descent followed by an increase, eventually stabilizing around 0.69 for the discriminator loss. The L1 loss and MSE loss showed exponential decay behavior.

We experimented with transferring weights from previously trained patches to initialize training for new patches. However, this approach did not yield a significant improvement in training time or loss behavior.

\section{Simulations and Results}
\label{sec:results-discussion}
We perform three sets of simulations to assess the performance of our CMB reconstruction pipeline: one that uses a simplified circular beam for the temperature only skymap, and two that incorporate the Planck satellite’s actual scanning strategy and beam pattern for temperature and E-mode polarization skymaps respectively. The goal is to assess how well the model generalizes under increasingly realistic observational conditions. We also analyze the effect of adversarial training and the impact of model size.

\subsection{Results with circular scan}

\begin{figure}[htbp]
    \centering
    \begin{subfigure}[b]{ 0.49\textwidth}
        \includegraphics[trim=0cm 0.1cm 1.9cm 1.8cm, clip=true, width=\textwidth]{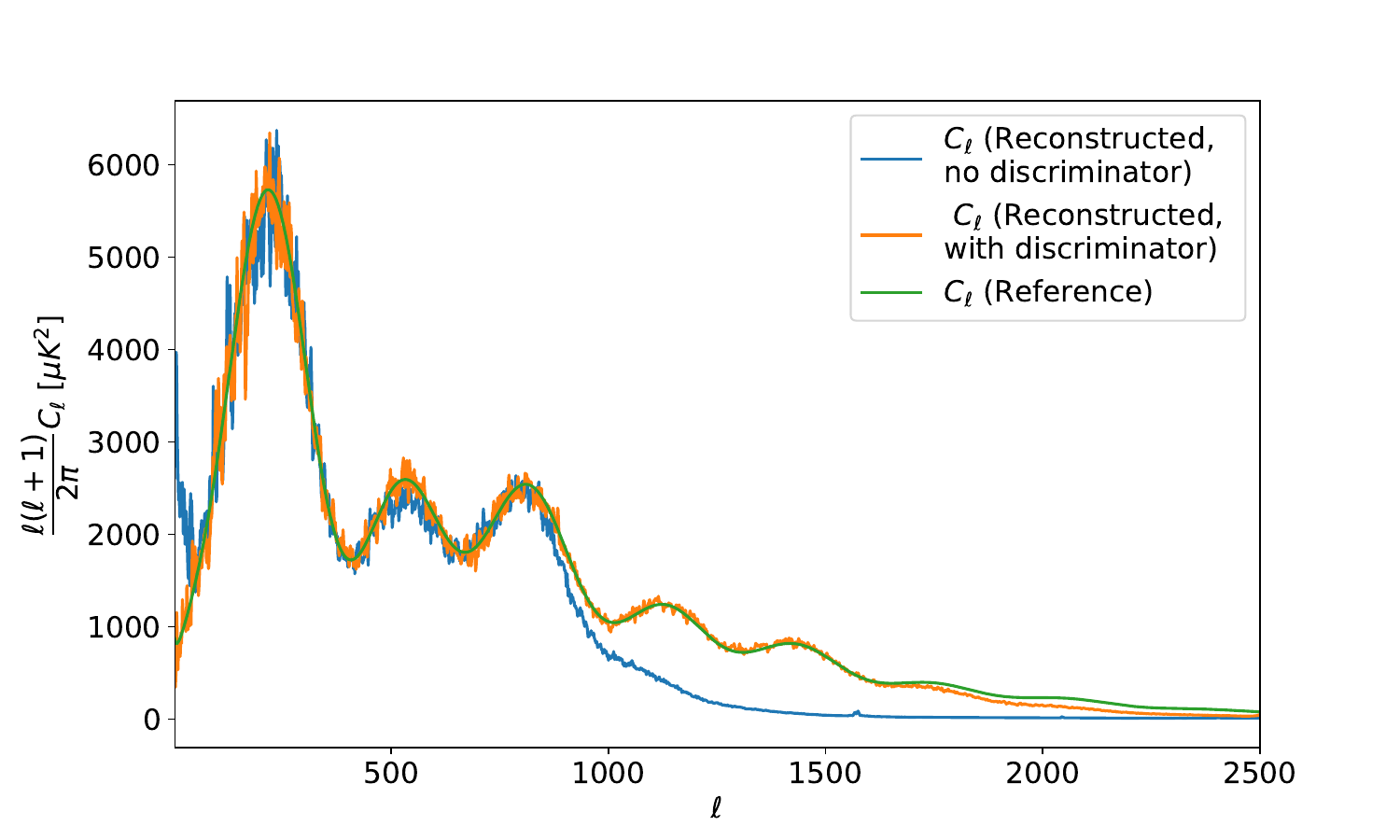} 
        \label{fig:linear_scale_comparison}
    \end{subfigure}
    \hfill
    \begin{subfigure}[b]{0.49\textwidth}
        \includegraphics[trim=0cm 0.1cm 1.9cm 1.8cm, clip=true,width=\textwidth]{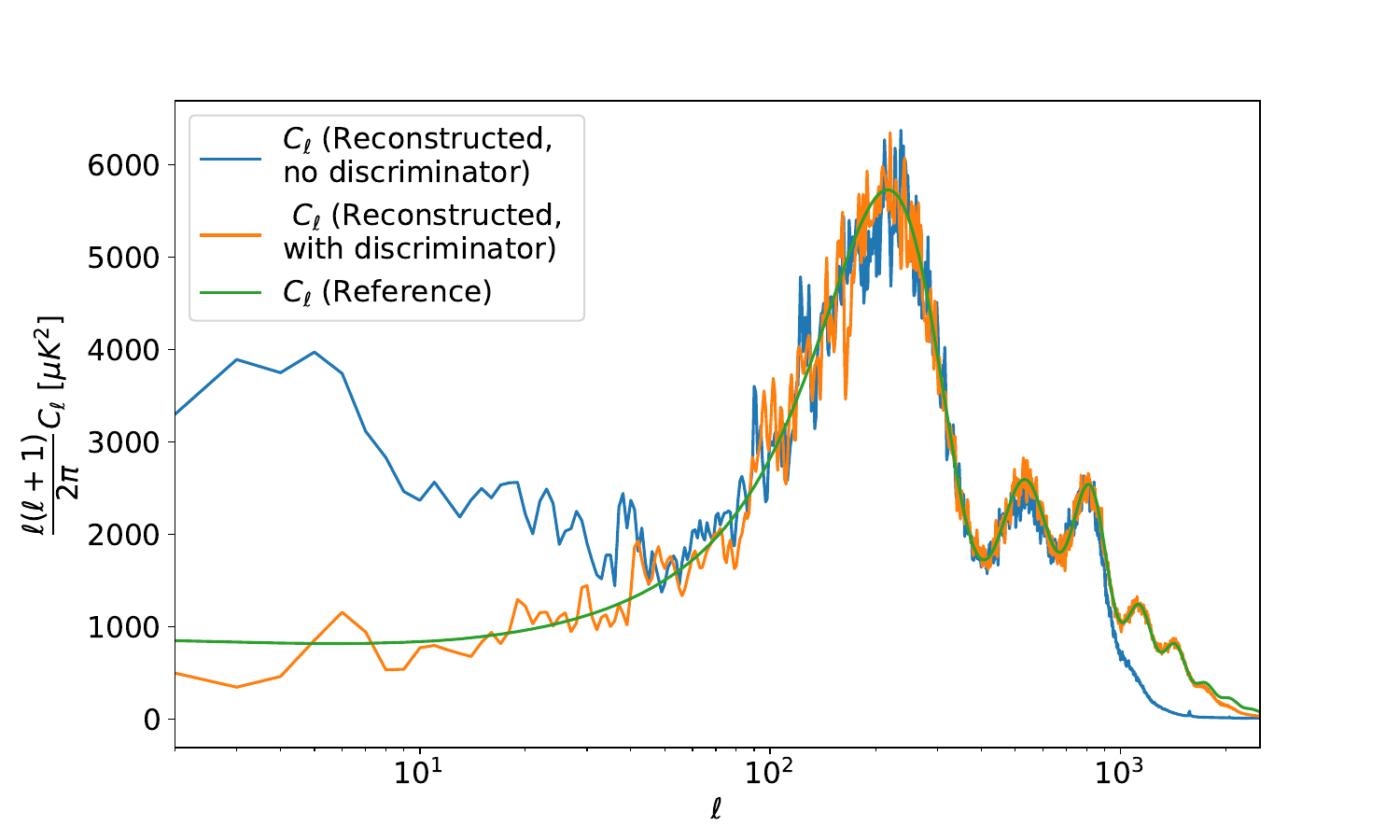} 
        \label{fig:log_scale_comparison}
    \end{subfigure}
    \caption{Comparison of power spectra of CMB maps reconstructed with and without the discriminator model. (left) Linear scale highlights differences at higher multipole moments (\textit{l}). (right) Logarithmic scale focuses on differences at lower multipole moments (\textit{l}). For this analysis we use $n_p = 8$.}
    \label{fig:power_spectra_comparison}
\end{figure}

In this section, we present our analysis based on the circular beam scan on CMB temperature skymap. For this convolution, no actual scanning pattern was considered; as a result, the noise variance is uniform across all pixels (see Sec.~\ref{Sec:NoiseBeam} for details). 
Figure~\ref{fig:power_spectra_comparison} shows the results from our analysis. Green plot shows the  reference $C_l$ that is used for generating the  simulated CMB maps. The orange curve shows the $C_l$ from the outputs  from our neural network. The comparison indicates that the network successfully recovers the input spectrum almost up to $l=1500$, beyond which the reconstructed power gradually drops. We show the results in both linear and logarithmic scales to allow readers to assess the agreement across both high and low multipole ranges. 
As mentioned earlier, $n_p$ is a modeling parameter, and for this analysis we set $n_p = 8$. 

\subsubsection{Effect of Adversarial Training}
\label{subsec:discriminator-comparison}

To evaluate the impact of including a discriminator in the training process, we compared the power spectra of CMB maps reconstructed using only a generator model versus those reconstructed using both a generator and a discriminator model. Figure~\ref{fig:power_spectra_comparison} illustrates these comparisons. The plot on the left presents the power spectra using a linear scale for the x-axis (multipole moment \textit{l}), highlighting the difference at higher multipole moments. We can see that the power spectra produced without a discriminator (shown in blue) start to deviate from the original data after approximately \textit{l} = 900. The inclusion of the discriminator helps maintain the power spectra closer to the original even at these higher multipole moments. The plot on the right provides the same comparison but uses a logarithmic scale for the x-axis to better visualize differences at lower \textit{l} values. Here, it is evident that the model without a discriminator yields higher values than the original data for low \textit{l} values. In contrast, the power spectra generated with a discriminator show a more accurate representation of the original data, particularly at these lower \textit{l} values. These plots clearly demonstrate that the inclusion of a discriminator in the training process helps improve the reconstruction quality by better aligning the power spectra with the original data, especially at higher multipole moments where the generator alone fails to maintain consistency. The discriminator acts as a regularizing mechanism, preventing the power spectra from deviating significantly from the original at critical points.

\begin{figure}[htbp]
    \centering
    \begin{subfigure}[b]{0.49\textwidth}
        \includegraphics[trim=0.0cm 0.1cm 2.1cm 1.4cm, clip, width=\textwidth]{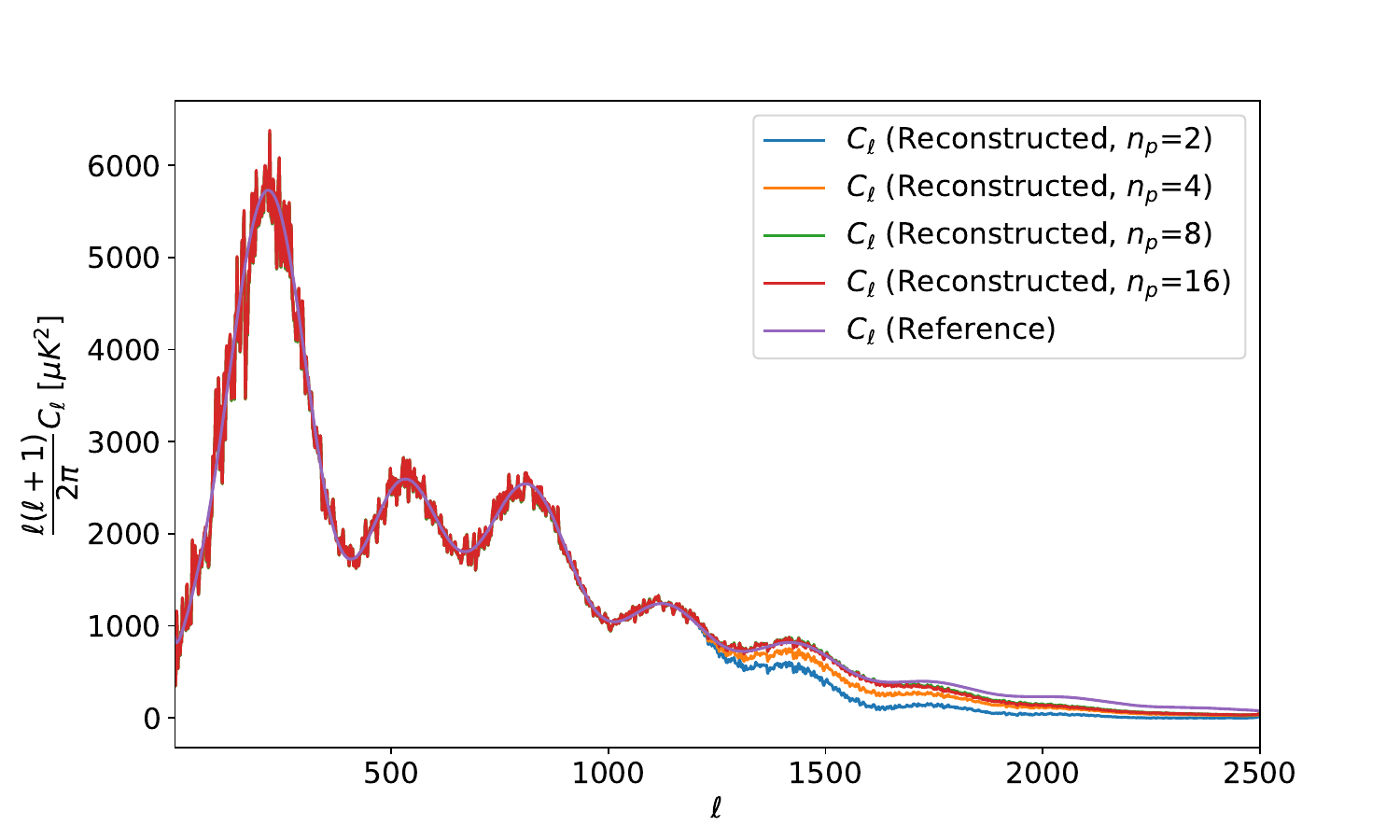}
        \label{fig:normal_scale_model_size}
    \end{subfigure}
    \hfill
    \begin{subfigure}[b]{0.49\textwidth}
        \includegraphics[trim=0.0cm 0.1cm 2.1cm 1.4cm, clip, width=\textwidth]{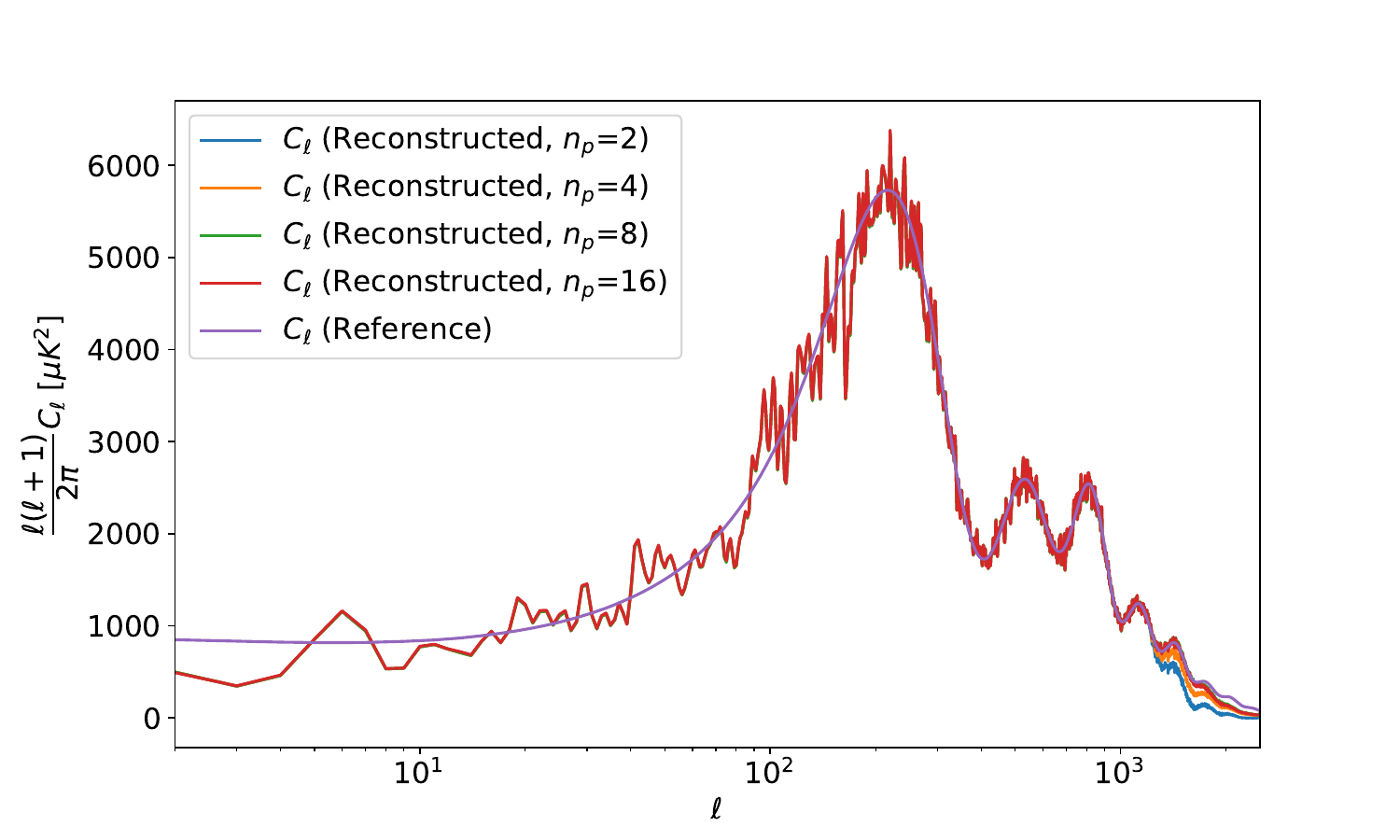}
        \label{fig:log_scale_model_size}
    \end{subfigure}
    \vspace{0.5cm} 
    \centering
    \begin{subfigure}[b]{0.49\textwidth} 
        \includegraphics[trim=0.0cm 0.1cm 2.1cm 1.4cm, clip, width=\textwidth]{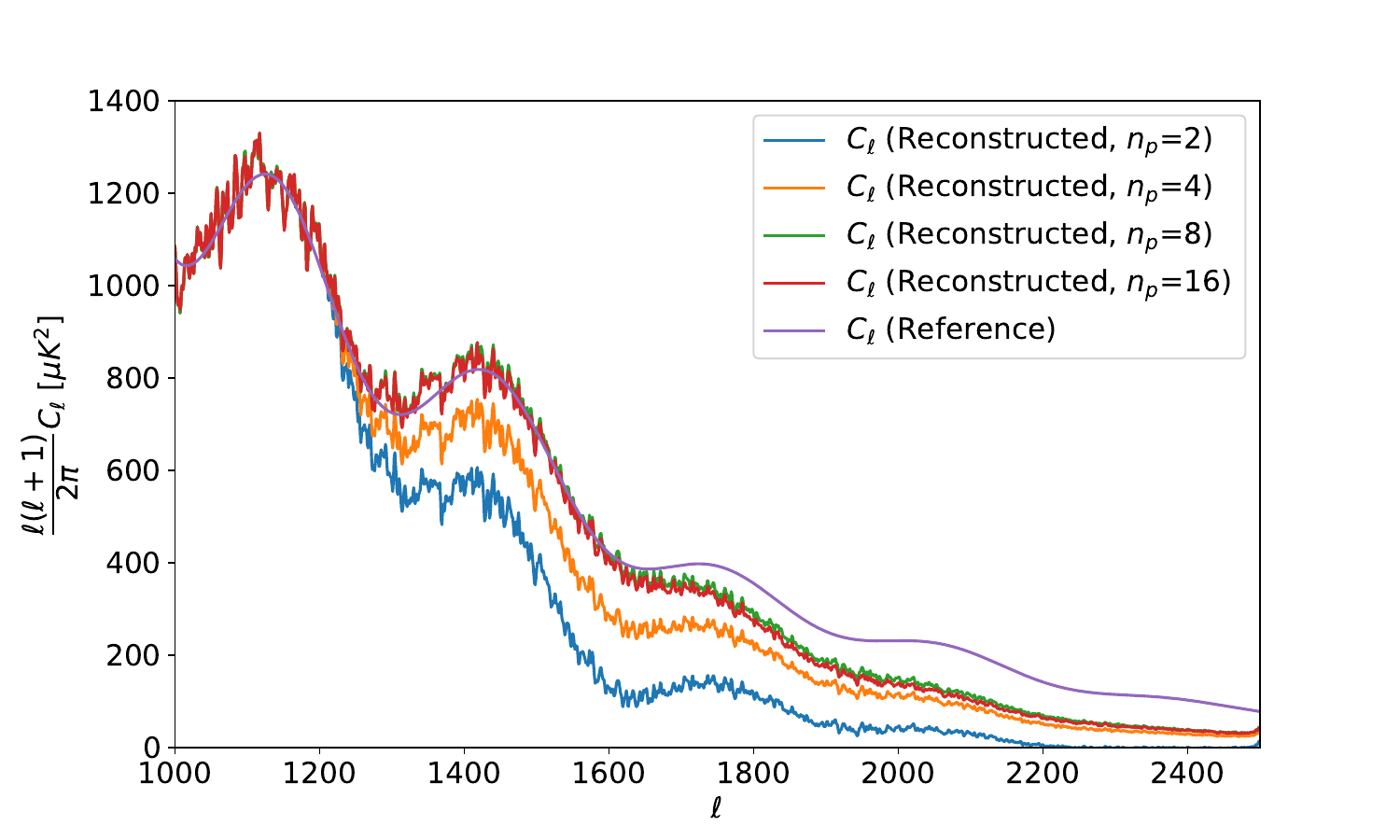} 
        \label{fig:high_multipole_model_size}
    \end{subfigure}
    \label{fig:model_size_comparison}
    \hfill
    \begin{subfigure}[b]{0.49\textwidth} 
        \includegraphics[trim=0.0cm 0.1cm 2.1cm 1.4cm, clip, width=\textwidth]{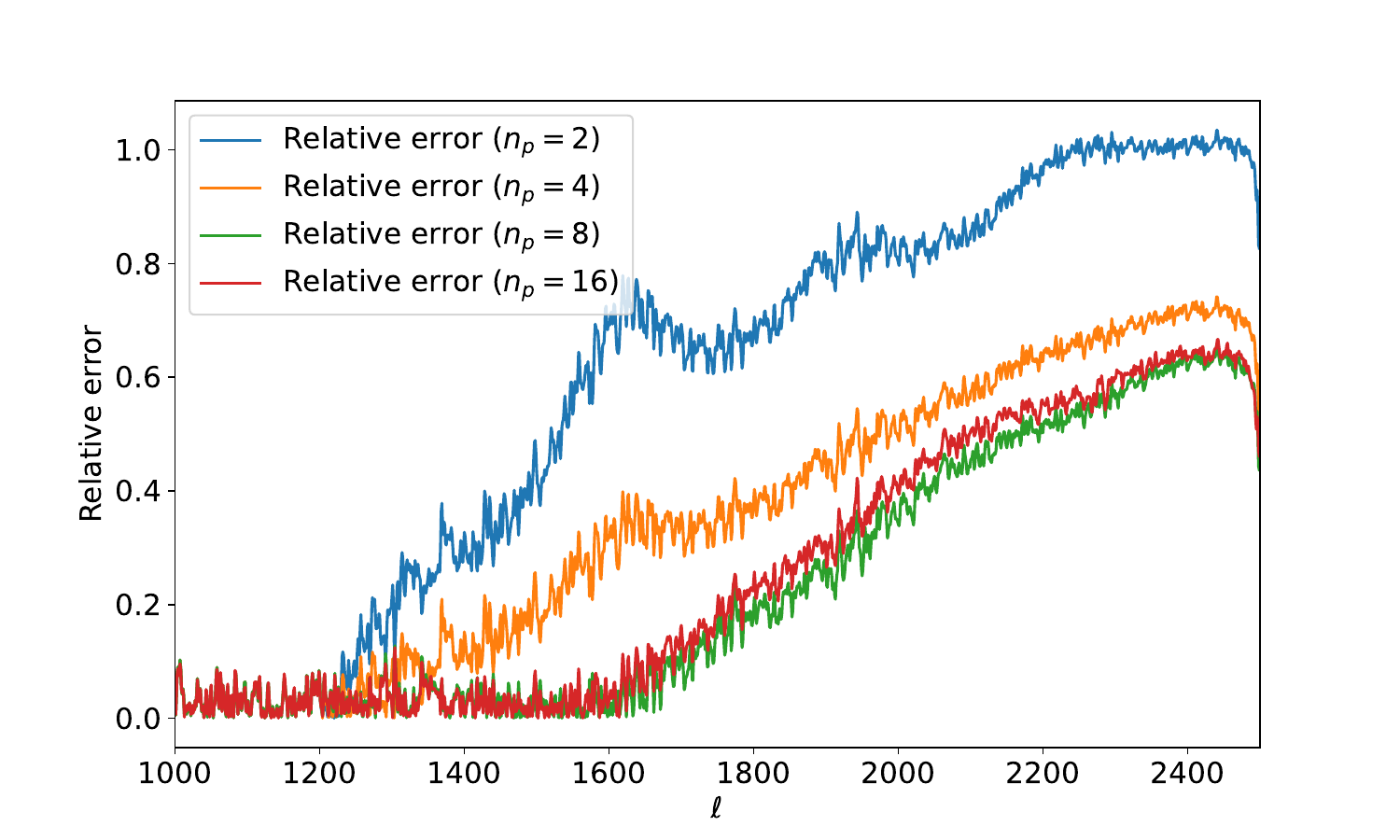}
        \label{fig:error}
    \end{subfigure}
    \caption{This figure compares the power spectra, $C_l$, from the  reconstructed CMB maps using $n_p = 2, 4, 8$, and $16$. At low multipoles, all models exhibit similar recovery. However, for smaller $n_p$, the reconstructed power drops off more rapidly at higher multipoles. The top-left plot compares the $C_l$ values across different model sizes. The top-right plot shows the same spectra with a logarithmic x-axis to emphasize the behavior at low multipoles. The bottom-left plot focuses on $C_l$ for $l > 1000$, clearly showing the loss of power at higher multipoles for various values of $n_p$. The bottom-right panel displays the relative error, calculated as $\left| \left(C_l^{\mathrm{Ref}} - C_l^{\mathrm{Recons}}\right) / C_l^{\mathrm{Ref}} \right|$.}   
    \label{fig:model_size_comparison_error}
\end{figure}

\subsubsection{Impact of Model Size on Reconstruction Performance}
\label{subsec:model-size-impact}

To explore the effect of model size on reconstruction performance under the circular beam assumption, we compared the power spectra of CMB maps reconstructed using different model sizes, corresponding to hyperparameter values $n_p$ of 2, 4, 8, and 16. Figure~\ref{fig:model_size_comparison_error} illustrates these comparisons across various scales. The logarithmic scale plot (top-right) shows that all model sizes perform similarly well for low \textit{l} values, suggesting they effectively capture the early, low-frequency features of the CMB power spectra. However, the linear scale plot (top-left) and the plot focusing on higher multipoles (bottom-left) reveal that as the model size increases, the accuracy of the power spectra improves for high \textit{l} values. Specifically, models with sizes n=8 and n=16 maintain a closer alignment with the original data up to approximately \textit{l} = 1700. In contrast, the smaller models diverge earlier: the \textit{$n_p$=2} model deviates significantly around \textit{l} = 1100, and the \textit{$n_p$=4} model diverges around \textit{l} = 1300. The bottom-right plot shows the relative error between the reference and the reconstructed $C_l$s, given by the absolute value of $(C_l^{Ref} - C_l^{Recons})/C_l^{Ref}$. The plot shows that the improvement is minimal when we change the hyperparameter $n_p=8$ to $n_p=16$. This analysis indicates that increasing model size allows the capture of more detailed features, leading to more accurate reconstructions at higher \textit{l} values, better capturing high-frequency components. Nevertheless, increasing model size introduces trade-offs regarding complexity, leading to longer training times and higher computational costs. While models \textit{$n_p$=8} and \textit{$n_p$=16} perform similarly well up to \textit{l} = 1700, the \textit{$n_p$=8} model offers a good balance between complexity and performance, making it an optimal choice in this context.

\subsection{Reconstruction Results with Real Beam Convolution for temperature map}
\label{subsec:real-beam-detailed}

\begin{figure}[ht]
    \centering
    \includegraphics[trim=0cm 0.1cm 1.8cm 1.5cm, clip=true,width=0.49\textwidth]{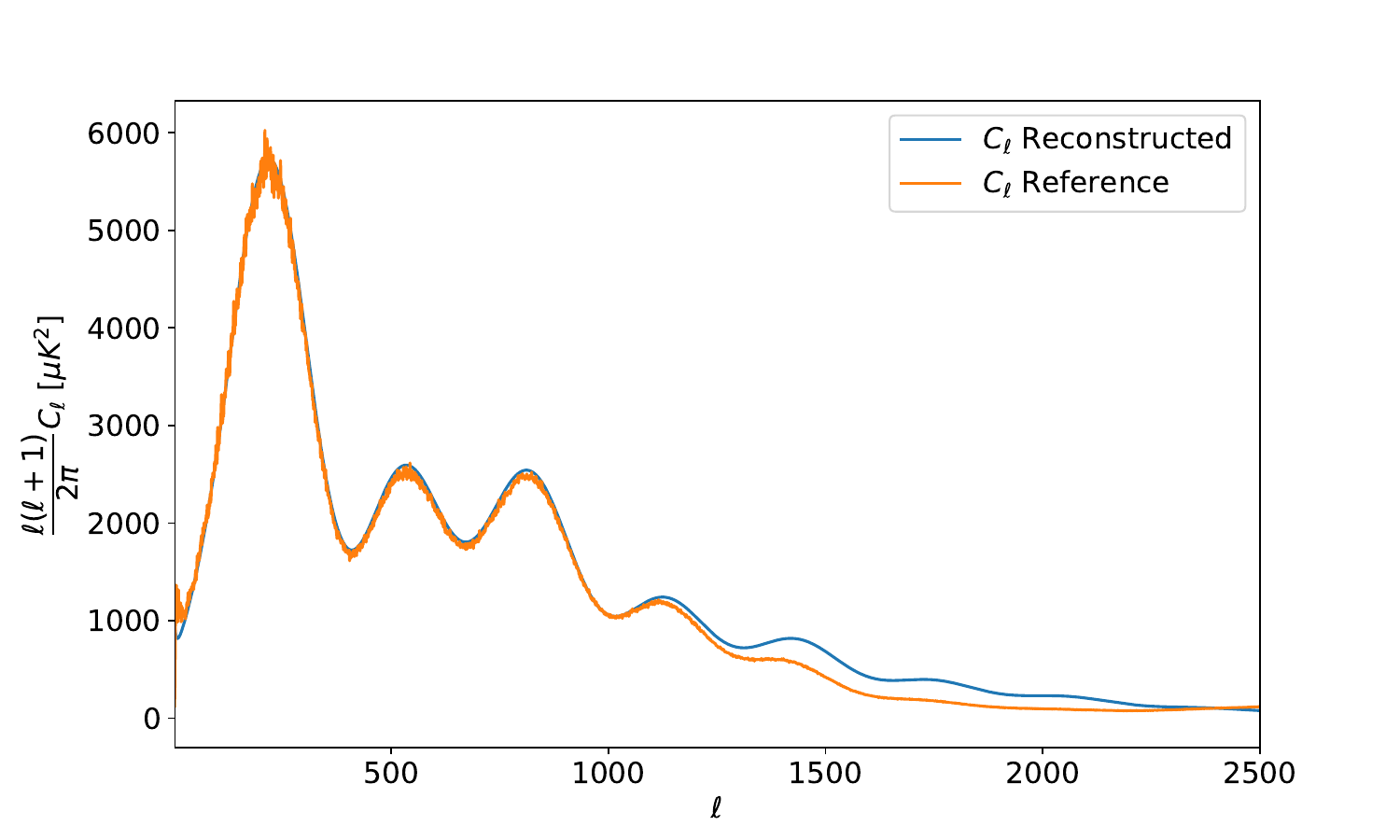}
    \hfill
    \includegraphics[trim=0cm 0.1cm 1.8cm 1.5cm, clip=true,width=0.49\textwidth]{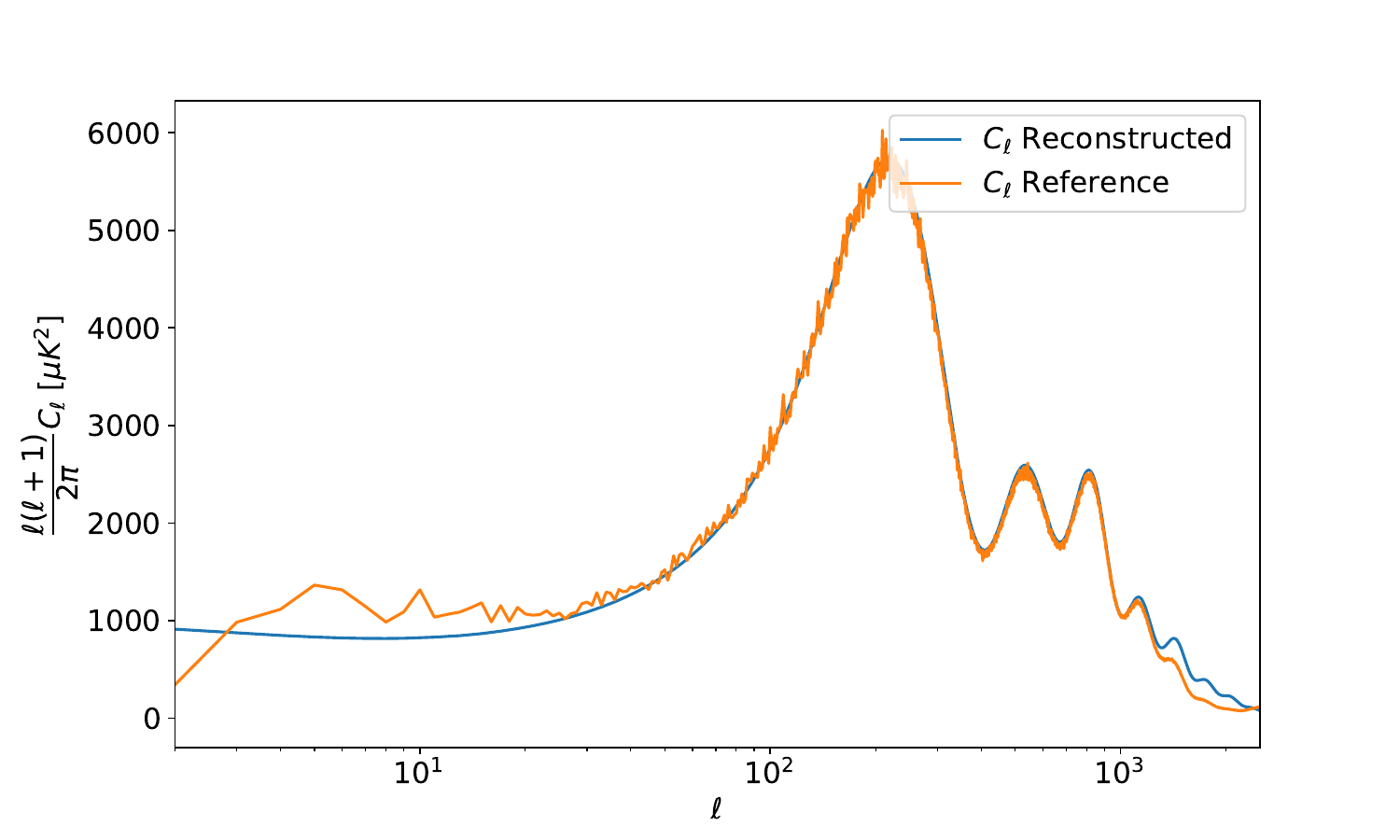}
    \includegraphics[trim=0cm 0.1cm 1.8cm 1.5cm, clip=true,width=0.49\textwidth]{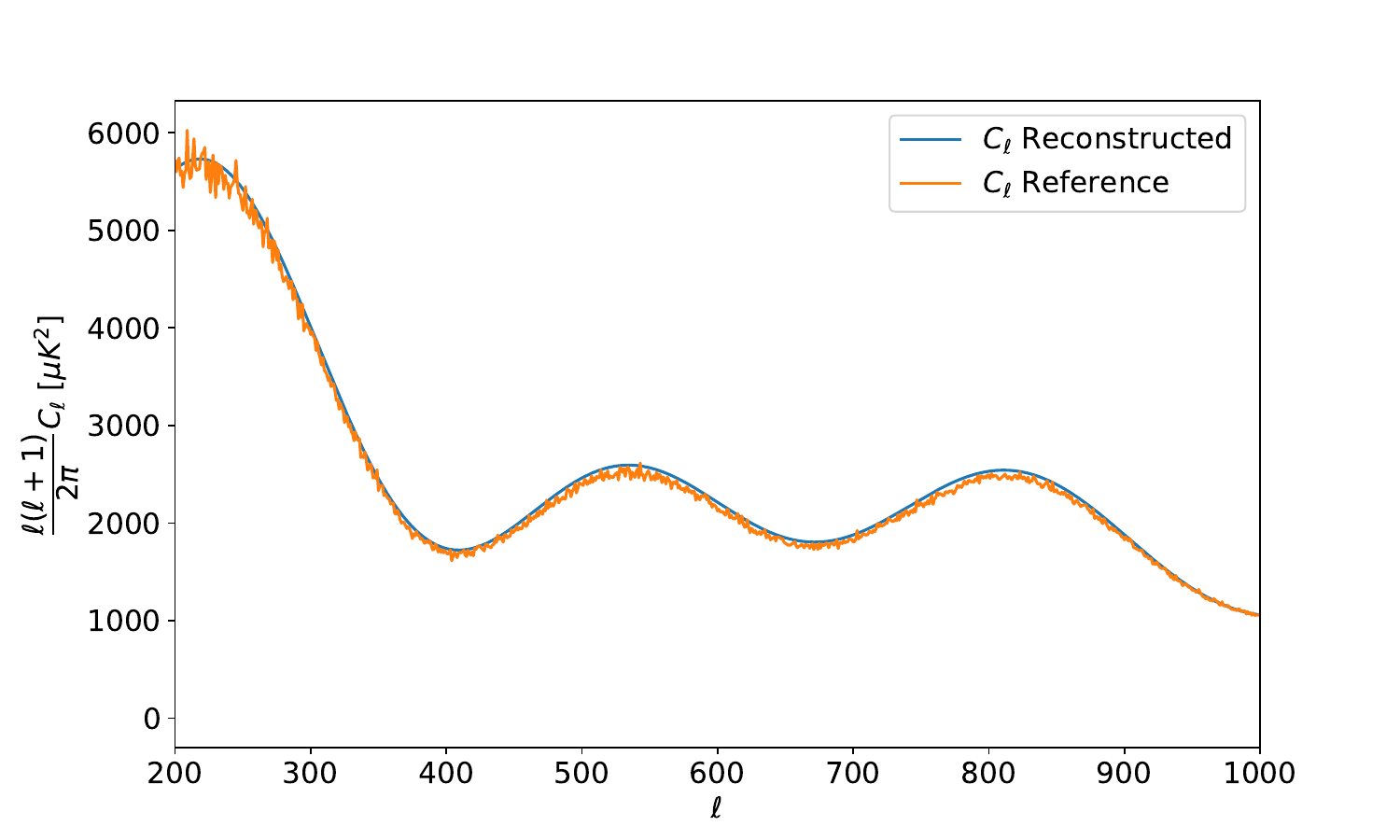}
    \hfill
    \includegraphics[trim=0cm 0.1cm 1.8cm 1.5cm, clip=true, width=0.49\textwidth]{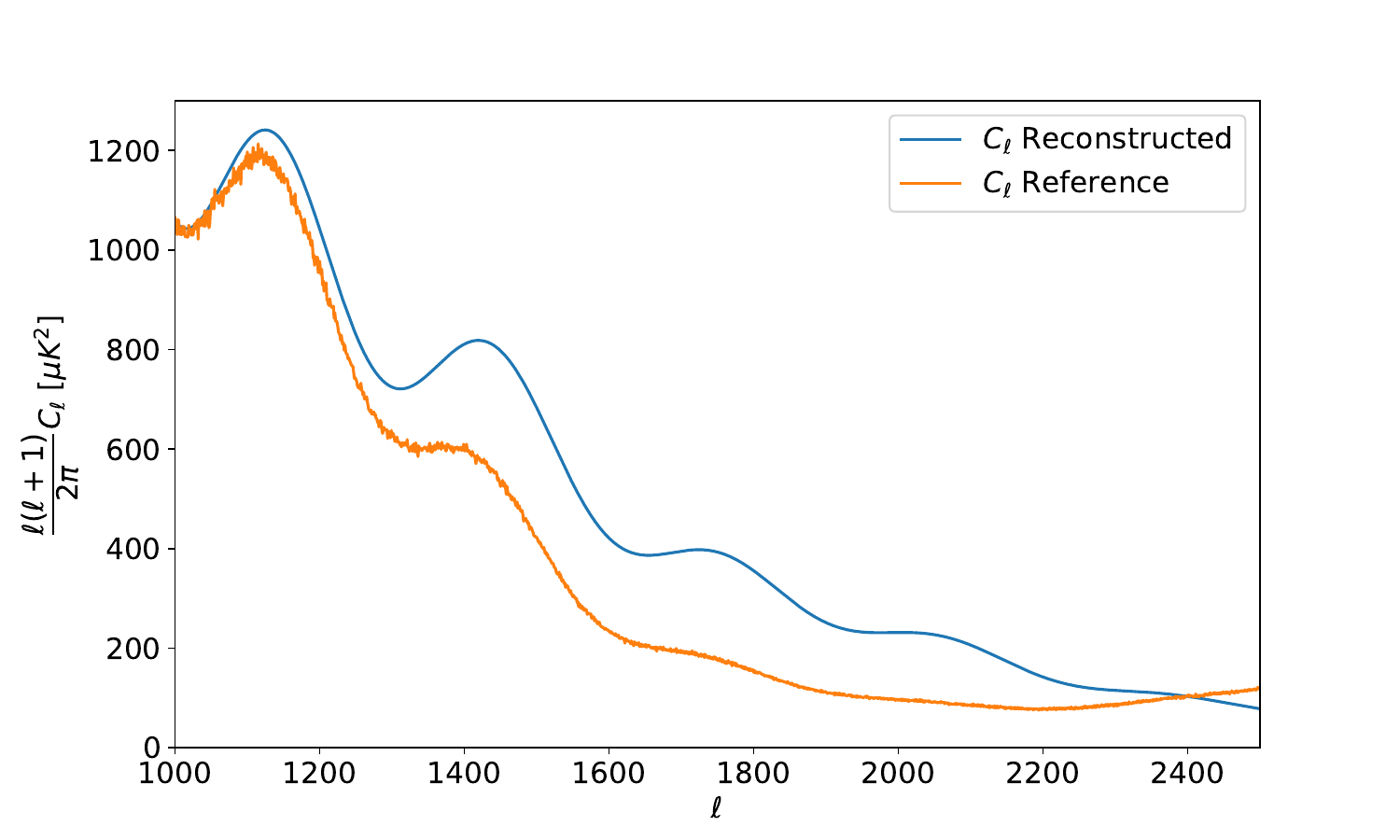}
    \caption{Top-Left: Full power spectrum of the reconstructed CMB maps using real beam convolution. Top-Right: Semilogx plot of the same power spectrum. Bottom-Left: Power spectrum in the multipole range $200 \leq l \leq 1000$. Bottom-Right: Power spectrum for $l > 1000$, illustrating the performance at higher multipole moments. }
    \label{fig:real-beam-l-ranges}
\end{figure}

In this analysis, we convolve the foreground contaminated CMB temperature maps using Planck beam and scan pattern as described in Sec.~\ref{Sec:SimulationPipeline} and then add the noise to generate the simulated realizations for all 6 frequency channels. 
Training proceeded using the discriminator-based approach with $n_p=8$. We have presented the results from our analysis in Figure~\ref{fig:real-beam-l-ranges}. The top-left plot shows the full power spectrum covering the entire multipole range ($l=2$ to $2500$). We can see that the network can recover the  spectra significantly well up-to multipole $l\sim 1000$, after which the power starts falling off.  
The next plot, i.e. the plot on the top-right plot shows the same power spectrum but the  x-axis is set to the log scale (semilogx plot). This allows the readers to understand the behavior of the power-spectrum at low multipoles. 
The third figure highlights the behavior of the reconstruction specifically within the multipole range $l = 200$ to $1000$. Lastly, the bottom-right figure focuses on the performance at higher multipole moments ($l > 1000$), a regime where the model faces significant loss.

These figures collectively demonstrate that the incorporation of the real beam convolution into the training pipeline, coupled with the discriminator, effectively preserves the CMB power spectrum characteristics across the full range of multipoles, while also ensuring accurate reconstruction in both the intermediate and high-l regimes.


Figures~\ref{fig:patches0-5} and \ref{fig:patches-6-11} present a detailed visual assessment of the network's CMB reconstruction performance for each of the 12 patches.
The first 4 patches are from the upper hemisphere, the next 4 patches are from the equatorial region and the final 4 are from the lower hemisphere. 
Each patch is displayed across four rows: the noisy input map containing foregrounds, the predicted CMB signal from the generator network, the true underlying CMB signal, and the difference between the predicted and true CMB (Predicted - True). Here we would like to point out that there are total 6 frequency channels that are used as inputs. In the plots we display only the $70$GHz channel. 
For plotting the input, true and predicted maps we set the temperature range between $-200\mu K$ and $200\mu K$. For plotting the differences no such range is set. The change in the color pallet is due to difference in range. Our plots show that the foreground removal algorithm works perfectly near the polar region. The difference in the signals are less than $2\mu K$, (approximately $1\%$) of the maximum pixel temperature in the true map. Near the Galactic center, some residual signals are visible, particularly in the plots for patches 5, 6, 7, and 8. However, even close to the Galactic plane, the residual signal remains below $\sim2$–$6,\mu\mathrm{K}$ (approximately $2\%$), except for a few isolated pixels where the difference reaches $8$–$10,\mu\mathrm{K}$.

\subsection{Reconstruction Results with Real Beam Convolution for Polarization Maps}

Training of the network proceeded using the same discriminator-based approach with $n_p=8$, targeting the reconstruction of the uncontaminated CMB polarization field. However, two key modifications were made for the polarization analysis. First, the generator architecture was adjusted to accept 3 input frequency maps, in contrast to the 6 channels used for the temperature reconstruction. Second, an additional preprocessing sanitization step was required to appropriately handle masked pixels present in the \texttt{healpy} maps prior to feeding the data into the network.

\begin{figure}[ht]
    \centering
    \includegraphics[trim=0cm 0.1cm 1.8cm 1.5cm, clip=true,width=0.49\textwidth]{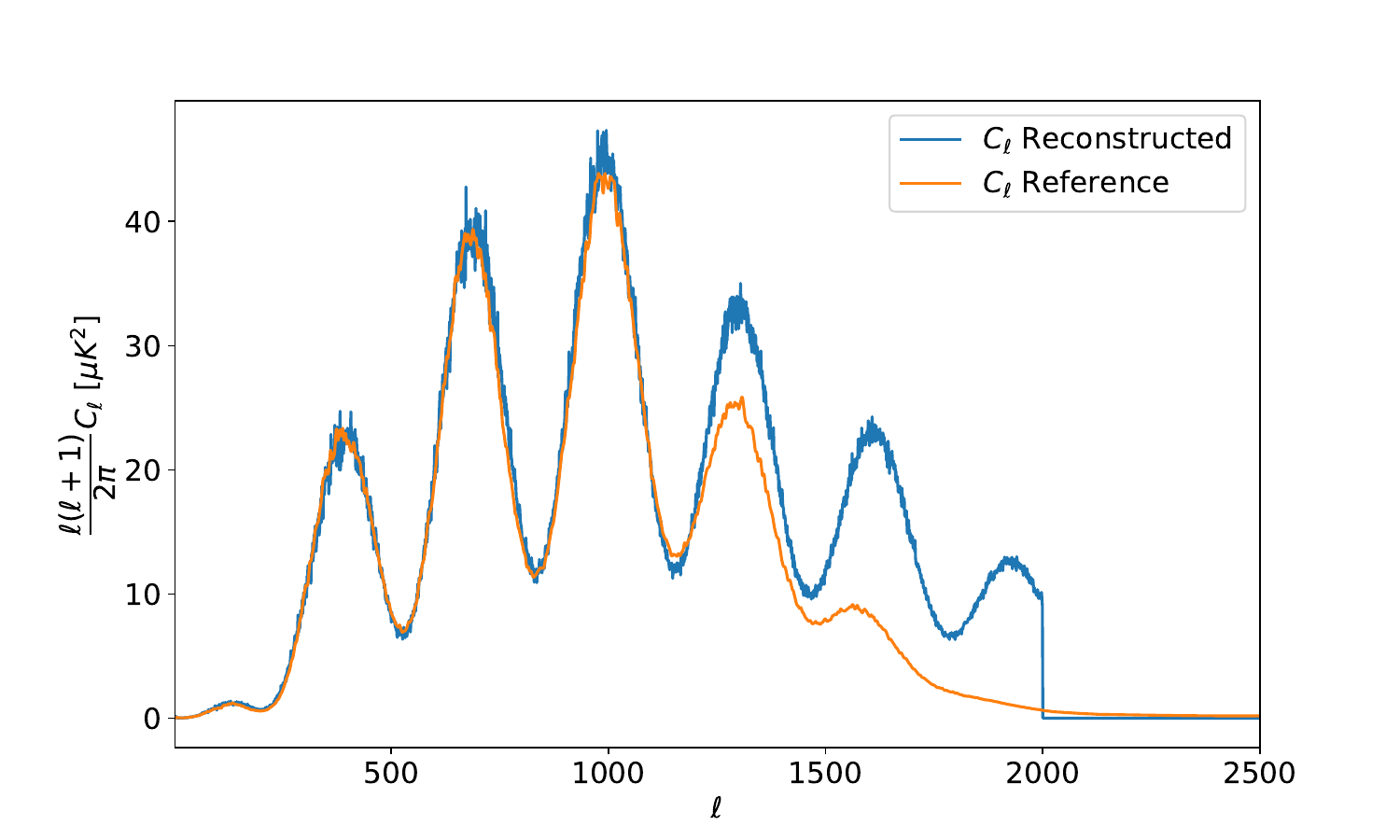}
    \hfill
    \includegraphics[trim=0cm 0.1cm 1.8cm 1.5cm, clip=true,width=0.49\textwidth]{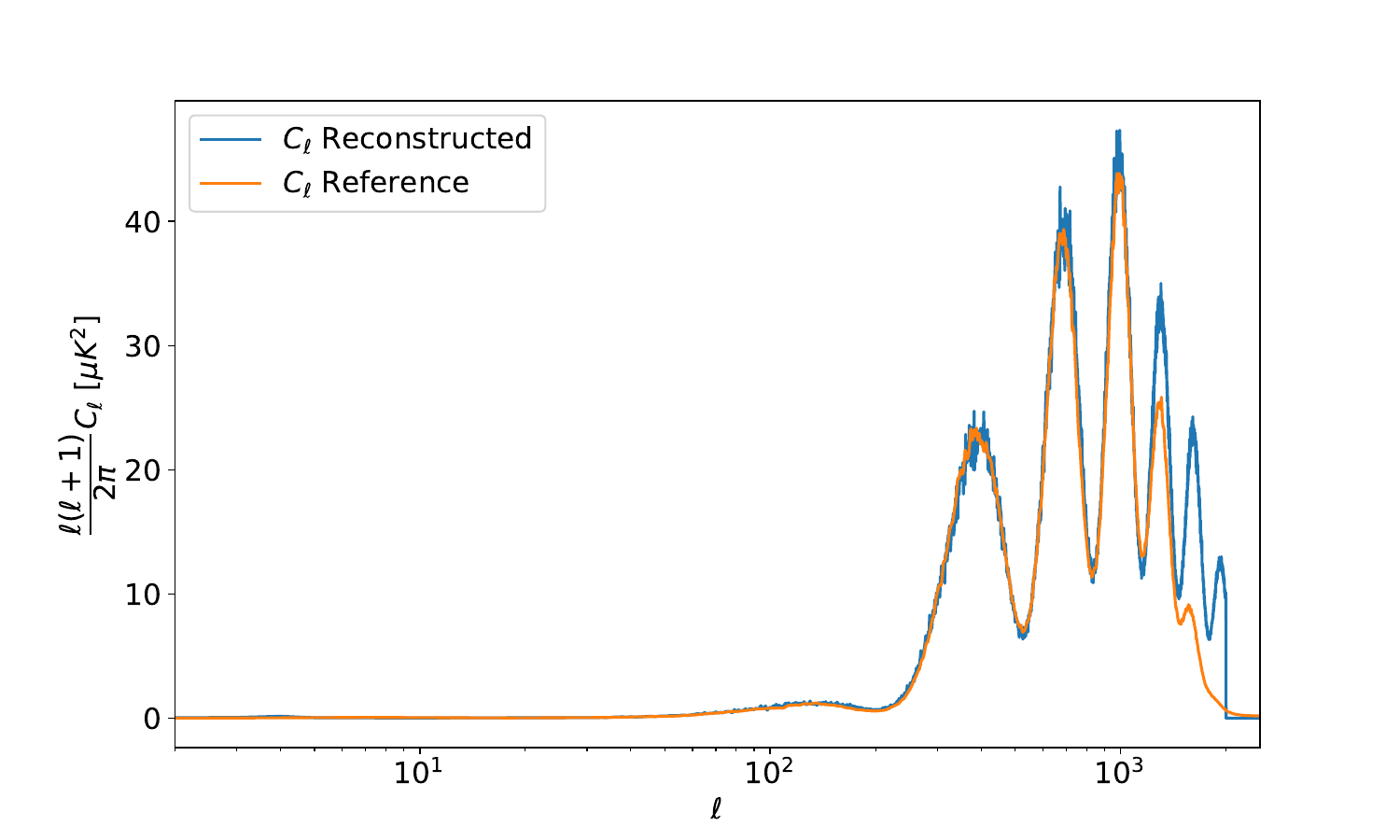}
    
    \vspace{0.2cm}
    
    \includegraphics[trim=0cm 0.1cm 1.8cm 1.5cm, clip=true,width=0.49\textwidth]{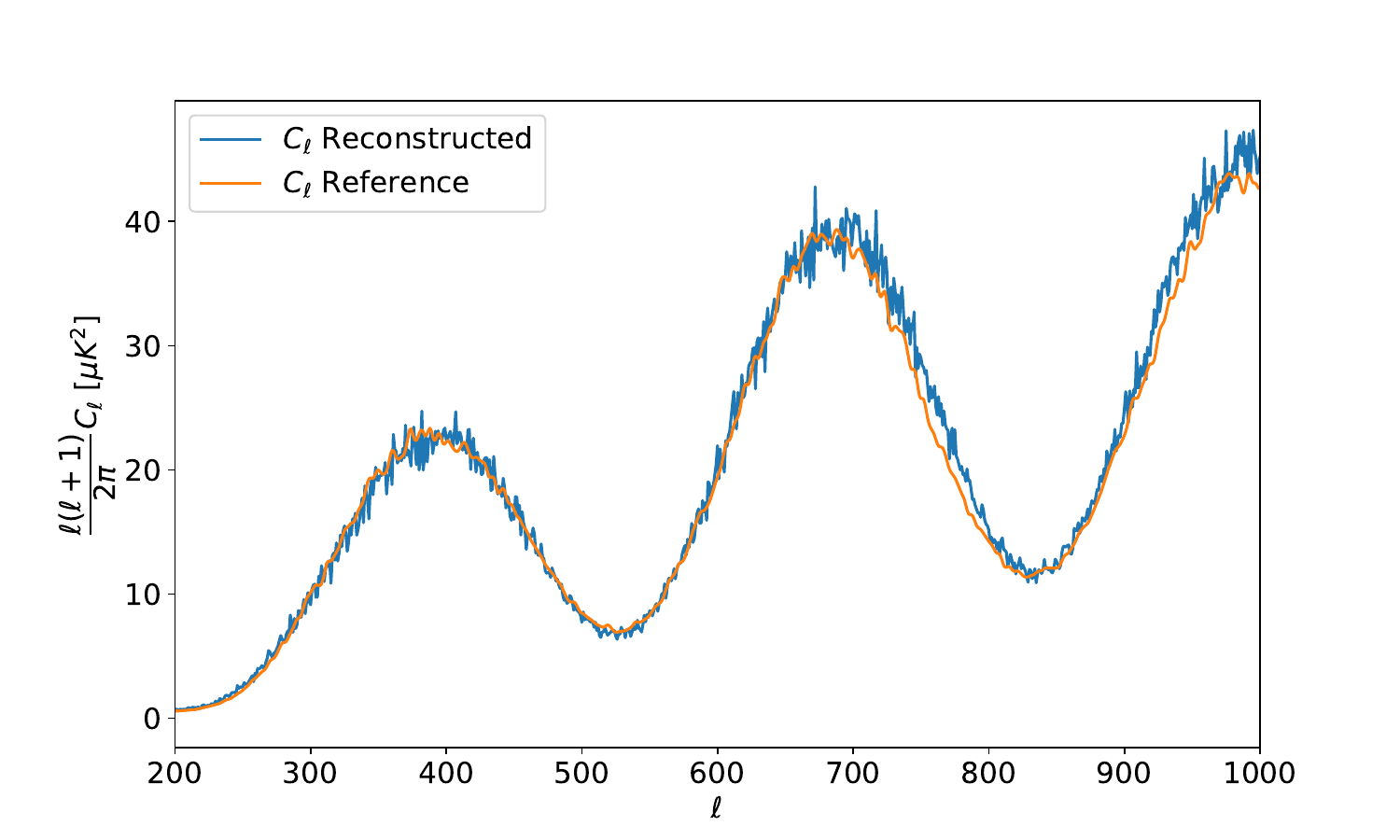}
    \hfill
    \includegraphics[trim=0cm 0.1cm 1.8cm 1.5cm, clip=true, width=0.49\textwidth]{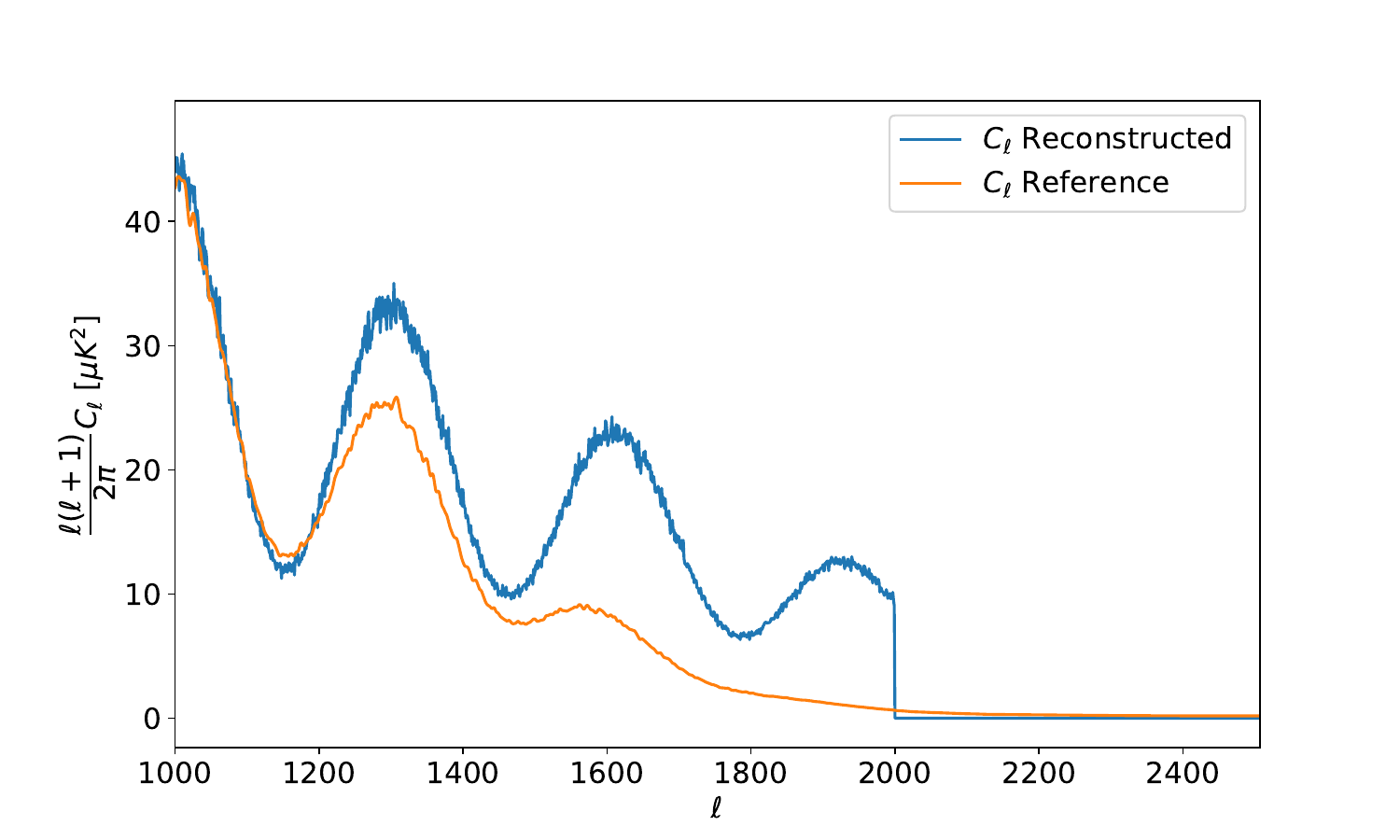}
    \caption{Top-Left: Full $EE$ power spectrum of the reconstructed CMB polarization maps using real beam convolution and realistic scan orientations. Top-Right: Semilogx plot of the same $EE$ power spectrum. Bottom-Left: $EE$ power spectrum in the multipole range $200 \leq l \leq 1000$. Bottom-Right: Power spectrum for $l > 1000$, illustrating the model's performance at higher multipole moments where instrument noise dominates.}
    \label{fig:pol-real-beam-l-ranges}
\end{figure}

We present the power spectra of the reconstructed polarization maps, focusing on the $E$-mode ($EE$) spectrum, in Figure~\ref{fig:pol-real-beam-l-ranges}. The top-left panel shows the full $EE$ power spectrum covering the entire multipole range ($l=2$ to $2500$). The network successfully recovers the polarization spectra up to multipole $l \sim 1100$. Beyond this scale, the reconstructed power begins to attenuate, 

To comprehensively evaluate the reconstruction quality under these realistic observational and scanning conditions, we decompose the spectra into specific multipole regimes. The top-right plot displays the same $EE$ power spectrum on a semilogarithmic scale (semilogx) to emphasize the behavior at low multipoles. The bottom-left figure highlights the intermediate regime within the multipole range $200 \leq l \leq 1100$. Lastly, the bottom-right figure focuses on the high-multipole performance ($l > 1000$), illustrating the regime where the reconstruction faces significant signal loss due to beam smearing and noise domination.

The raw difference in the signals are less than $0.5\mu K$, (approximately $2\%$) of the maximum pixel temperature in the true map. The residual signals Near the Galactic center, are lower compared to the temprature maps, except for a few isolated pixels where the difference reaches $1\mu\mathrm{K}$.

\begin{figure}[ht]
    \centering
    \includegraphics[trim=0cm 0.0cm 0.8cm 0.5cm, clip=true,width=0.49\textwidth]{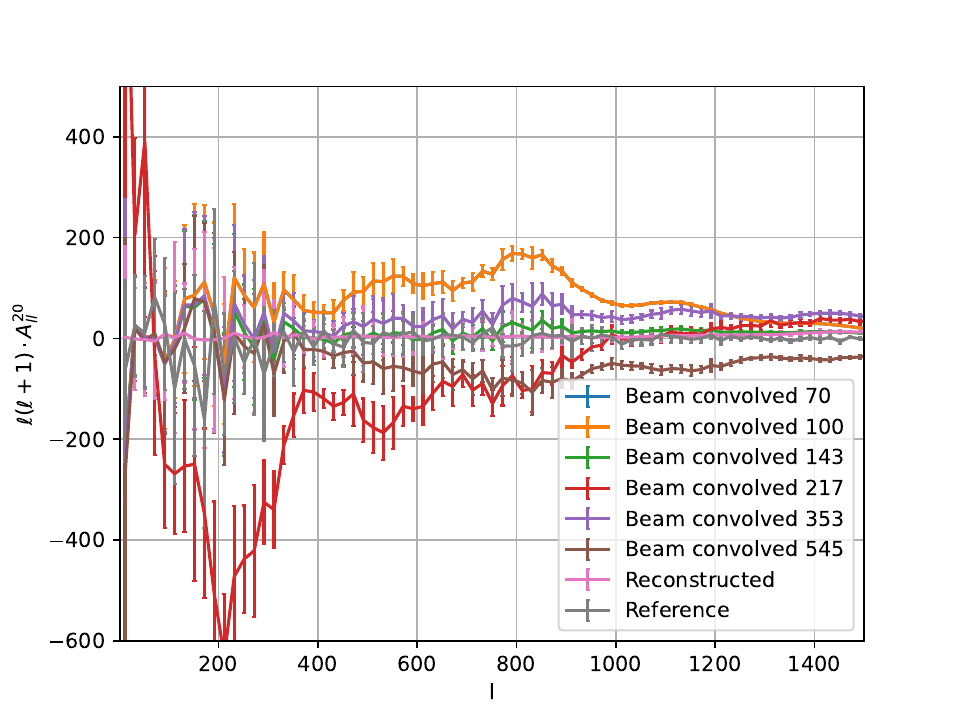}
    \hfill
    \includegraphics[trim=0cm 0.0cm 0.8cm 0.5cm, clip=true,width=0.49\textwidth]{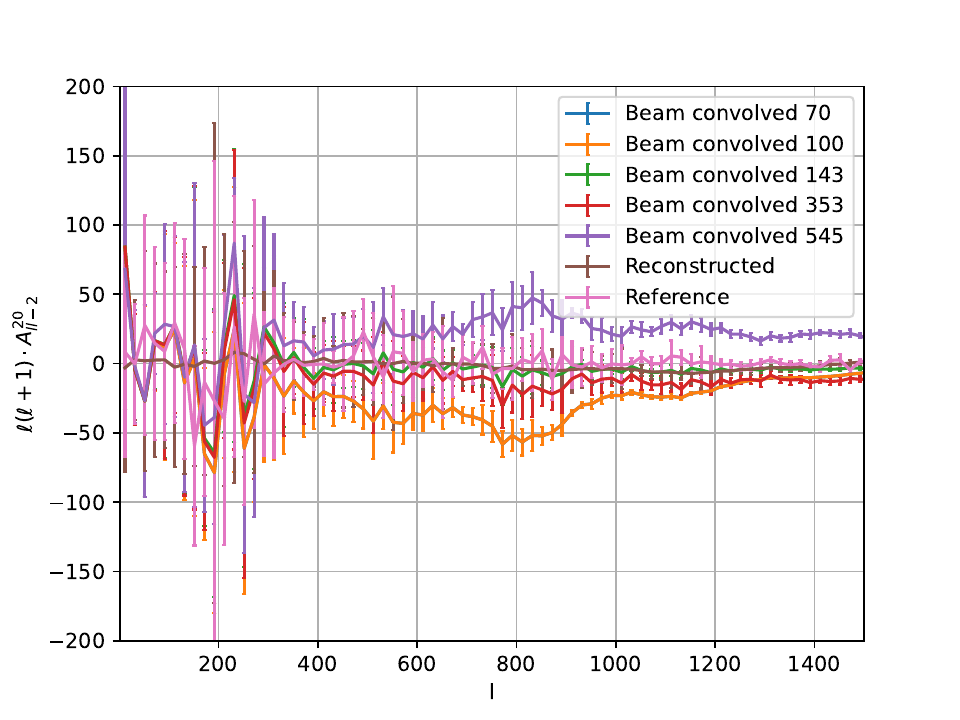}    
    \includegraphics[trim=0cm 0.0cm 0.8cm 0.5cm, clip=true,width=0.49\textwidth]{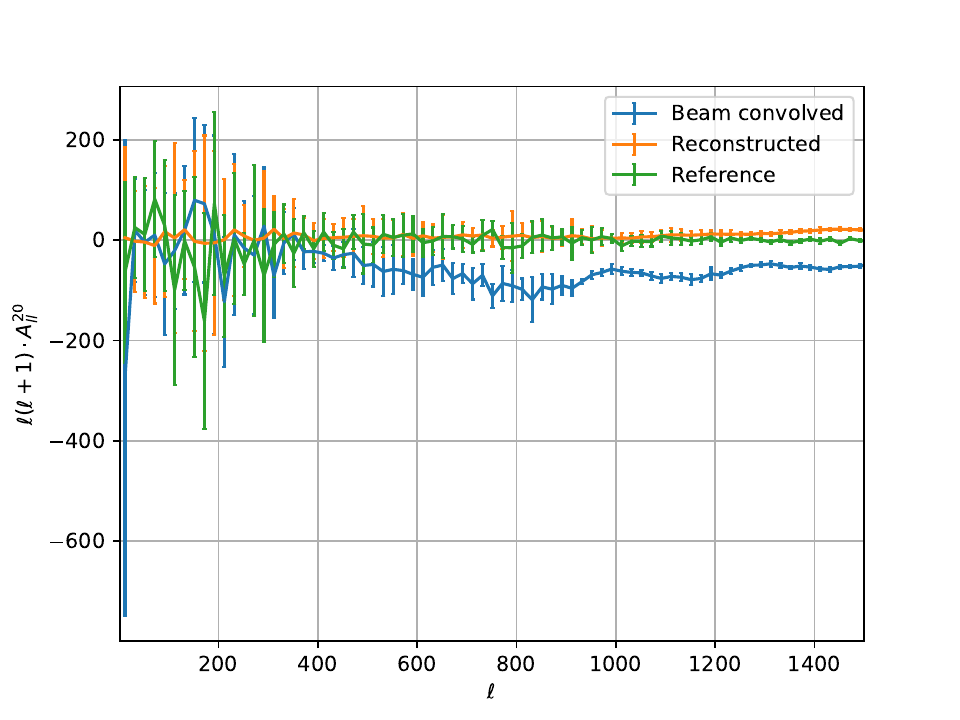}
    \hfill
    \includegraphics[trim=0cm 0.0cm 0.8cm 0.5cm, clip=true,width=0.49\textwidth]{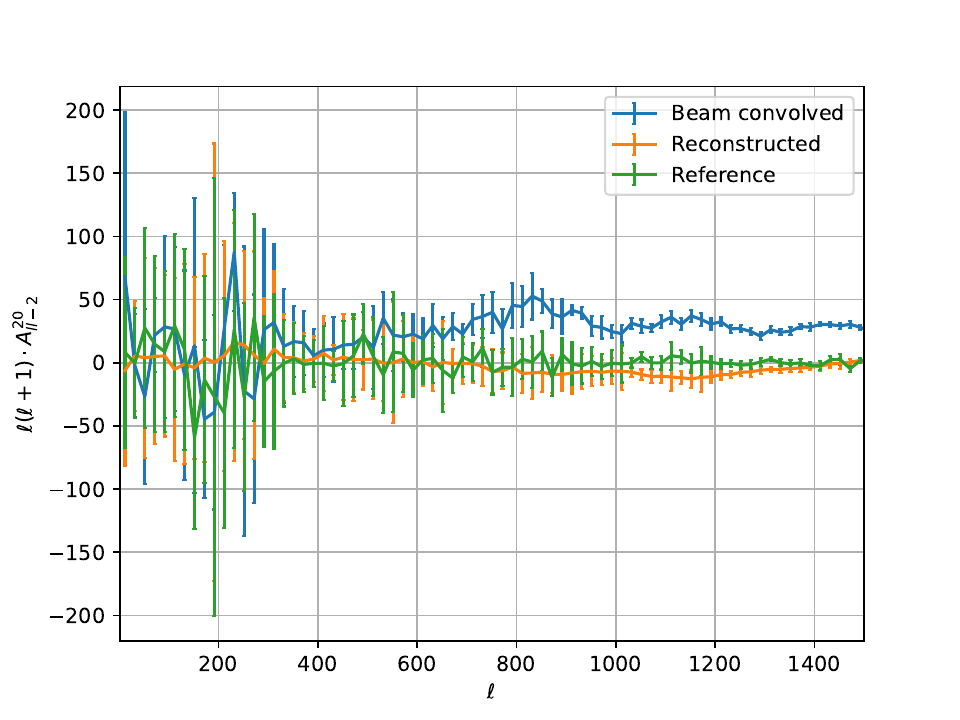}
    \caption{Bipolar Spherical Harmonic (BipoSH) coefficients evaluating the Statistical Isotropy (SI) of the CMB maps. \textbf{Top panels:} An overlay of the BipoSH coefficients across all six frequency channels for the input maps with reference, and reconstructed maps, illustrating the comprehensive distribution of the SI violation and its subsequent correction. \textbf{Bottom panels:} A detailed view isolating the $A^{20}_{ll}$ and $A^{20}_{ll-2}$ BipoSH coefficients for one input frequency.  the reference maps (green) are SI by construction, yielding coefficients consistent with zero. Convolution with the instrumental beam (blue) induces an SI violation, reflected by the non-zero coefficient values. The neural network reconstruction (orange) effectively suppresses this induced signal, restoring the coefficients to values consistent with zero and demonstrating that the reconstructed sky maps preserve statistical isotropy.
    }
    \label{fig:BipoSHtemp}
\end{figure}

\subsection{Exploring BipoSH coefficients }

In the standard cosmological model, the Universe is assumed to be statistically isotropic (SI). The CMB temperature anisotropies can be expanded in spherical harmonics as  
\begin{equation}
T(\hat{n}) = \sum_{l=0}^{\infty} \sum_{m=-l}^{l} a_{l m} \, Y_{l m}(\hat{n}),
\end{equation}

\begin{figure}[ht]
    \centering
    \includegraphics[trim=0cm 0.0cm 0.8cm 0.5cm, clip=true,width=0.49\textwidth]{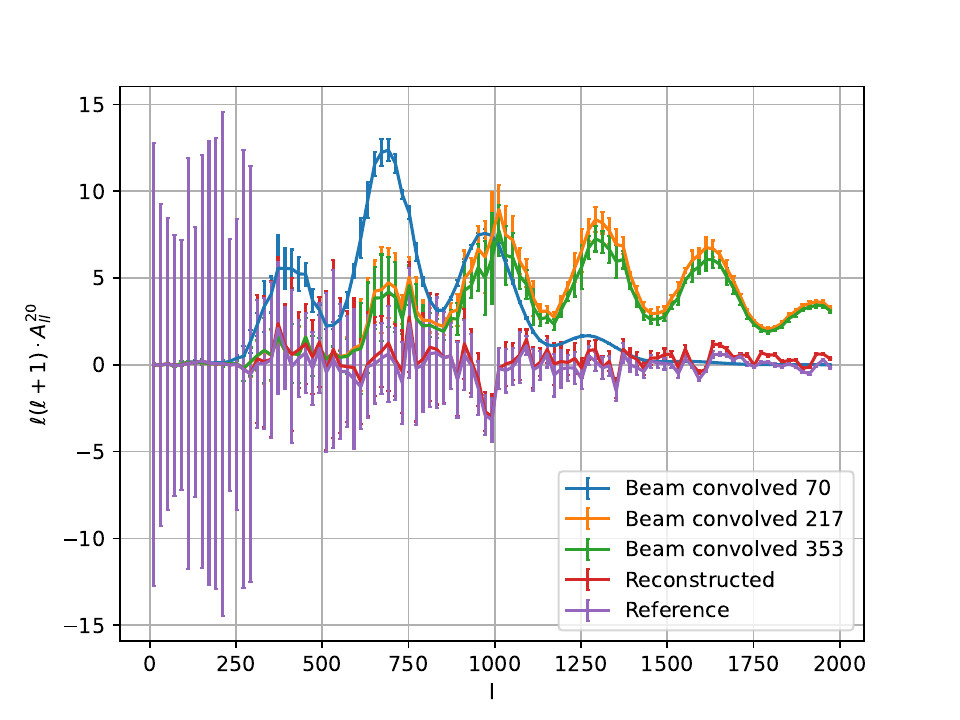}
    \hfill
    \includegraphics[trim=0cm 0.0cm 0.8cm 0.5cm, clip=true,width=0.49\textwidth]{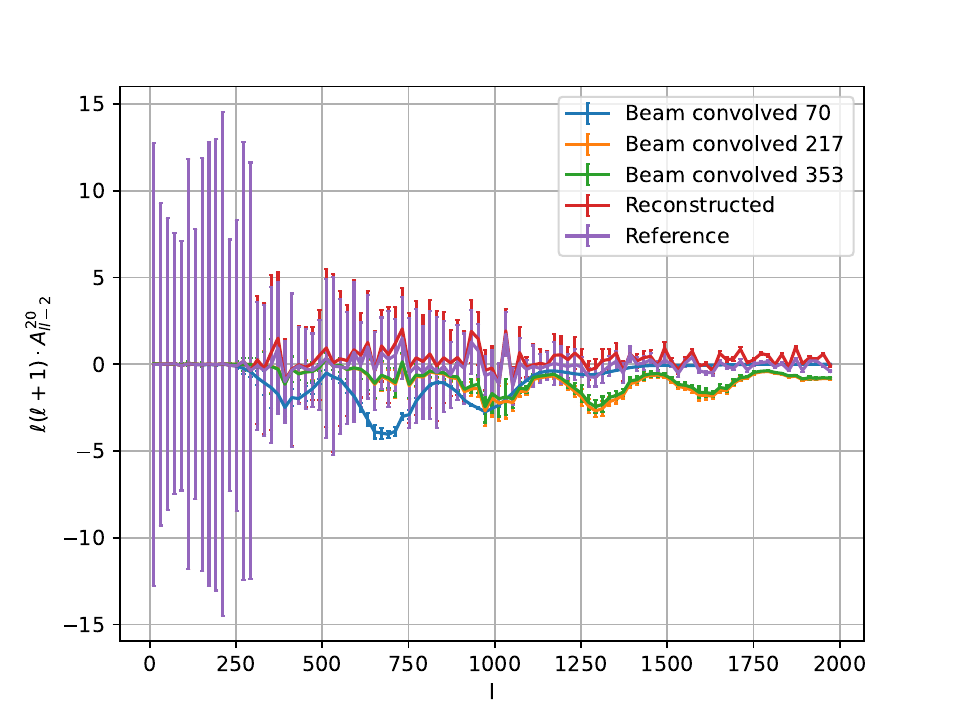}
    \caption{BipoSH coefficients evaluating Statistical Isotropy (SI) for the polarization ($E$-mode) maps. Similar to the analysis in Figure \ref{fig:BipoSHtemp}, this plot displays the coefficients across the three input frequency channels. The neural network reconstruction  effectively suppresses the beam-induced SI violation , restoring the coefficients to zero in alignment with the reference maps .}

    \label{fig:BipoSH}
\end{figure}

\noindent where $a_{l m}$ are the corresponding expansion coefficients. If these anisotropies constitute a Gaussian random field---consistent with high-precision Planck measurements ---their statistical properties are fully determined by the two-point correlation function. Under SI, this is equivalently expressed in harmonic space as  
\begin{equation}
\langle a_{l m} \, a_{l' m'}^{*} \rangle = C_{l} \, \delta_{l l'} \, \delta_{m m'},
\end{equation}
\noindent where $\langle \cdots \rangle$ denotes an ensemble average over CMB realizations.

However, in a non-SI universe, to capture the full description of the two-point correlation function, the BipoSH spectra are required~\cite{Hajian_2003}. The BipoSH spectra $A_{l l}^{L M}$ are defined by the expression~\cite{Bennett_2011},

\begin{equation}
\left\langle a_{l m} a_{l^{\prime} m^{\prime}}^*\right\rangle=\sum_{L M} A_{l l}^{L M} C_{l m l^{\prime} m^{\prime}}^{L M} C_{N r^{\prime} 0}^{L 0}\left(\Pi_l \Pi_{l^{\prime}} / \Pi_L\right),
\end{equation}

\noindent where $\Pi_L:=(2 L+1)^{1 / 2}$ and $C_{l m l^{\prime} m^{\prime}}^{L M}$ are the Clebsch-Gordan coefficients. While the BipoSH spectrum $A_{l l}^{00}=C_l$ is the standard angular power spectrum, detecting non-vanishing power in the remaining BipoSH spectra $A_{l l}^{L M}, L, M \neq 0$ forms the basic criteria for probing SI violation.

Non-circular beams, non-uniform scan patterns, and anisotropic noise introduces BipoSH coefficients in the observed CMB sky. Thus, if residual foregrounds, anisotropic noise, or uncorrected beam and scan effects remains, they can lead to isotropy violation in the reconstructed skymaps~\cite{Das:2018hnr, Das:2015gca, Kumar:2014nda, Pant_2016,Adam:2024kgs,Book_2012,aich2015weaklensing}.

Figure~\ref{fig:BipoSHtemp} shows the $A^{20}_{ll}$ and $A^{20}_{ll-2}$ BipoSH coefficients for the maps in ecliptic coordinates from the temperature skymaps. Because the scan pattern is symmetric across the upper and lower ecliptic hemispheres, these two coefficients are expected to capture the dominant contributions in the scanned sky maps. In the top plots we have the BipoSH coefficients from all the maps after beam convolution (before adding any noise and foreground). The reference maps, generated with HEALPix, are statistically isotropic and therefore yield BipoSH coefficients consistent with zero. In the plot we can see that all the data points are consistent with $0$. In contrast, the scanned and beam convolved maps display clear nonzero values, reflecting isotropy violations induced by the noncircular beam and scan strategy. The amplitude of the BipoSH coefficients is determined by the beam shape~\cite{Pant_2016}. To emphasize the BipoSH coefficients from both the reference and reconstructed maps in the bottom panel, we show the BipoSH coefficients for the reference (green) and reconstructed (orange) maps, together with those corresponding to the smallest beam, i.e., the 545 GHz channel. 

The addition of anisotropic noise and Galactic foregrounds should, in principle, further enhance such violations. However, the reconstructed maps show no such signatures, with the BipoSH coefficients returning to values consistent with zero. This demonstrates that the neural reconstruction not only suppresses residual foregrounds and anisotropic noise but also effectively removes the distortions introduced by the noncircular beam and scanning pattern—something conventional methods are unable to achieve. This makes the neural reconstruction approach significantly more powerful than existing techniques.

In Figure~\ref{fig:BipoSH}, we present similar plots the E-mode polarization maps at different frequencies. We observe that the BipoSH signatures induced by the beam shapes and the scan strategy are completely removed. However, the standard component-separation algorithm fails to eliminate the SI signal introduced by the beam convolution and the scan pattern.

Various previous studies have also noted that splitting maps into patches and training a neural network on each patch can introduce edge effects, potentially leading to anomalous signals in the final skymap~\cite{Yan:2025csf}. Any such effect would leave an imprint in the BipoSH coefficients. As the BipoSH coefficients from the final maps are consistent with zero, we conclude that no statistically significant signal due to patch-wise reconstruction is present in the final maps.


\begin{table}[htbp]
  \centering
  \caption{Reconstruction results for patches 0-5. The columns show: Input map with foregrounds, the network's predicted CMB, the true CMB, and the difference between the predicted and true CMB.}
  \label{fig:patches0-5} 
  \begin{tabular}{@{}*{4}{m{\imgcolwidth}}@{}}

    \toprule
    \textbf{Input} & \textbf{Predicted} & \textbf{True} & \textbf{Diff} \\
    \midrule
    \includegraphics[trim=1.6cm 0.8cm 1.8cm 0.8cm, clip, width=\graphicwidth]{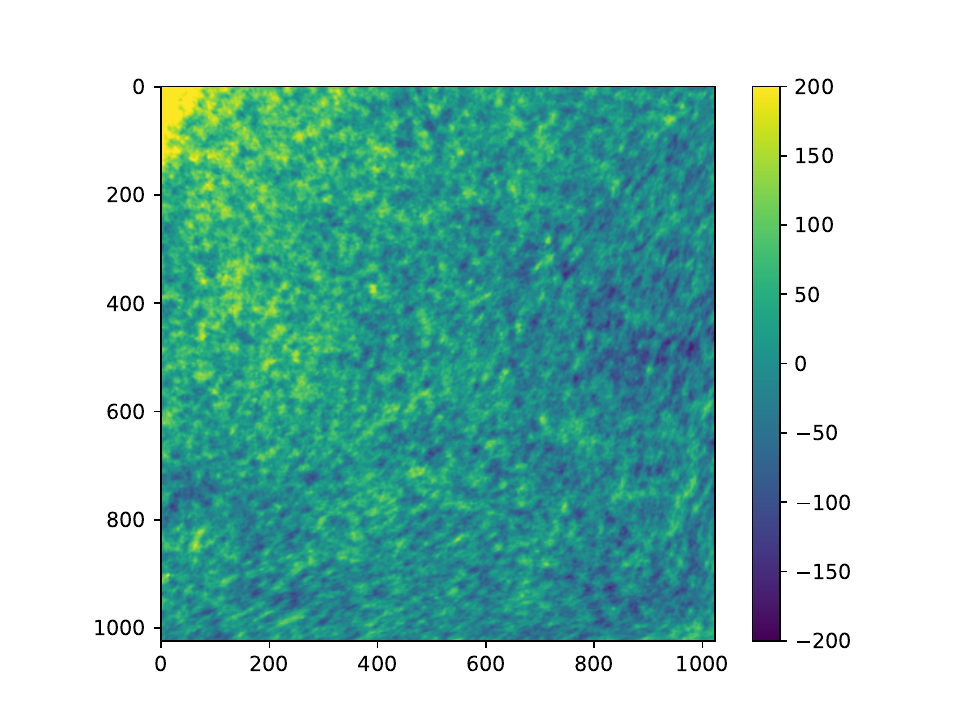} &
    \includegraphics[trim=1.6cm 0.8cm 1.8cm 0.8cm, clip, width=\graphicwidth]{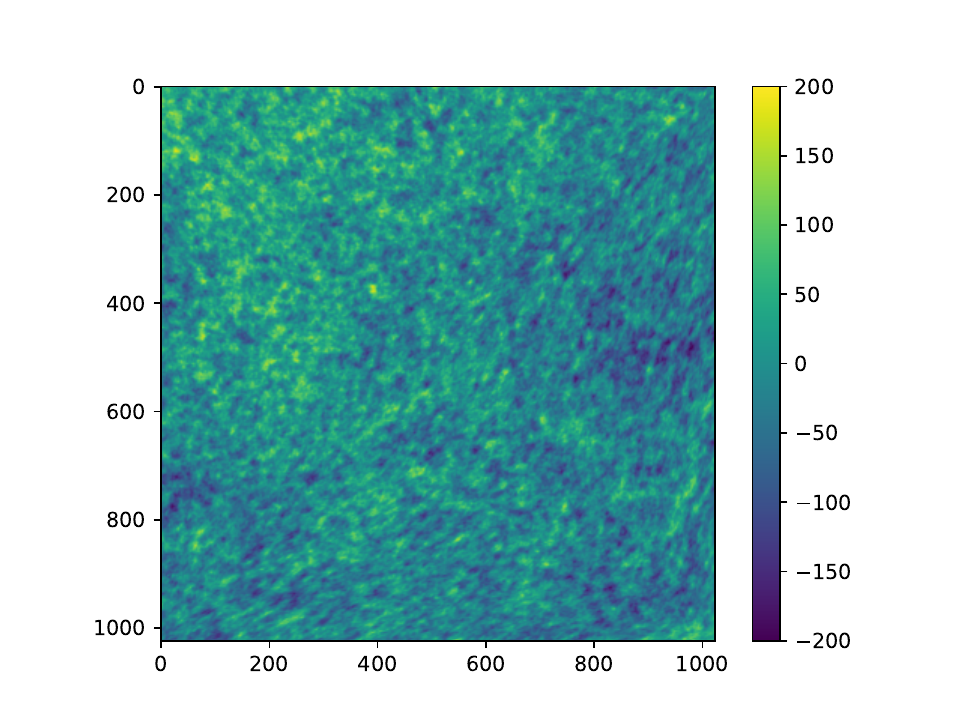} &
    \includegraphics[trim=1.6cm 0.8cm 1.8cm 0.8cm, clip, width=\graphicwidth]{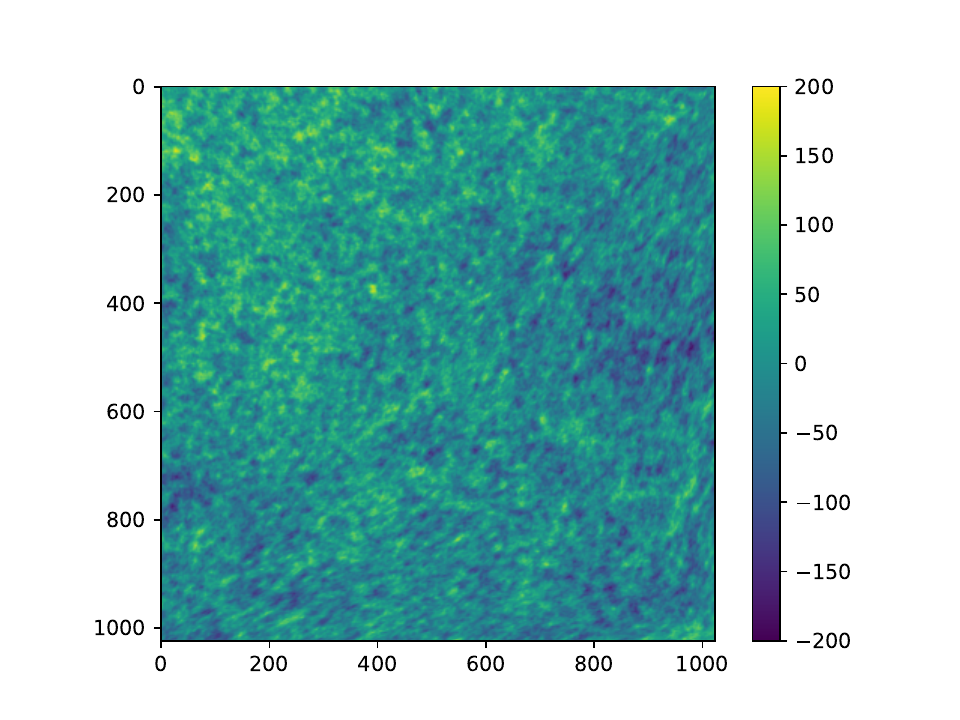} &
    \includegraphics[trim=1.6cm 0.8cm 1.8cm 0.8cm, clip, width=\graphicwidth]{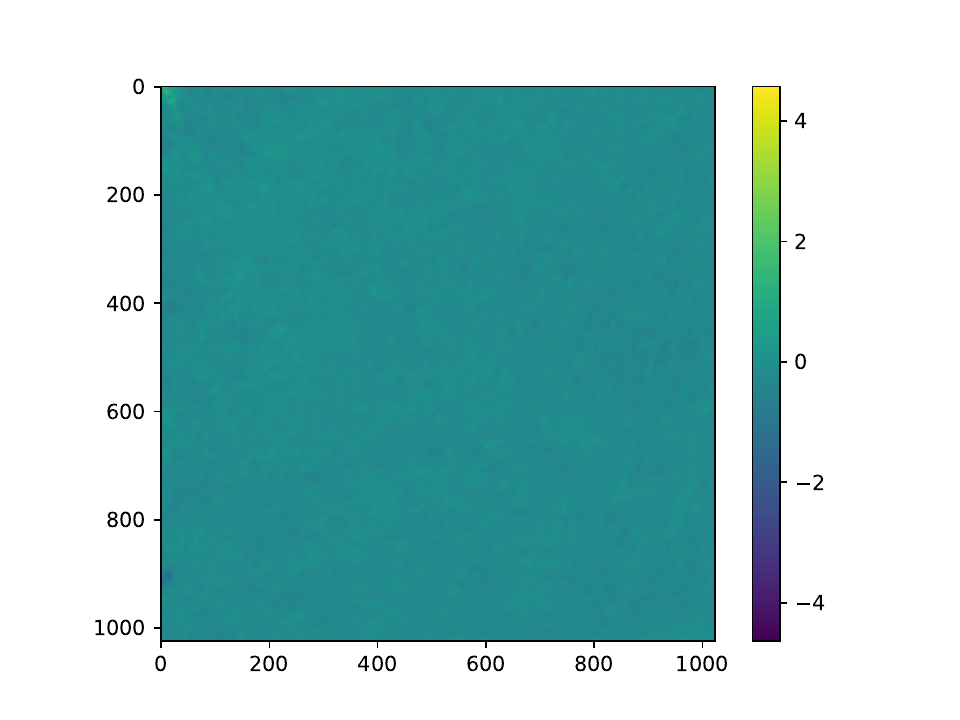} \\
    \addlinespace[1ex]        
    \includegraphics[trim=1.6cm 0.8cm 1.8cm 0.8cm, clip, width=\graphicwidth]{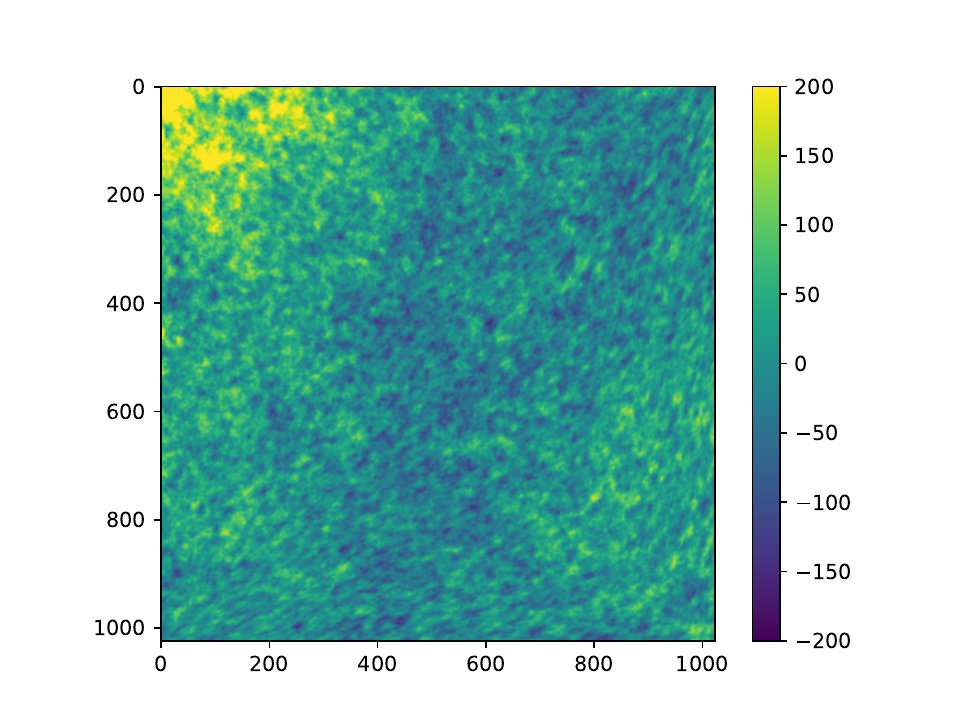} &
    \includegraphics[trim=1.6cm 0.8cm 1.8cm 0.8cm, clip, width=\graphicwidth]{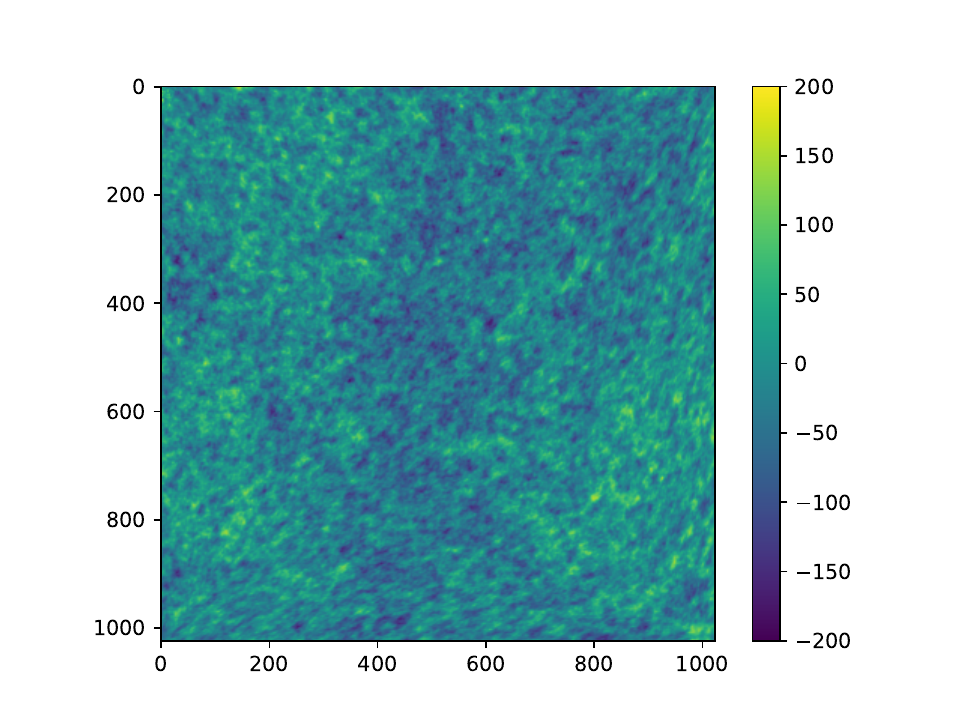} &
    \includegraphics[trim=1.6cm 0.8cm 1.8cm 0.8cm, clip, width=\graphicwidth]{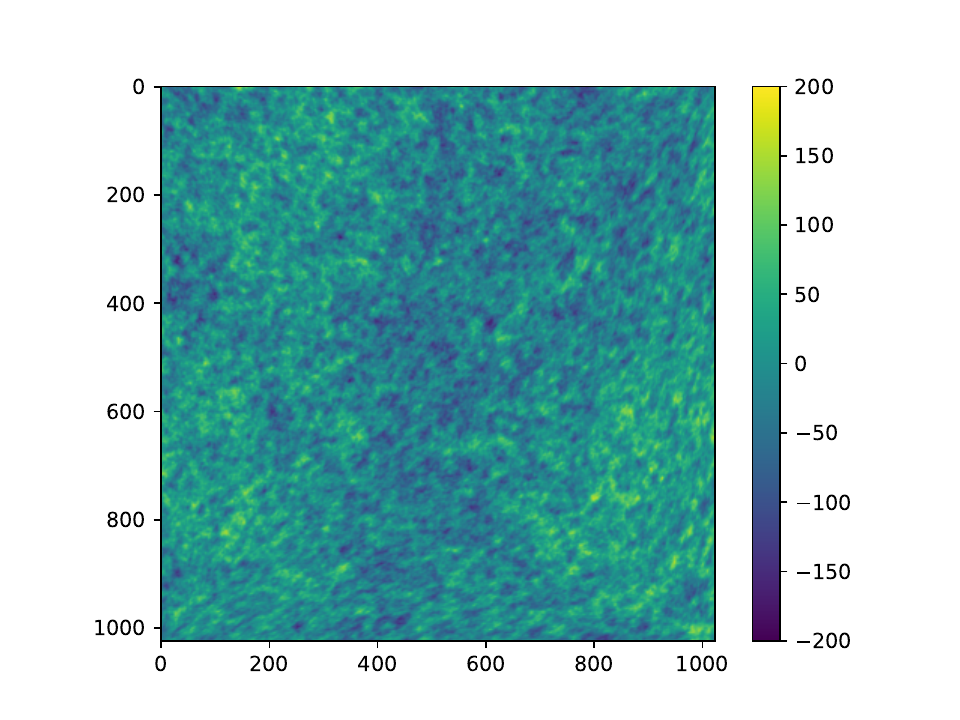} &
    \includegraphics[trim=1.6cm 0.8cm 1.8cm 0.8cm, clip, width=\graphicwidth]{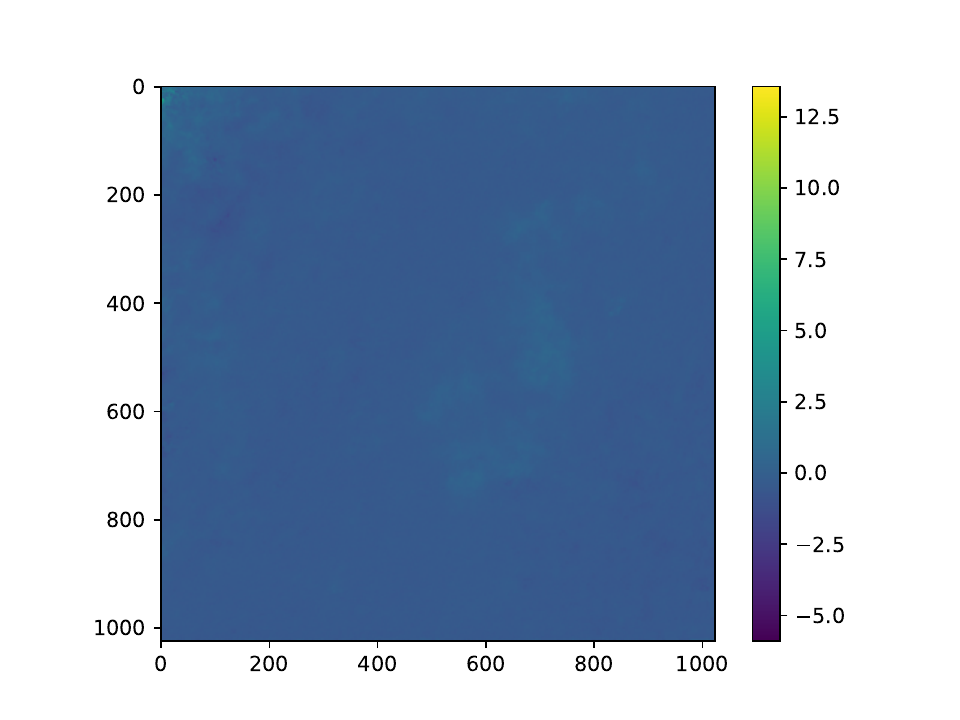} \\
    \addlinespace[1ex]
    \includegraphics[trim=1.6cm 0.8cm 1.8cm 0.8cm, clip, width=\graphicwidth]{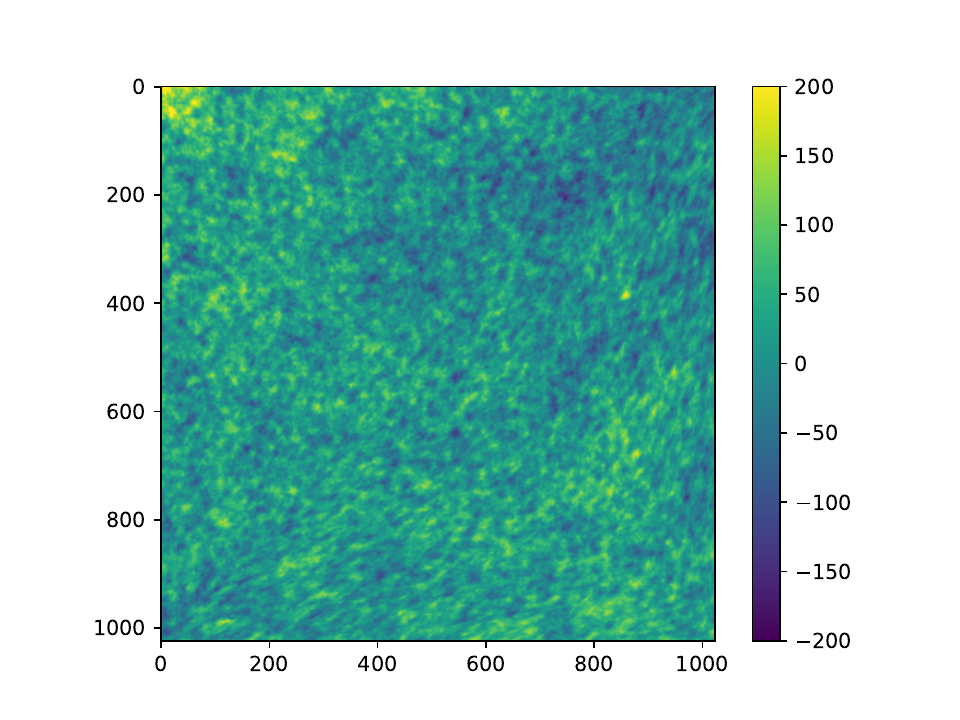} &
    \includegraphics[trim=1.6cm 0.8cm 1.8cm 0.8cm, clip, width=\graphicwidth]{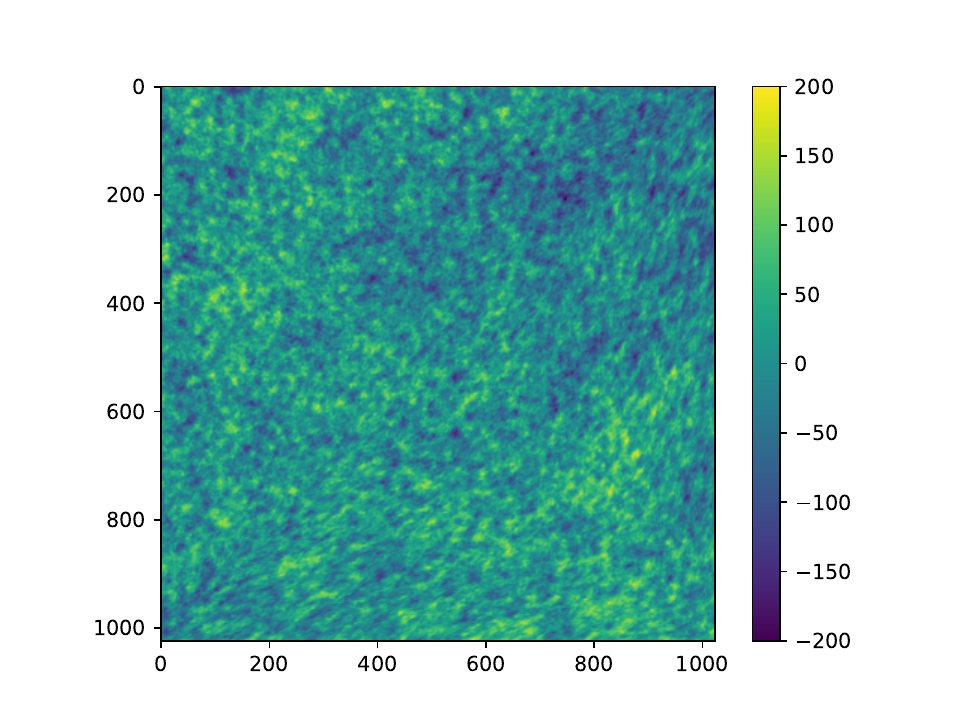} &
    \includegraphics[trim=1.6cm 0.8cm 1.8cm 0.8cm, width=\graphicwidth]{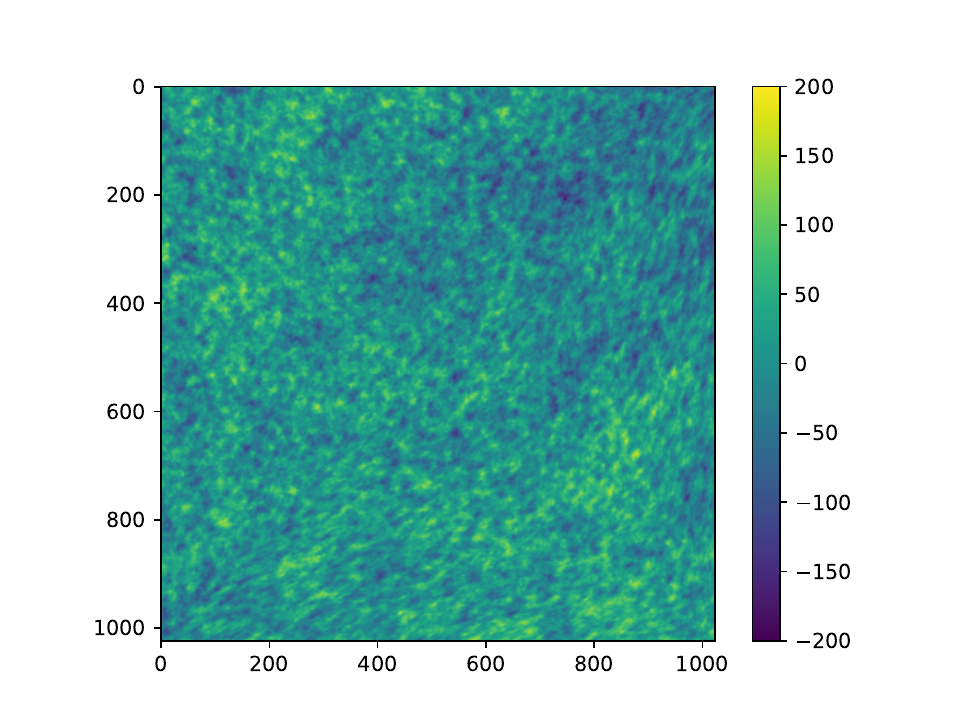} &
    \includegraphics[trim=1.6cm 0.8cm 1.8cm 0.8cm, clip, width=\graphicwidth]{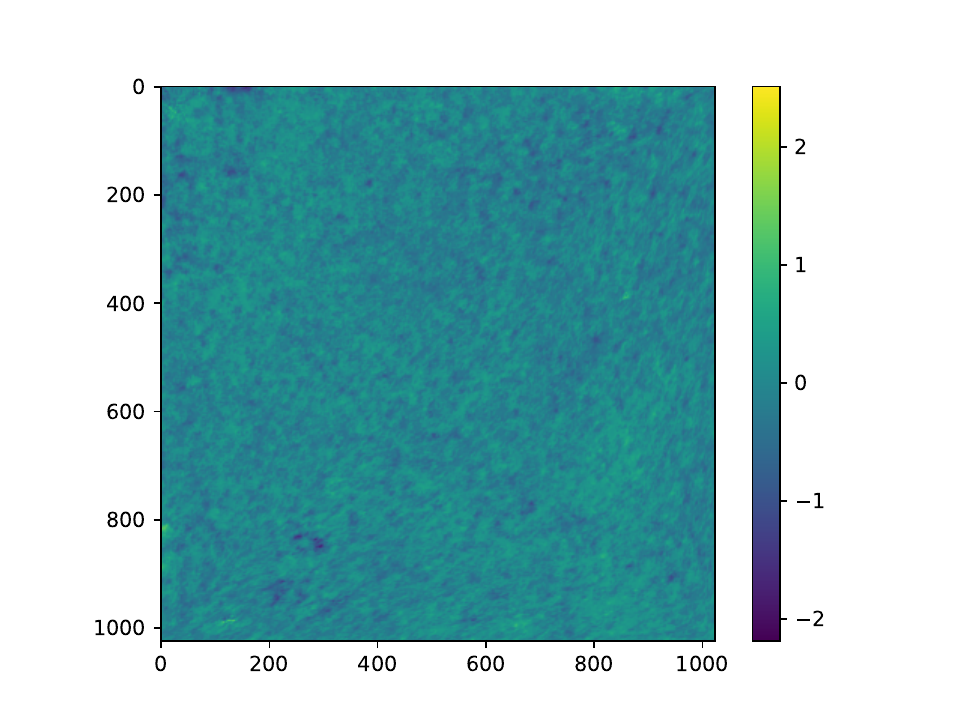} \\
    \addlinespace[1ex]
    \includegraphics[trim=1.6cm 0.8cm 1.8cm 0.8cm, clip, width=\graphicwidth]{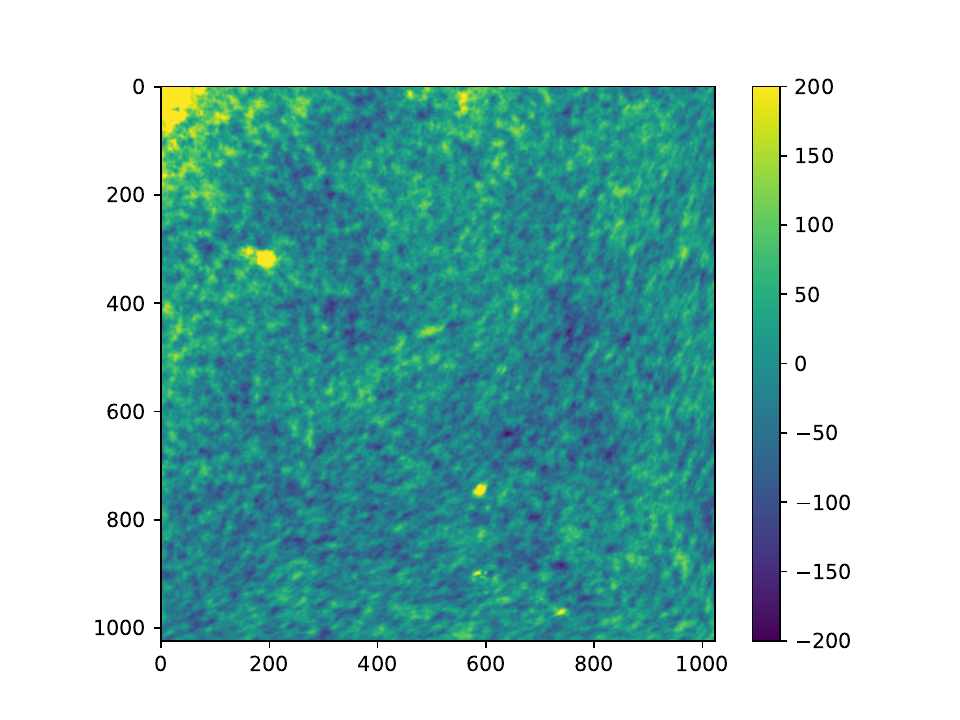} &
    \includegraphics[trim=1.6cm 0.8cm 1.8cm 0.8cm, clip, width=\graphicwidth]{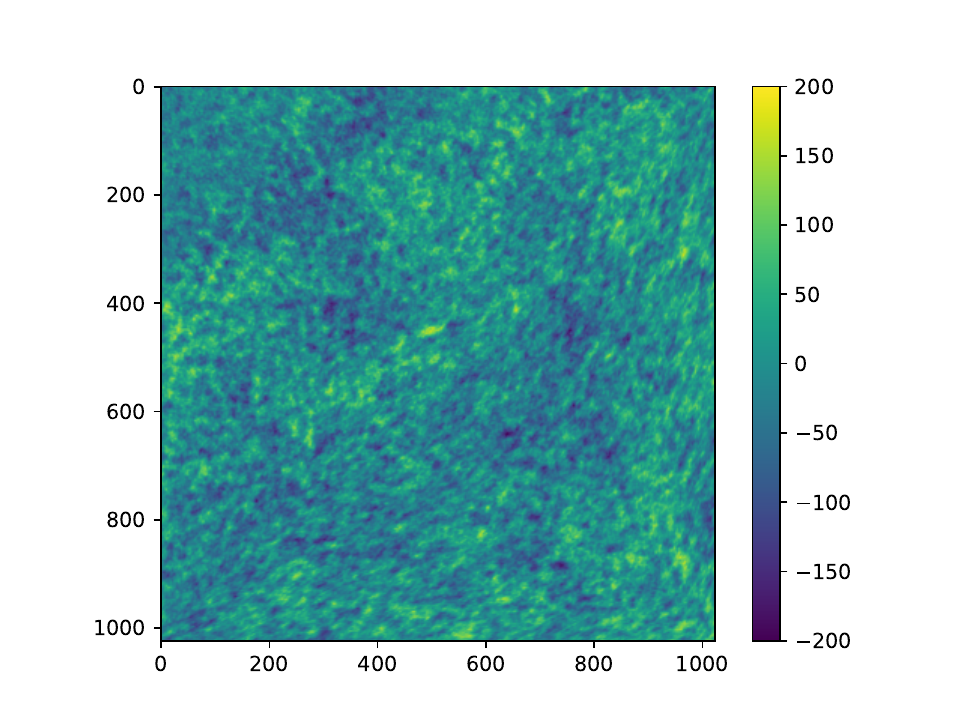} &
    \includegraphics[trim=1.6cm 0.8cm 1.8cm 0.8cm, clip, width=\graphicwidth]{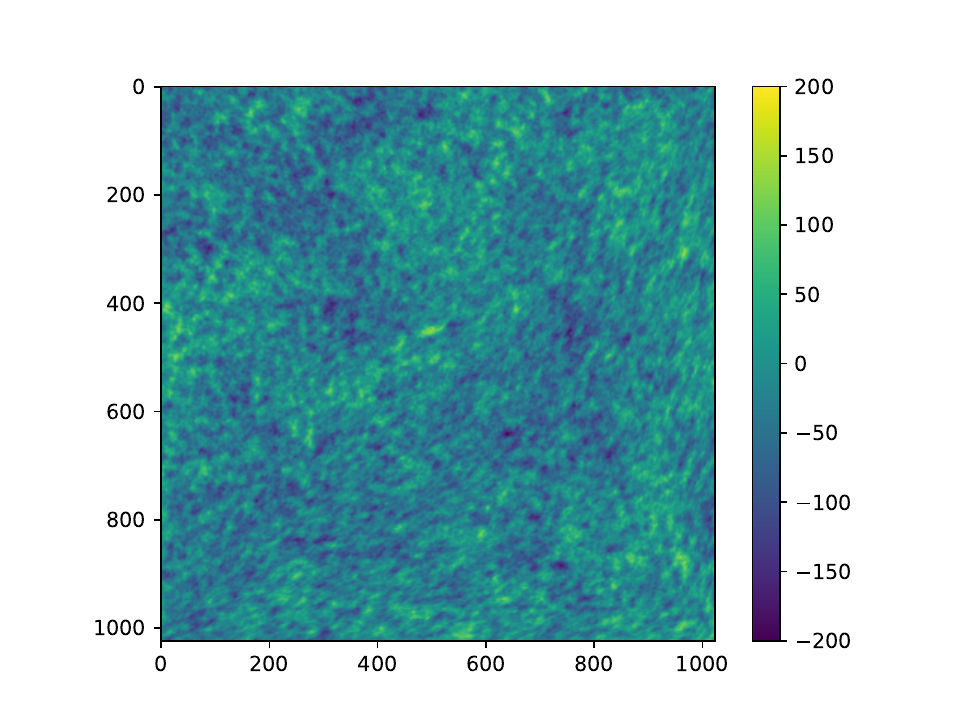} &
    \includegraphics[trim=1.6cm 0.8cm 1.8cm 0.8cm, clip, width=\graphicwidth]{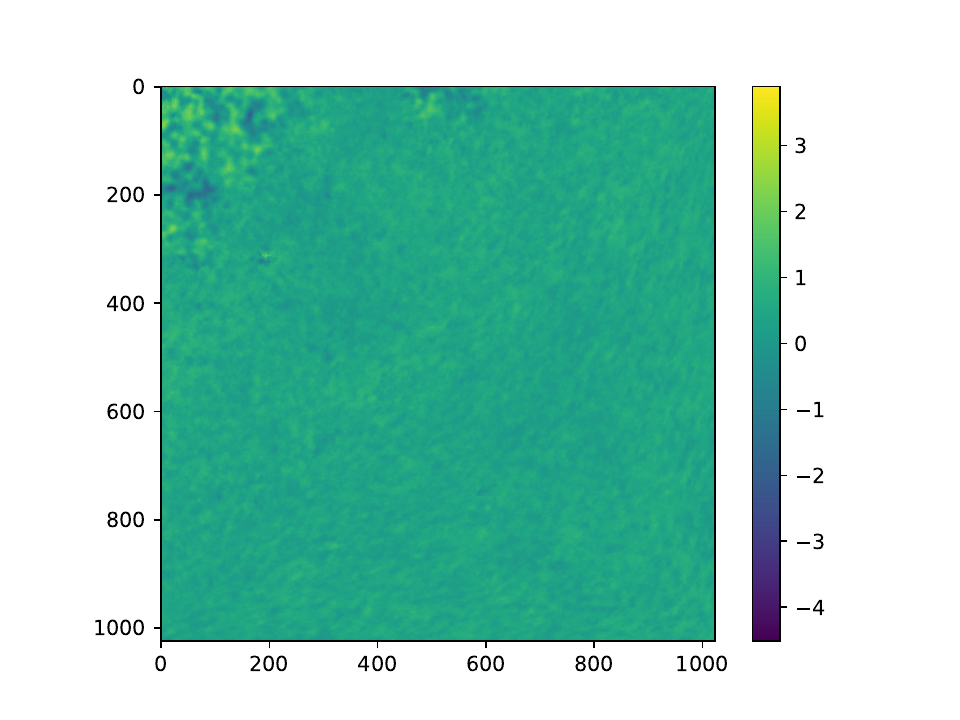} \\
    \addlinespace[1ex]
    \includegraphics[trim=1.6cm 0.8cm 1.8cm 0.8cm, clip, width=\graphicwidth]{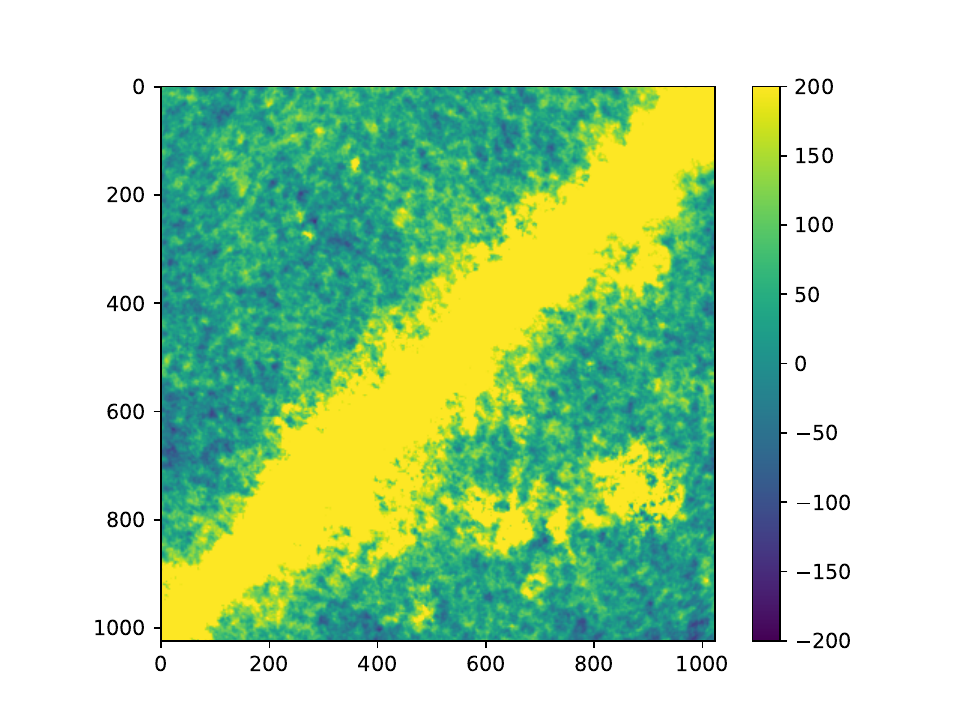} &
    \includegraphics[trim=1.6cm 0.8cm 1.8cm 0.8cm, clip, width=\graphicwidth]{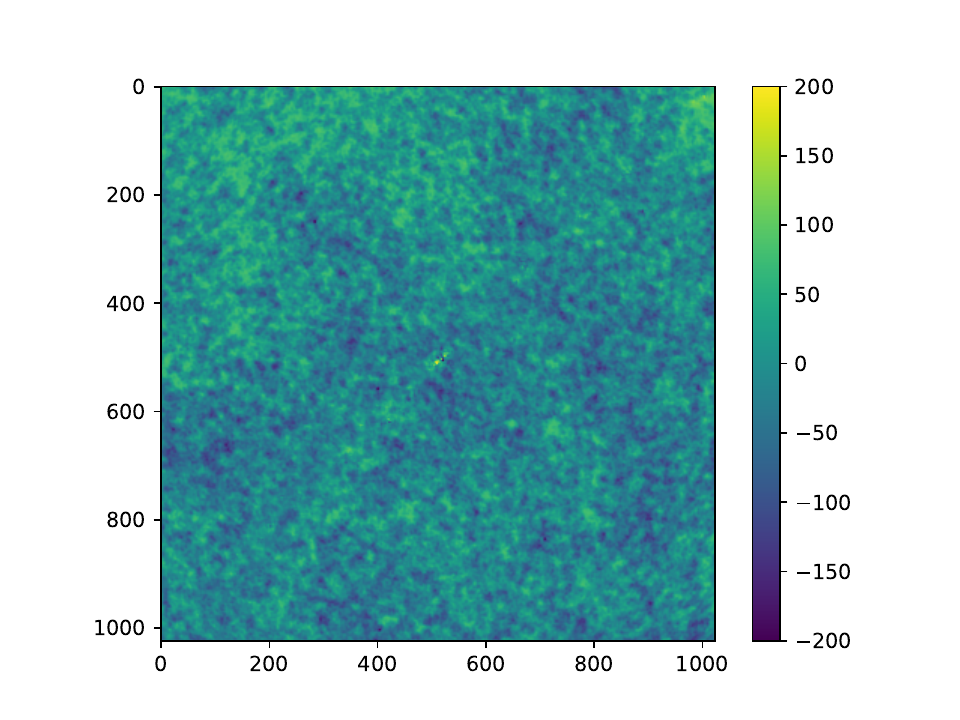} &
    \includegraphics[trim=1.6cm 0.8cm 1.8cm 0.8cm, clip, width=\graphicwidth]{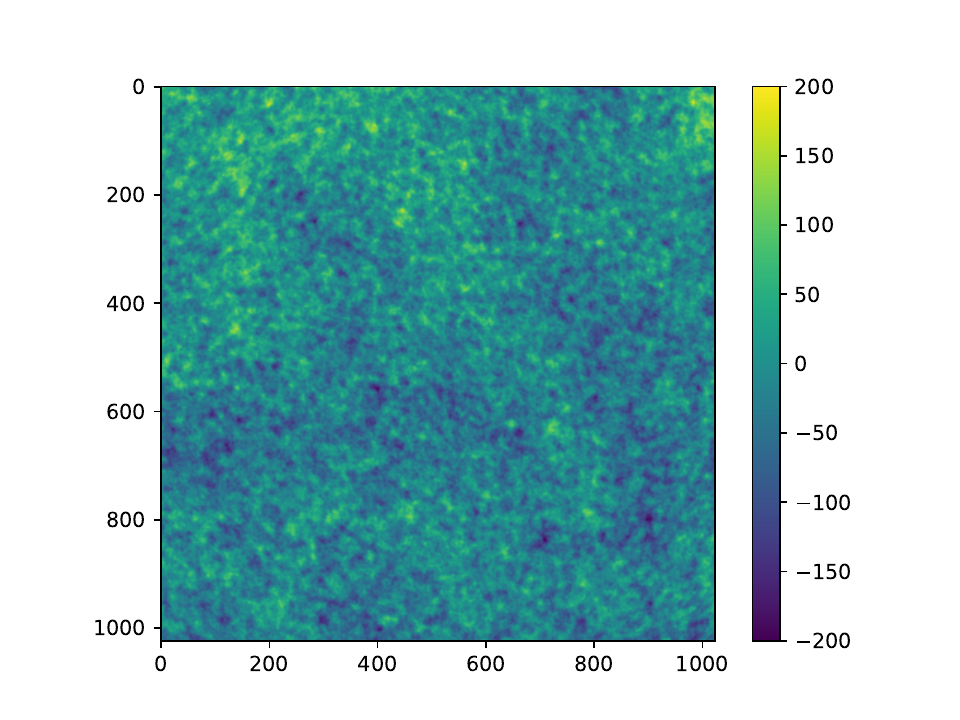} &
    \includegraphics[trim=1.6cm 0.8cm 1.8cm 0.8cm, clip, width=\graphicwidth]{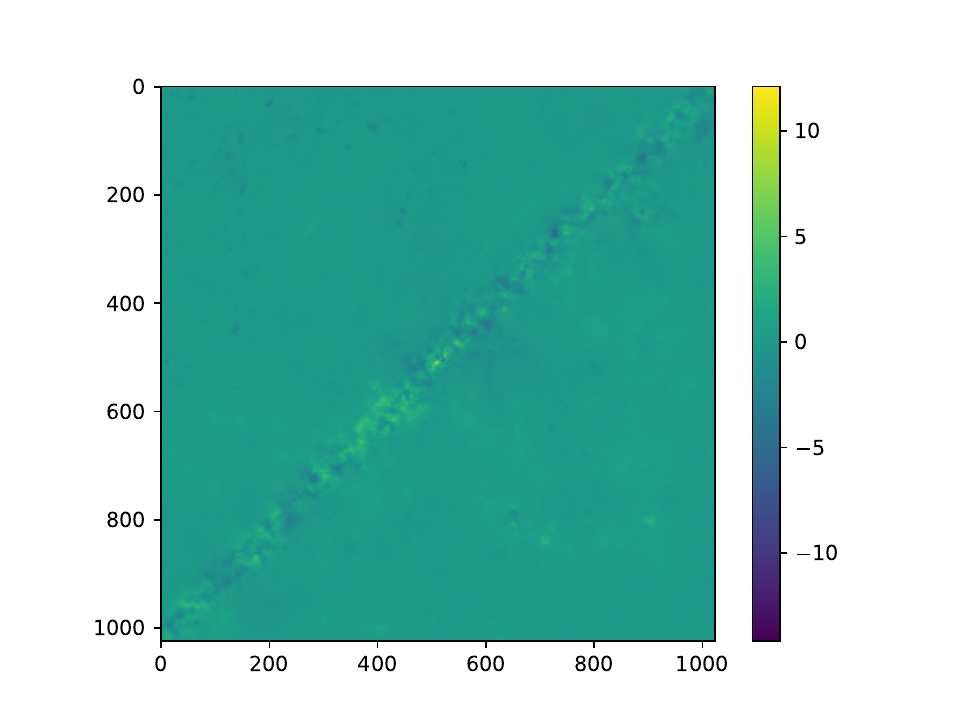} \\
    \addlinespace[1ex]    
    \includegraphics[trim=1.6cm 0.8cm 1.8cm 0.8cm, clip, width=\graphicwidth]{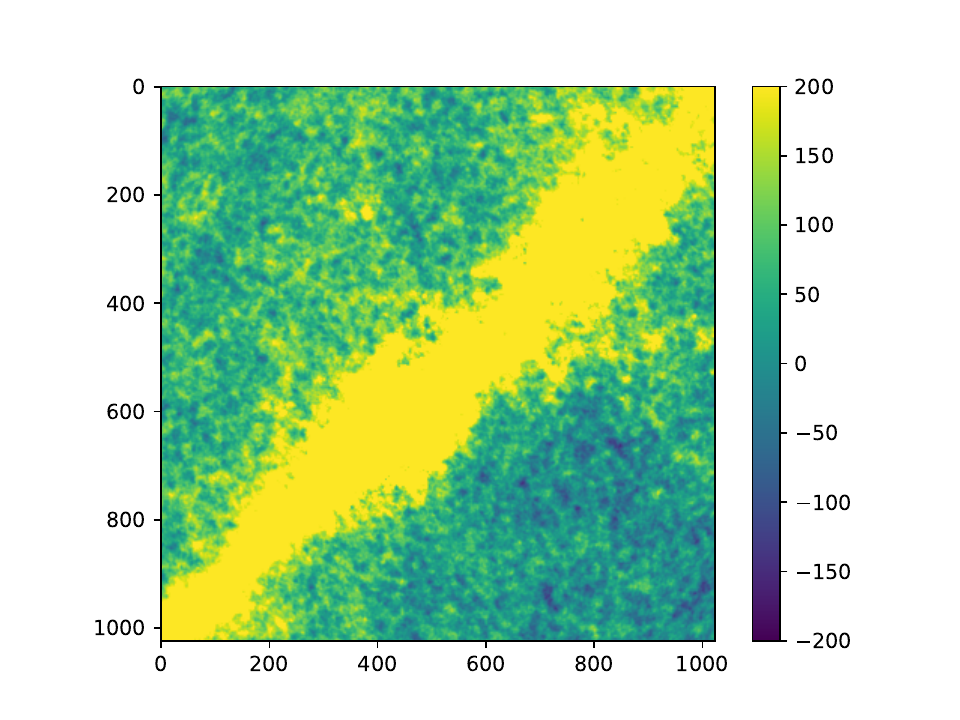} &
    \includegraphics[trim=1.6cm 0.8cm 1.8cm 0.8cm, clip, width=\graphicwidth]{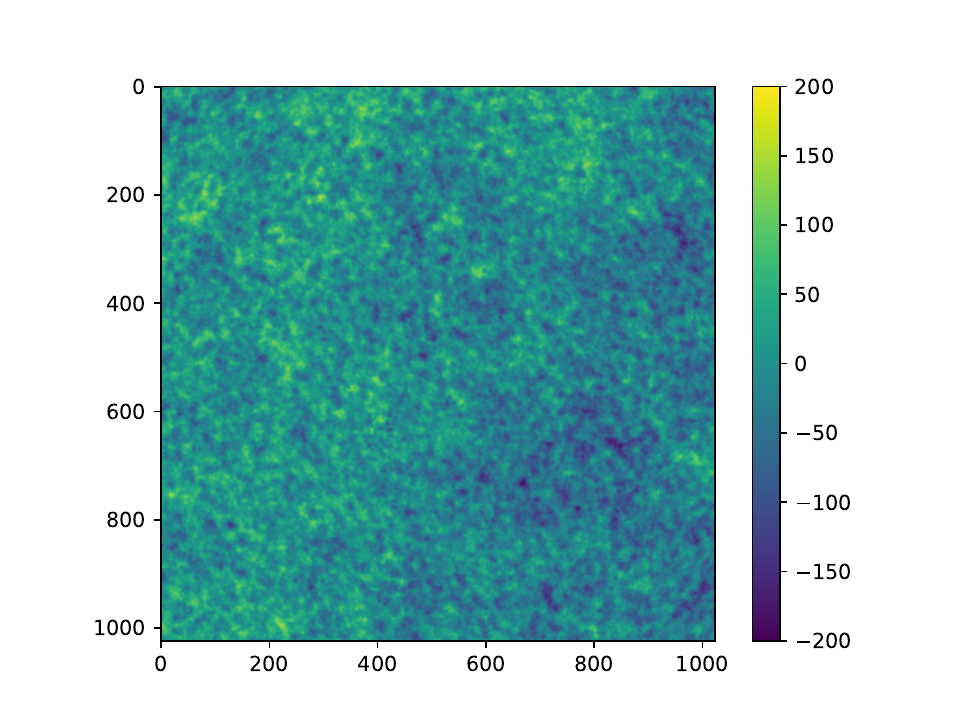} &
    \includegraphics[trim=1.6cm 0.8cm 1.8cm 0.8cm, clip, width=\graphicwidth]{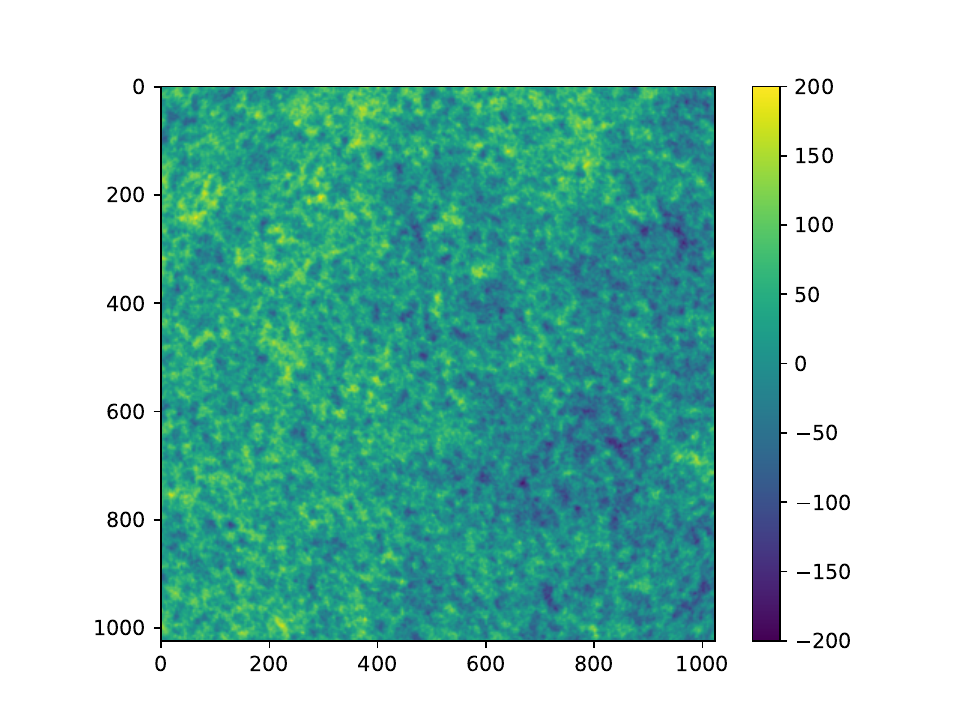} &
    \includegraphics[trim=1.6cm 0.8cm 1.8cm 0.8cm, clip, width=\graphicwidth]{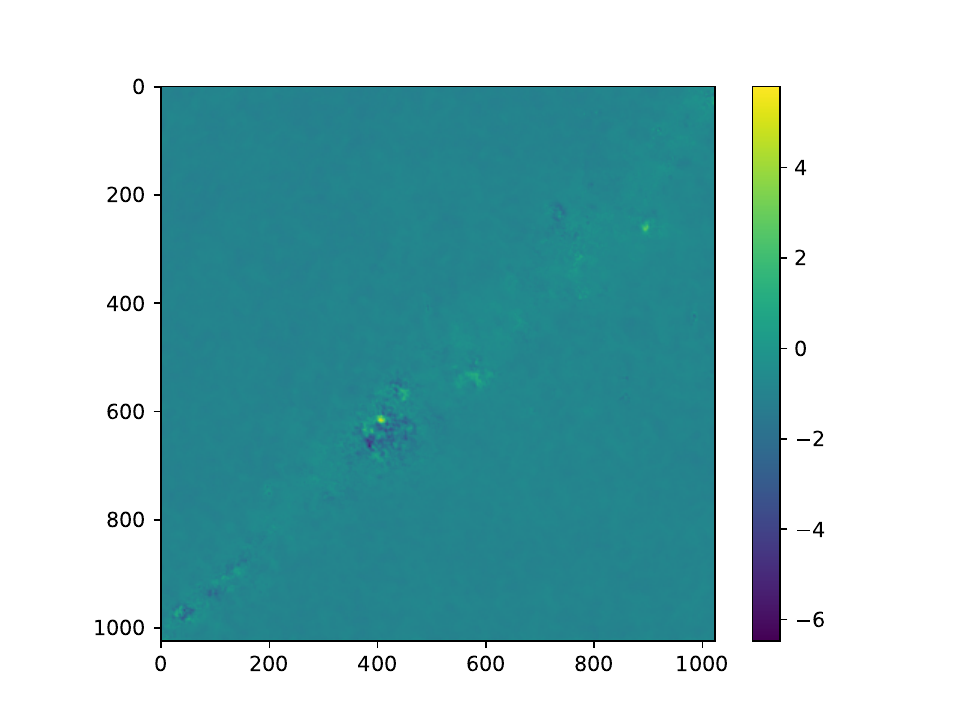} \\
    \bottomrule
  \end{tabular}
\end{table}
\begin{table}[htbp]
  \centering
  \caption{Reconstruction results for patches 6-11. The columns show: Input map with foregrounds, the network's predicted CMB, the true CMB, and the difference between the predicted and true CMB.}
  \label{fig:patches-6-11}
  \begin{tabular}{@{}*{4}{m{\imgcolwidth}}@{}}
    \toprule
    \textbf{Input} & \textbf{Predicted} & \textbf{True} & \textbf{Diff} \\
    \midrule
    \includegraphics[trim=1.6cm 0.8cm 1.8cm 0.8cm, clip, width=\graphicwidth]{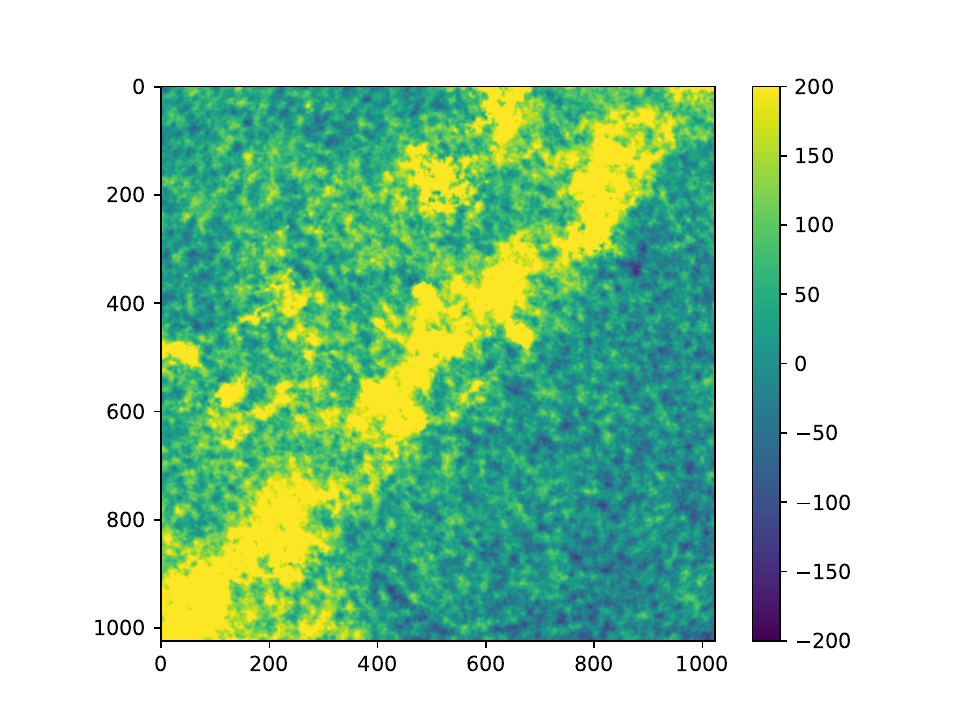} &
    \includegraphics[trim=1.6cm 0.8cm 1.8cm 0.8cm, clip, width=\graphicwidth]{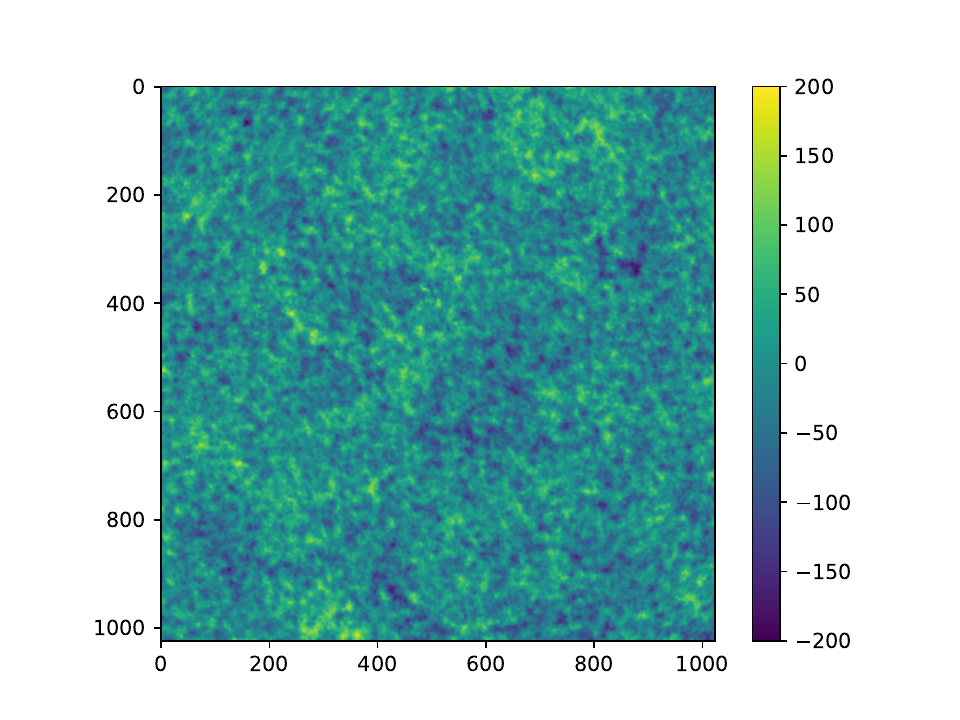} &
    \includegraphics[trim=1.6cm 0.8cm 1.8cm 0.8cm, clip, width=\graphicwidth]{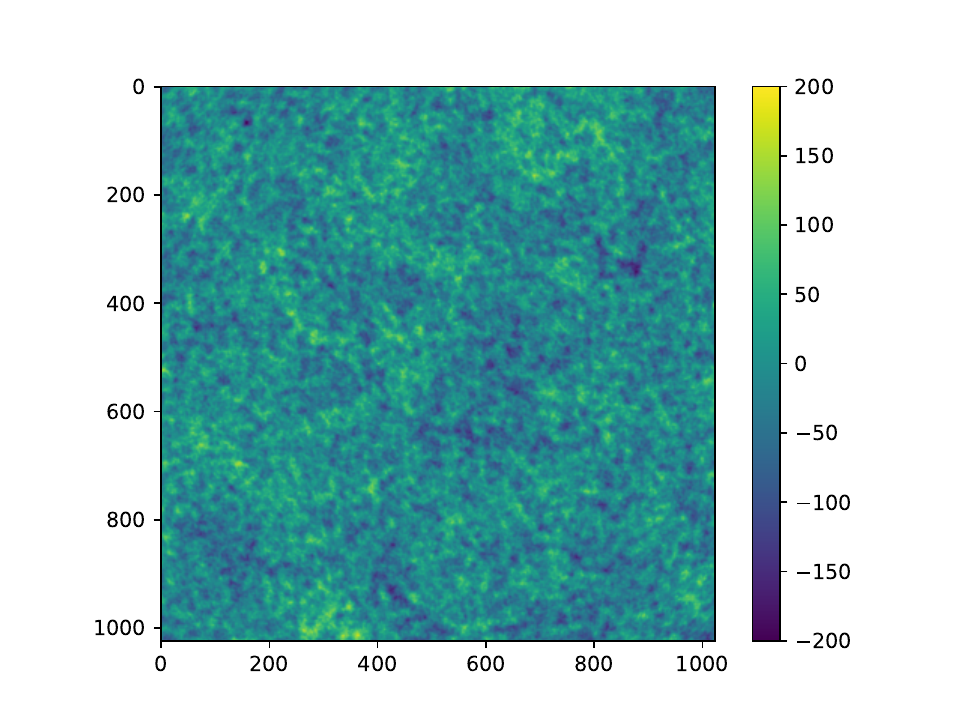} &
    \includegraphics[trim=1.6cm 0.8cm 1.8cm 0.8cm, clip, width=\graphicwidth]{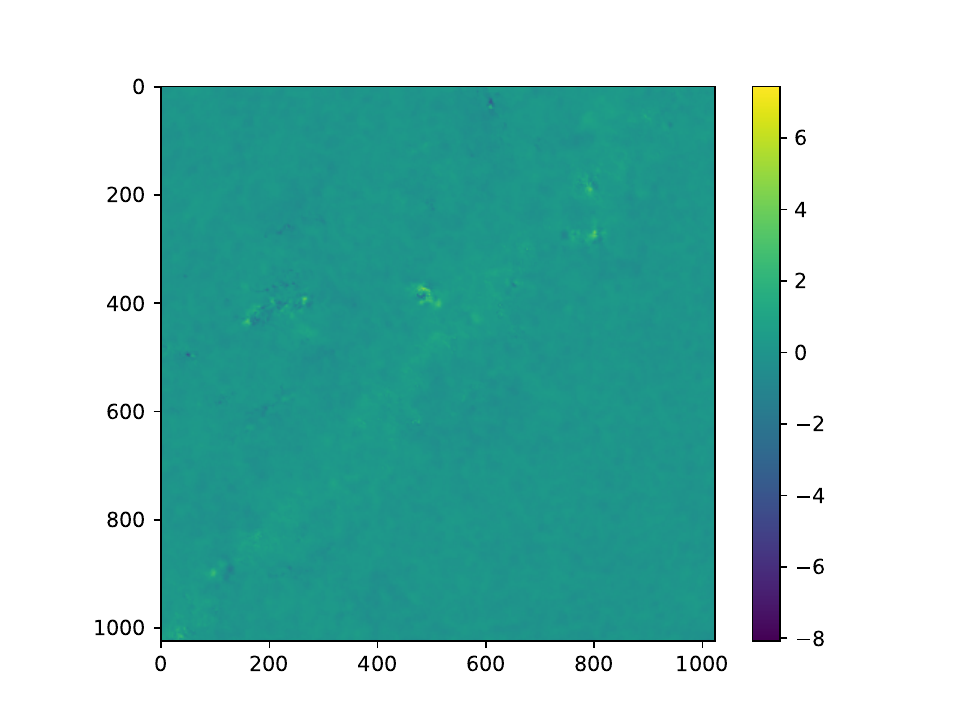} \\
    \addlinespace[1ex]        
    \includegraphics[trim=1.6cm 0.8cm 1.8cm 0.8cm, clip, width=\graphicwidth]{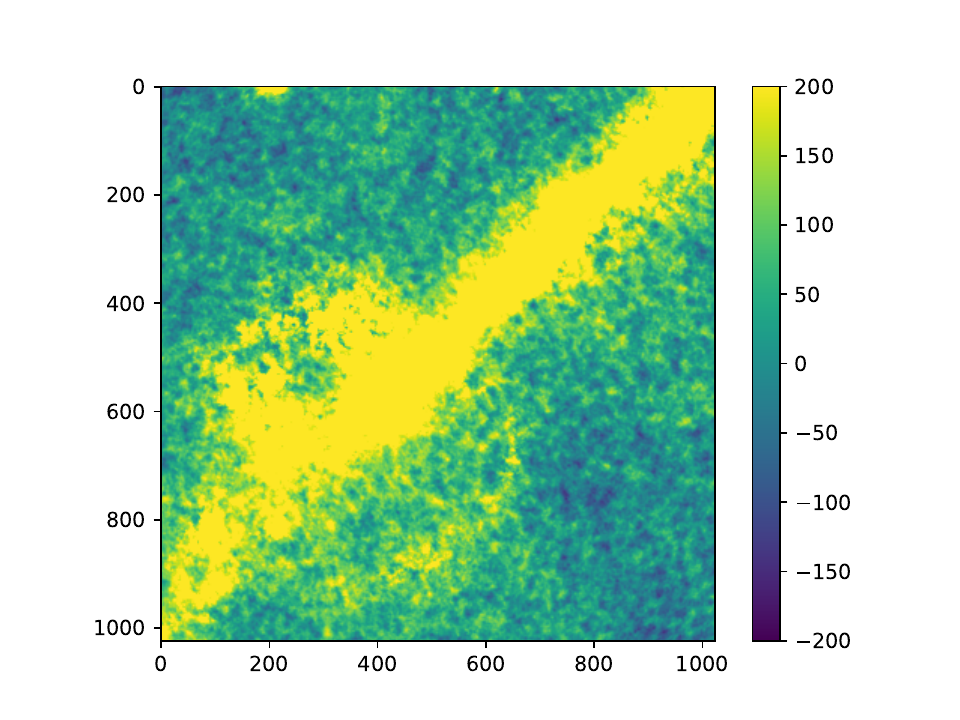} &
    \includegraphics[trim=1.6cm 0.8cm 1.8cm 0.8cm, clip, width=\graphicwidth]{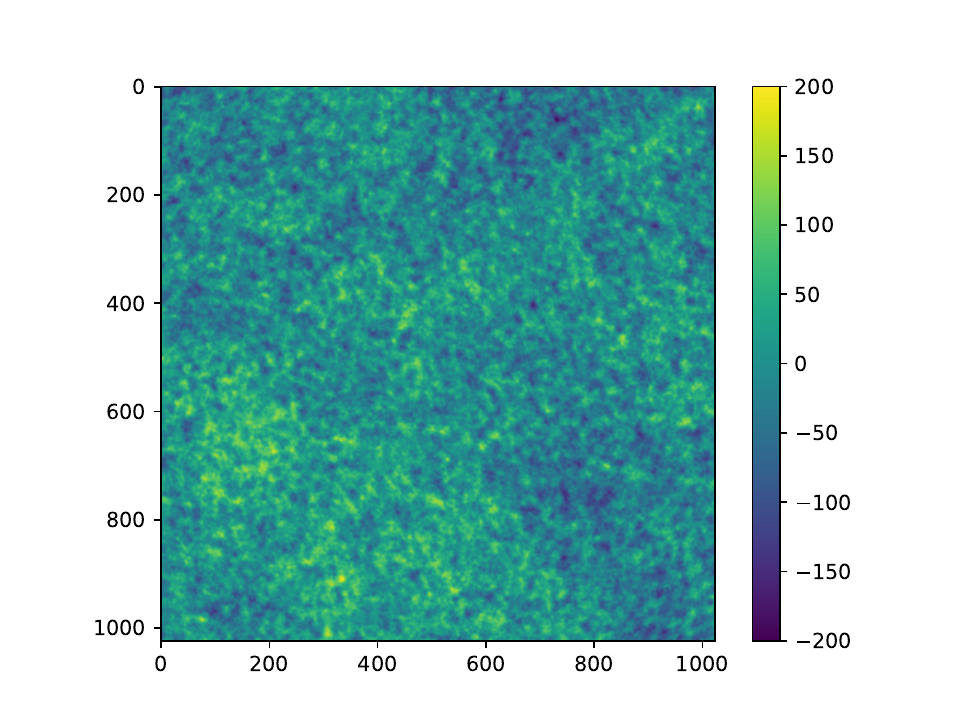} &
    \includegraphics[trim=1.6cm 0.8cm 1.8cm 0.8cm, clip, width=\graphicwidth]{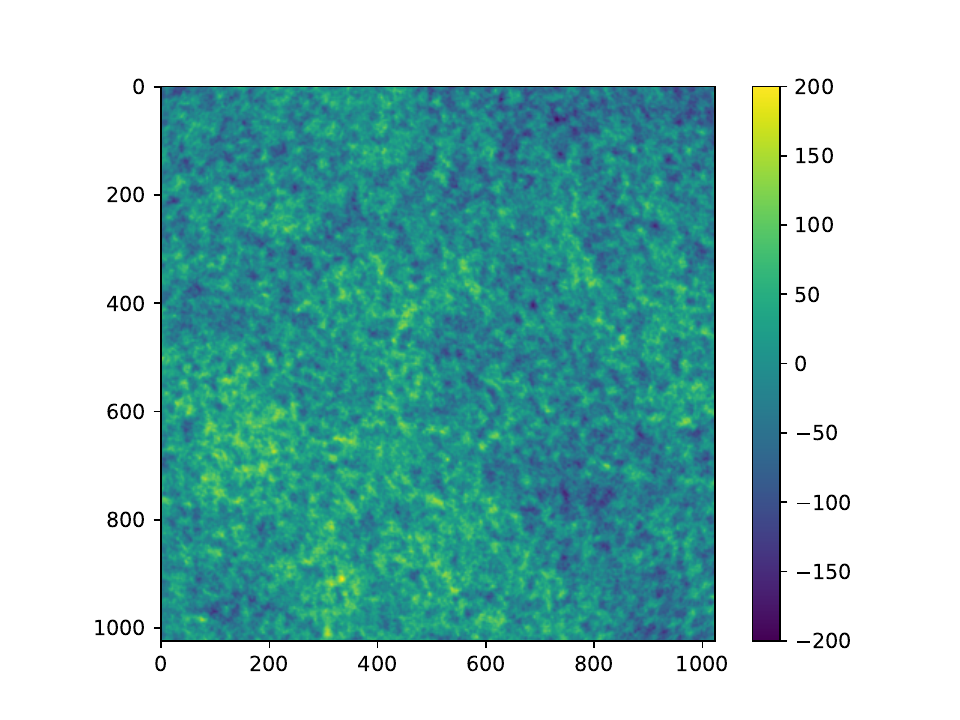} &
    \includegraphics[trim=1.6cm 0.8cm 1.8cm 0.8cm, clip, width=\graphicwidth]{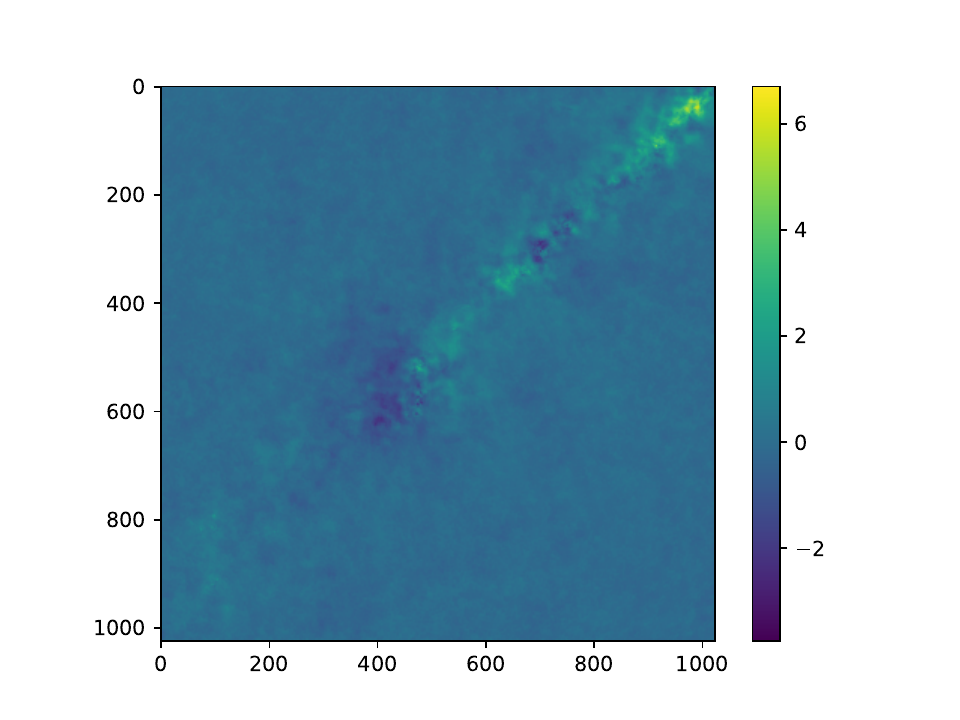} \\
    \addlinespace[1ex]
    \includegraphics[trim=1.6cm 0.8cm 1.8cm 0.8cm, clip, width=\graphicwidth]{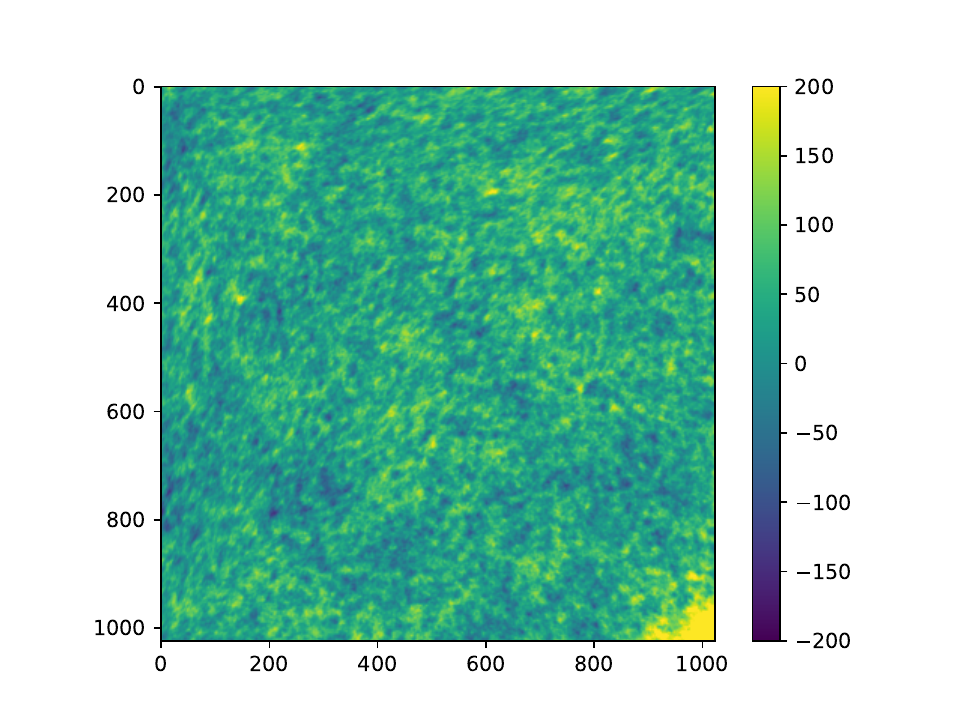} &
    \includegraphics[trim=1.6cm 0.8cm 1.8cm 0.8cm, clip, width=\graphicwidth]{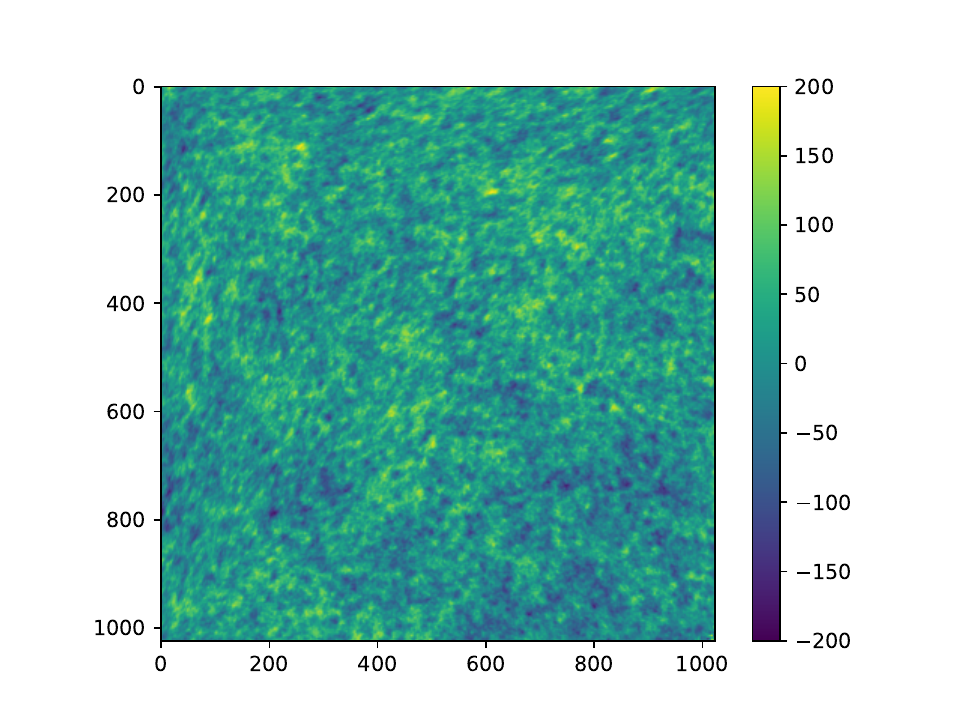} &
    \includegraphics[trim=1.6cm 0.8cm 1.8cm 0.8cm, clip, width=\graphicwidth]{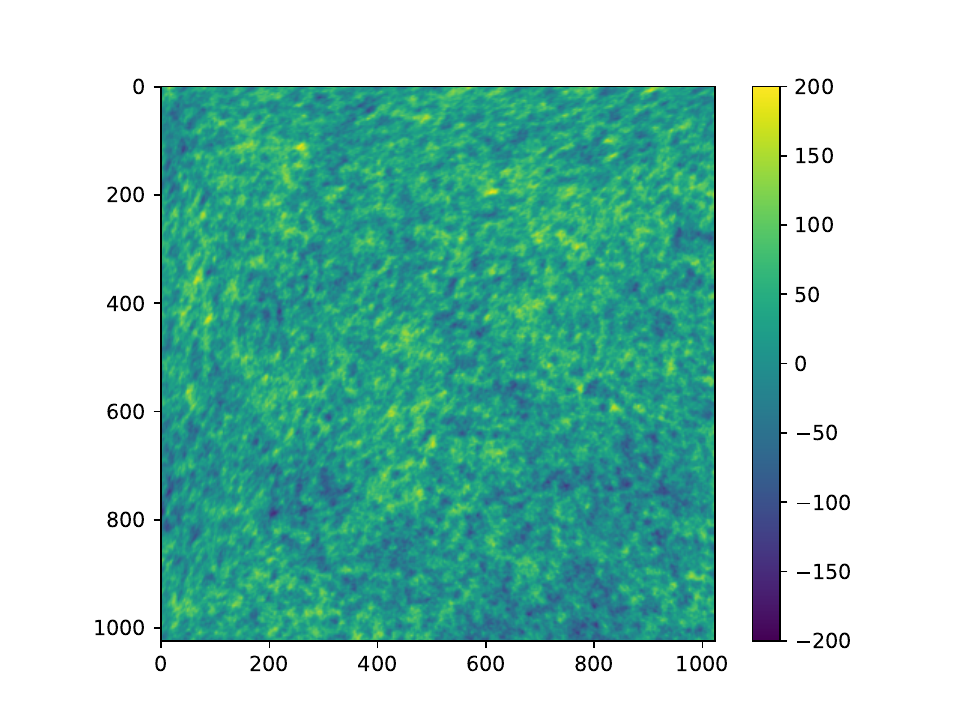} &
    \includegraphics[trim=1.6cm 0.8cm 1.8cm 0.8cm, clip, width=\graphicwidth]{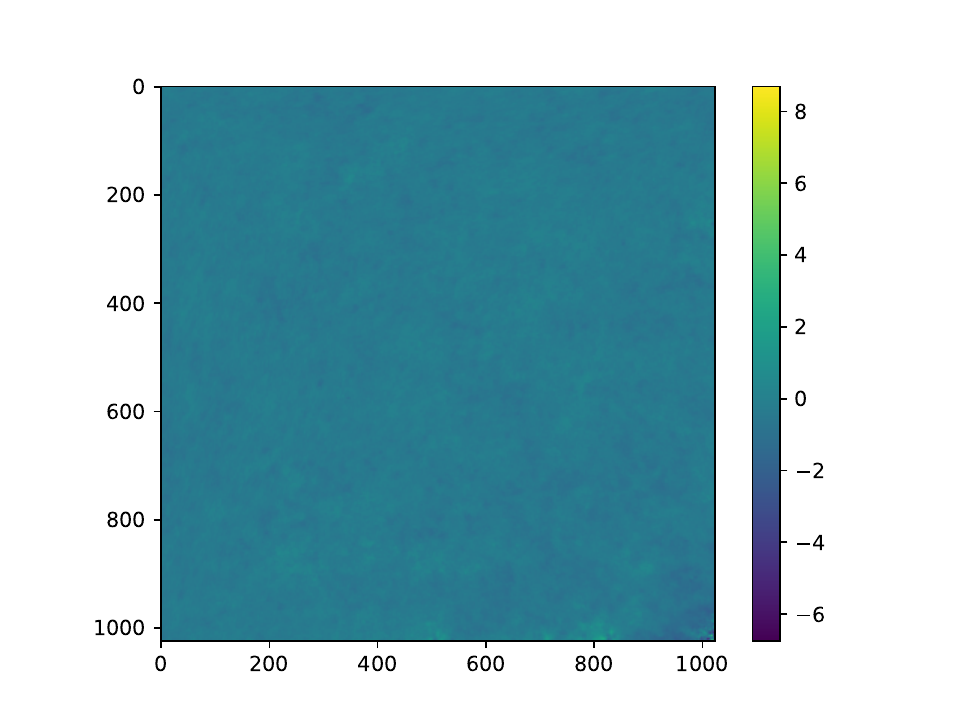} \\
    \addlinespace[1ex]
    \includegraphics[trim=1.6cm 0.8cm 1.8cm 0.8cm, clip, width=\graphicwidth]{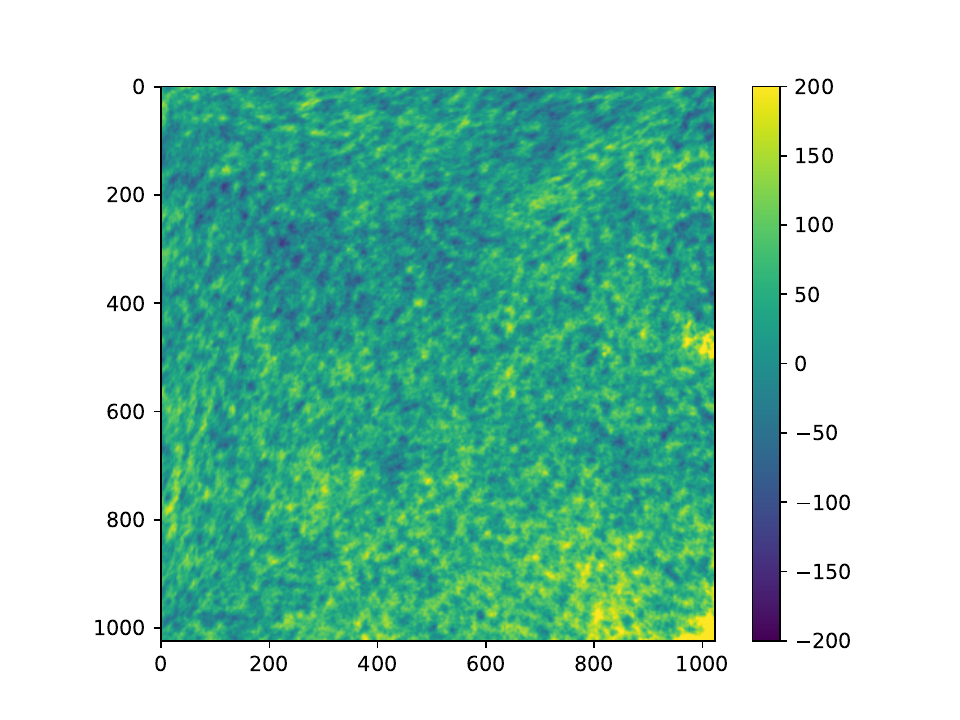} &
    \includegraphics[trim=1.6cm 0.8cm 1.8cm 0.8cm, clip, width=\graphicwidth]{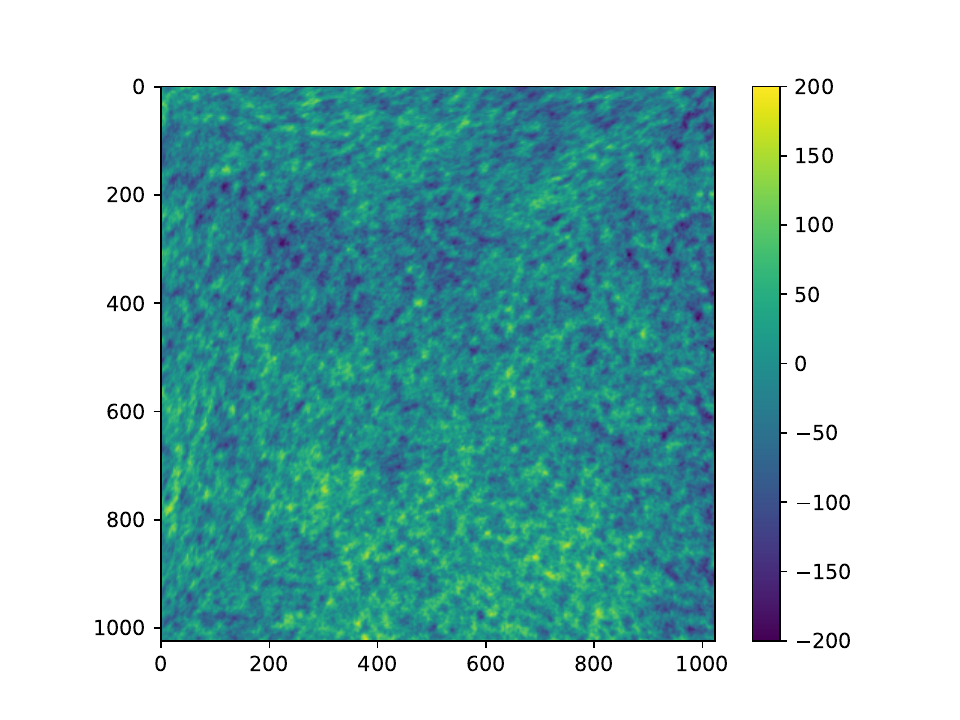} &
    \includegraphics[trim=1.6cm 0.8cm 1.8cm 0.8cm, clip, width=\graphicwidth]{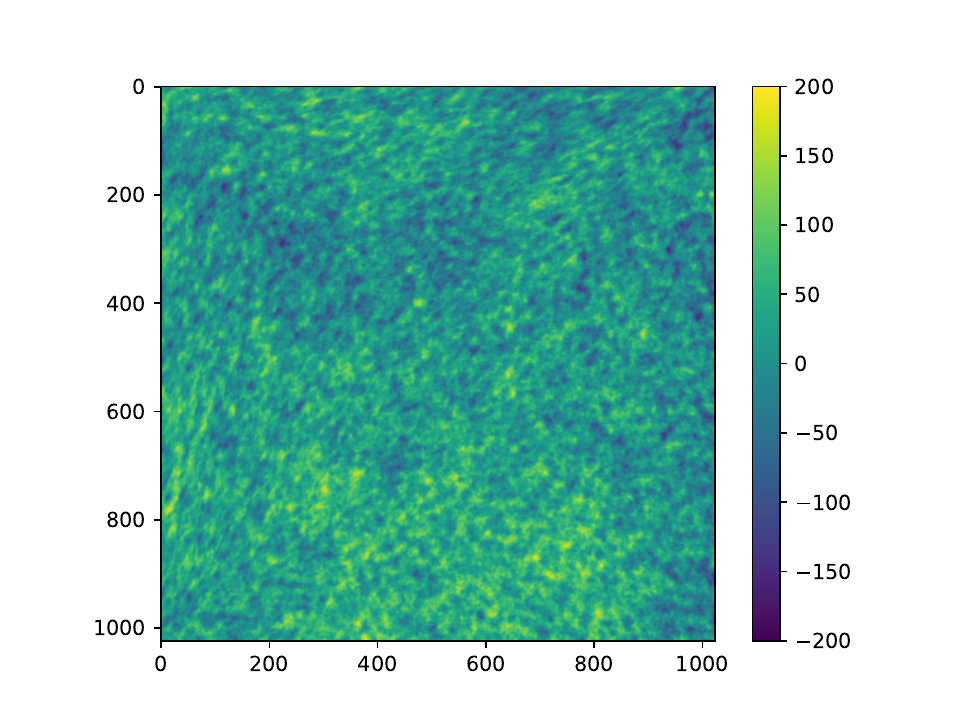} &
    \includegraphics[trim=1.6cm 0.8cm 1.8cm 0.8cm, clip, width=\graphicwidth]{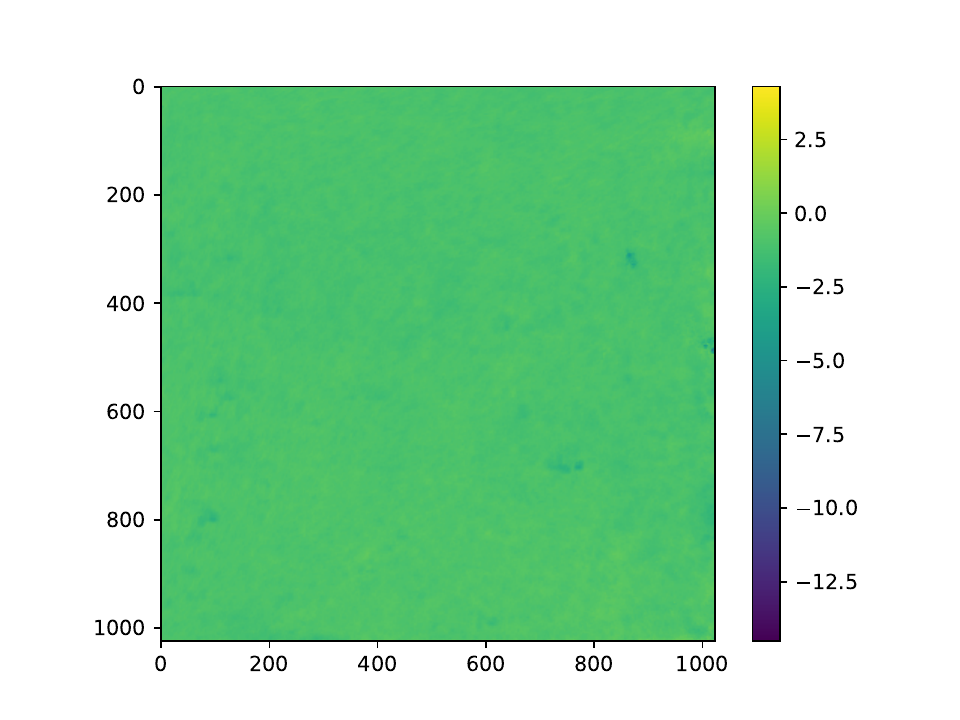} \\
    \addlinespace[1ex]
    \includegraphics[trim=1.6cm 0.8cm 1.8cm 0.8cm, clip, width=\graphicwidth]{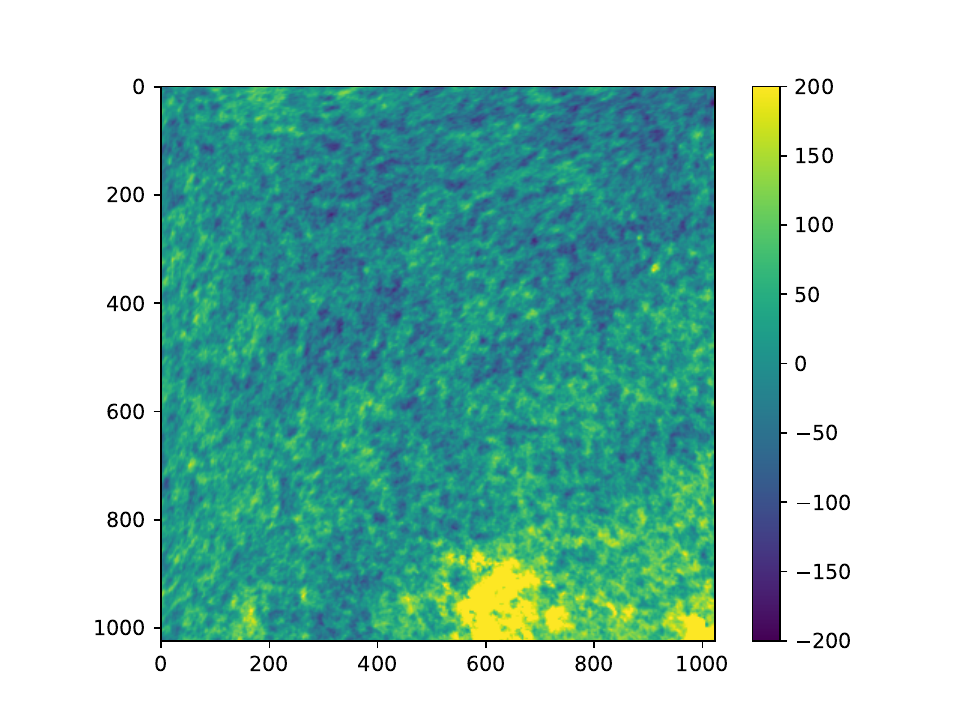} &
    \includegraphics[trim=1.6cm 0.8cm 1.8cm 0.8cm, clip, width=\graphicwidth]{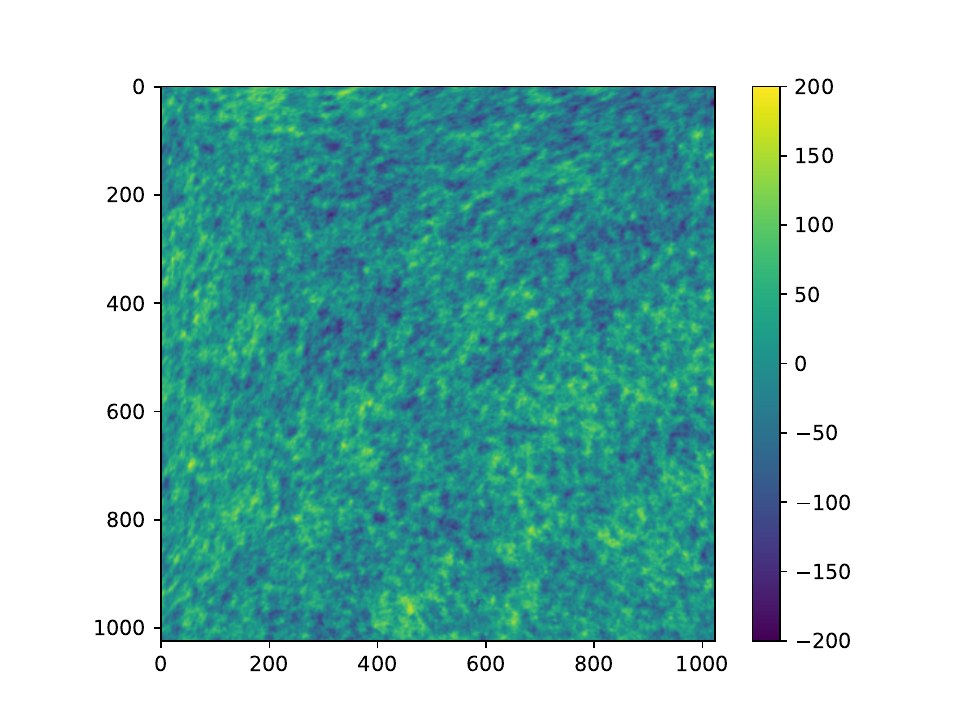} &
    \includegraphics[trim=1.6cm 0.8cm 1.8cm 0.8cm, clip, width=\graphicwidth]{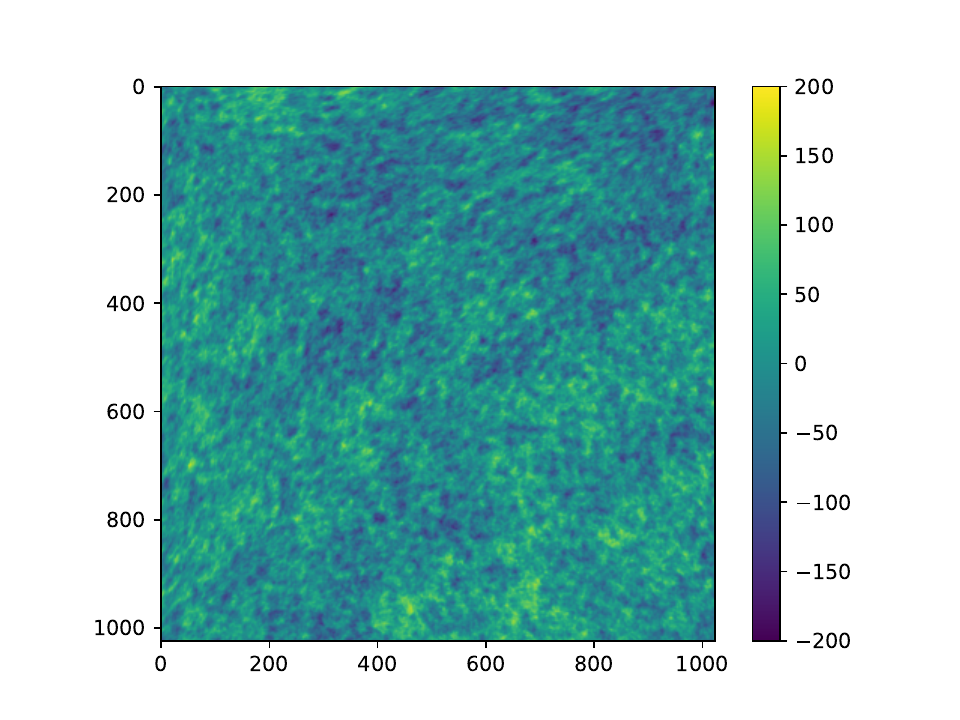} &
    \includegraphics[trim=1.6cm 0.8cm 1.8cm 0.8cm, clip, width=\graphicwidth]{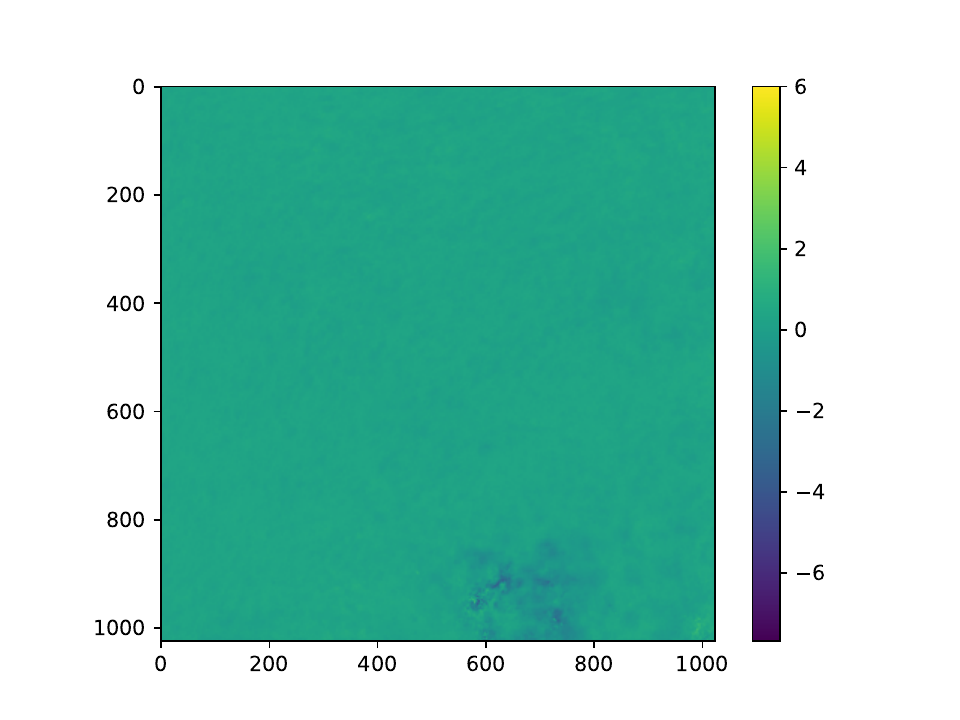} \\
    \addlinespace[1ex]    
    \includegraphics[trim=1.6cm 0.8cm 1.8cm 0.8cm, clip, width=\graphicwidth]{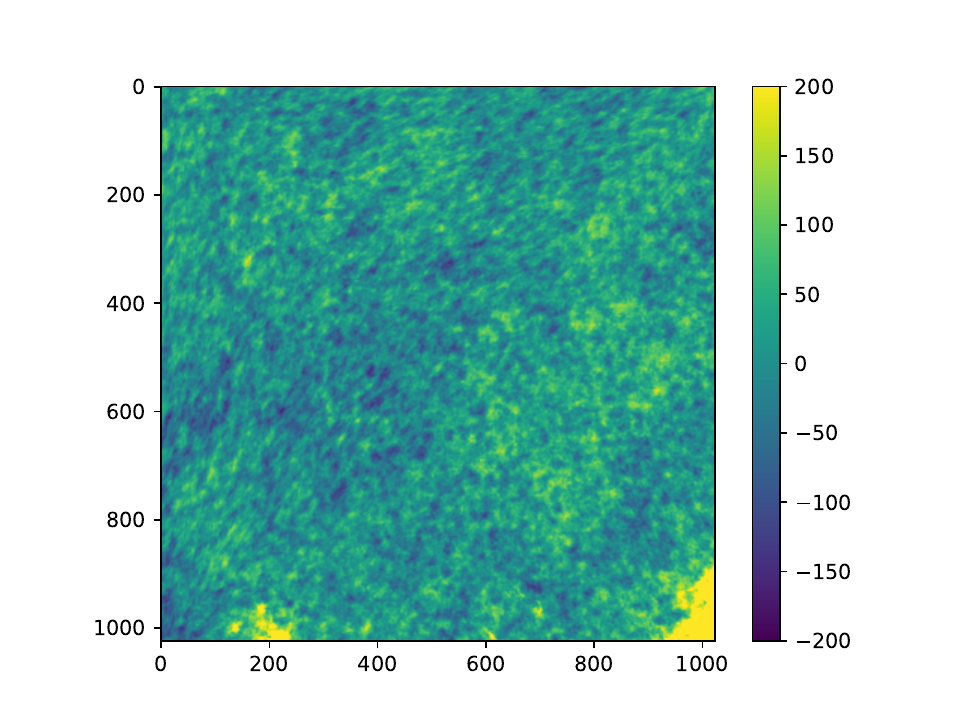} &
    \includegraphics[trim=1.6cm 0.8cm 1.8cm 0.8cm, clip, width=\graphicwidth]{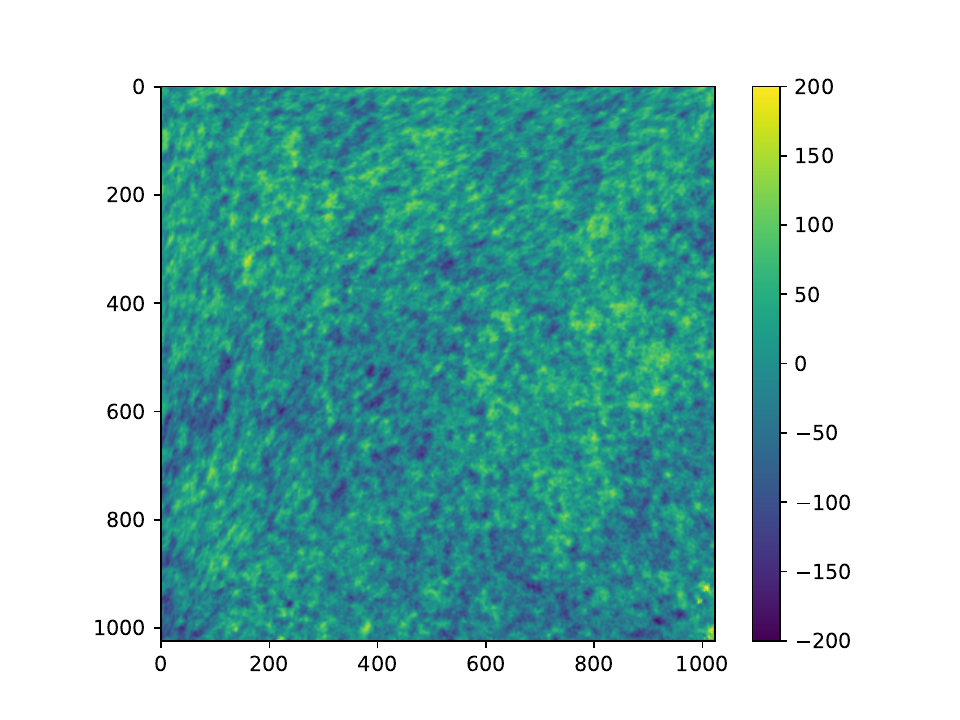} &
    \includegraphics[trim=1.6cm 0.8cm 1.8cm 0.8cm, clip, width=\graphicwidth]{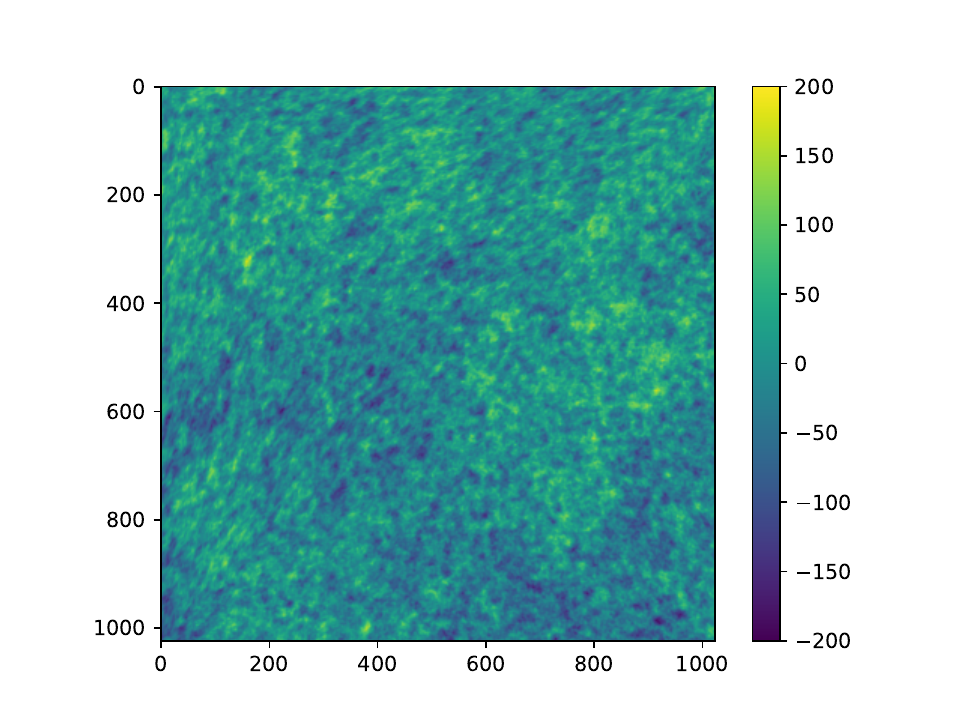} &
    \includegraphics[trim=1.6cm 0.8cm 1.8cm 0.8cm, clip, width=\graphicwidth]{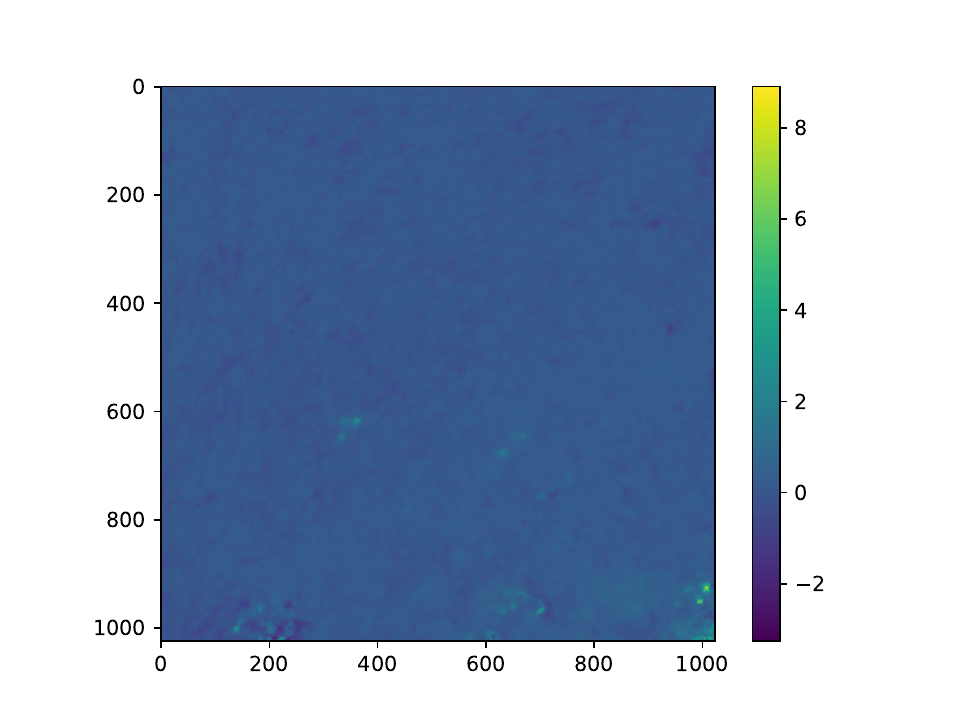} \\
    \bottomrule
  \end{tabular}
\end{table}

\section{Discussion}
\label{sec:discussion} 

The results presented provide important insights into the performance of the U-Net architecture, particularly when augmented with a discriminator and subjected to realistic observational conditions. Key findings emerge regarding the discriminator's role, the impact of model size, and the challenges posed by real beam convolution.

Our comparisons clearly demonstrate the benefit of including a discriminator during training (Section \ref{subsec:discriminator-comparison}). It significantly enhances reconstruction quality by improving the consistency of the reconstructed power spectra with the original data. Figure~\ref{fig:power_spectra_comparison} shows noticeable improvements at low multipole (up to $l \approx 100$) as well as at high multipoles ($l > 900$) compared to the generator-only model.

Concerning model size (Section \ref{subsec:model-size-impact}), initial experiments with Gaussian beam convolution suggested that performance gains diminish beyond $n=8$, with models like $n=8$ and $n=16$ maintaining good accuracy up to $l \approx 1700$ (Figure~\ref{fig:model_size_comparison_error}). The analysis incorporating a real beam convolution (Section \ref{subsec:real-beam-detailed}) reveal a more stringent performance limit under these observational conditions. Using the $n=8$ model, chosen based on the earlier trade-off analysis, we found that fidelity with the ground truth power spectrum is maintained reasonably well only up to $l \approx 1000$ (Figure \ref{fig:real-beam-l-ranges}). Beyond this multipole, the reconstructed spectrum shows increasing deviation, indicating that the model struggles to accurately recover the CMB signal at smaller angular scales when faced with beam smoothing effects. 

It is important to note that for circular beam convolutions, we use different FWHM values for different frequency channels, as listed in Table~\ref{tab:beam_sensitivity}. 
The noise variance is uniform across the sky. In contrast, for the real scan, the noise variance varies spatially, depending on the hit count distribution—further increasing the complexity of accurate reconstruction.


One limitation of this analysis is that the generation of realistic sky maps is computationally intensive, particularly for larger model sizes or higher-resolution maps. In this work, we use maps with a resolution of $n_{\text{side}} = 1024$, whereas the native Planck resolution is $n_{\text{side}} = 2048$. Generating maps at the full Planck resolution would substantially increase both computation time and storage requirements, primarily because of the beam convolution and map synthesis steps. In addition, training the neural network on data at this resolution would further intensify the computational load. For the polarization analysis, we have restricted ourselves to E-mode polarization and to three frequency bands, constrained by available computational resources. Including all frequency maps and both polarization modes would enable us to investigate polarization leakage and assess whether our neural network–based method can effectively mitigate it. However, such comprehensive studies are computationally expensive and have therefore been postponed to future work.

Future improvements to the neural network should focus on enhancing performance beyond $l \approx 1000$ under realistic beam conditions. While implementing model quantization or pruning could address computational efficiency, and advanced training algorithms might reduce runtime, the core challenge lies in improving high-$l$ fidelity. This could involve enhancing the model architecture (e.g., incorporating attention mechanisms, exploring different convolutional approaches suited for deconvolution tasks), refining the loss function to be more sensitive to high-$l$ power spectrum features, or developing multi-stage refinement processes.

These findings underscore the crucial need to balance reconstruction quality with computational feasibility and, critically, to develop deep learning models robust to realistic observational effects like beam convolution. Achieving accurate recovery of cosmological information encoded at smaller angular scales ($l > 1000$) remains a key area for future work in applying neural networks to CMB analysis.

\section{Conclusion}\label{sec:conclusion}

In this work, we demonstrate that a U-Net–based generator combined with a discriminator network offers a powerful framework for reconstructing CMB maps from simulated data. Several recent studies have applied machine learning and neural networks to foreground cleaning in temperature and polarization skymaps~\cite{Petroff_2020,Farsian_2020,yadav2024perceptronbasedilcmethod,casas2025recoveringcmbpolarizationmaps}. However, these works do not take into account the effects of non-circular beams and the instrumental scan pattern. Others use simulated skymaps from Planck~\cite{casas2022cenn}, which inherently include the beam effect, although little attention has been given to understanding its impact. The instrumental effects can leave imprints in statistical measures such as the Bipolar Spherical Harmonic (BipoSH) coefficients~\cite{Das_2016,Pant_2016}. In contrast, our analysis fully incorporates both the non-circular beam effects and the scan strategy, resulting in a more sophisticated and realistic treatment of the data. For the first time, we demonstrate that a neural network–based algorithm can not only clean the foregrounds but also effectively remove the distortions introduced by the noncircular beam and scan pattern. Moreover, most researchers rely on simple U-Net–based architectures for foreground cleaning. In contrast, our approach incorporates a discriminator network alongside the generator, which enhances training efficiency and accelerates convergence. 

To reduce computational cost, we adopt a patching strategy in which the full-sky map is divided into smaller, manageable segments. This approach significantly reduces the memory and time requirements compared to training on the entire sky at once. As map resolution increases, full-sky convolutional training becomes increasingly time-intensive and impractical. However, the patching method helps mitigate this issue. For example, a high-resolution map with $n_{side} = 2048$ can be partitioned into 48 patches of size $1024\times 1024$, allowing parallel training of each patch if sufficient computational resources are available. By forgoing full-sky spherical convolutional techniques like DeepSphere~\cite{defferrard2020deepspheregraphbasedsphericalcnn} in favor of this patchwise approach, we achieve a practical compromise between computational feasibility and foreground cleaning accuracy. Importantly, our analysis shows that the neural network developed in this work does not introduce any statistically significant edge effects, countering concerns raised in previous studies and reinforcing the robustness of the patching strategy~\cite{Yan:2025csf}.

Previous studies on neural-network–based CMB foreground cleaning have primarily relied on the mean squared error (MSE) or $L_2$ loss function alone. Although combining multiple loss functions is a common practice in machine learning, it has not been explored in this context. In our analysis, we incorporate a combination of loss functions and find that including an $L_1$ term alongside the $L_2$ loss leads to more stable training convergence.

While applying neural networks for foreground removal on the CMB sky may not always offer significant speed advantages over traditional statistical methods, particularly given that there is only one CMB sky, the approach holds substantial promise in other domains. In areas such as HI intensity mapping, where large volumes of data across multiple frequency channels are involved, neural networks can drastically accelerate data processing. For instance, tasks like RFI removal, calibration, and solar contamination mitigation in Tianlai observations currently demand considerable computational resources, often taking a full day to process a single day's data using hundreds of processors~\cite{Wu:2020jwm,Zuo:2020tym,Phan:2021xug}. Training a neural network for such tasks is computationally expensive upfront, but once trained, the inference time is minimal, enabling fast, large-scale data cleaning.

Beyond cosmology, similar architectures can be adapted for a variety of applications, including weather data analysis and other time-series or imaging datasets where traditional methods are limited by processing time~\cite{Hanslope2024UsingNN}. Thus, the proposed neural network framework is not only valuable for CMB analysis but also offers a scalable, efficient solution for broader scientific data challenges.

\subsection*{Data and Code Availability}

The codes for generating simulated sky maps by convolving the real beam and Planck scan pattern, can be found in \url{https://github.com/SJaiswal711/cmbnn}. The codes for training and testing the U-Net GAN model is available at 
\url{https://github.com/Obasho10/cmbnn}.

\acknowledgments
Computational resources for this study were provided by the Param Himalaya Supercomputing Facility at IIT Mandi. S.J. acknowledges financial support from Dr. Nirmalya Kajuri’s SERB-funded project (Project No. IITM/SERB/NK/395) and Dr. Vikram Khaire’s SETI-funded project (Project No. PHY2425003SETIKHAI) during the course of this work. S.D. thanks Aditya Rotti and Nidhi Pant for letting us use their code for calculating BipoSH spectrum. 

\bibliographystyle{JHEP}
\bibliography{reference}








\end{document}